\documentclass[traditabstract,doublecolumn]{aa}

\usepackage{subfig}
\usepackage{graphicx}
\usepackage{epstopdf}
\usepackage{deluxetable}
\usepackage{times,epsfig}
\usepackage{mathrsfs}
\usepackage{natbib}
\usepackage{amssymb}
\bibpunct{(}{)}{;}{a}{}{,}
\usepackage{caption}
\usepackage{mathrsfs}

\usepackage[english]{babel}
\usepackage{longtable}
\usepackage{url}
\usepackage{hyperref}

\newcommand{\nosne}{34}
\newcommand{\nosnenir}{26}

\authorrunning{Stritzinger et al.}
\titlerunning{CSP-I Photometry of Stripped Supernovae.}

\begin{document}

\title{The Carnegie Supernova Project~I: \\ photometry data release of low-redshift stripped-envelope supernovae\thanks{Based on observations collected at  Las Campanas Observatory.}}

\author{M.~D. Stritzinger\inst{1,2},
J. P. Anderson\inst{3},
C.~Contreras\inst{2,1},
E.~Heinrich-Josties\inst{4,5},
N.~Morrell\inst{2},
M.~M.~Phillips\inst{2},
J. Anais\inst{2},
L. Boldt\inst{2},
L. Busta\inst{6},
C.~R.~Burns\inst{5},
A. Campillay\inst{2},
C. Corco\inst{7,2},
S.~Castellon\inst{2},
G.~Folatelli\inst{8,2},
C. Gonz{\'a}lez\inst{2},
S. Holmbo\inst{1},
E. Y. Hsiao\inst{9,1,2},
W.~Krzeminski\inst{10,2},
F.~Salgado\inst{11,2},
J. Ser{\'o}n\inst{7,2},
S. Torres-Robledo\inst{12,2},
W.~L.~Freedman\inst{13,5},
M.~Hamuy\inst{14,2},
K.~Krisciunas\inst{15},
B.~F.~Madore\inst{5,16},
S.~E.~Persson\inst{5},
M.~Roth\inst{2},
N.~B.~Suntzeff\inst{15},
F. Taddia\inst{17},
W. Li\inst{18,19}, and
A. V. Filippenko\inst{18,20}
}

\institute{Department of Physics and Astronomy, Aarhus University, Ny Munkegade 120, DK-8000 Aarhus C, Denmark\\ (\email{max@phys.au.dk}) 
\and
Carnegie Observatories, Las Campanas Observatory, Casilla 601, La Serena, Chile
  \and
European Southern Observatory, Alonso de Cordova 3107, Vitacura, Casilla 19001, Santiago, Chile
 \and
Las Cumbres Observatory, 6740 Cortona Dr., Suite 102, Goleta, CA 93117, USA
\and
Observatories of the Carnegie Institution for Science, 813 Santa Barbara St., Pasadena, CA 91101, USA
\and
 Institut f{\"u}r Astro- and Particle Physics, University of Innsbruck, A-6020 Innsbruck, Austria
 \and
 Cerro Tololo Inter-American Observatory, Casilla 603, La Serena, Chile
 \and
 Facultad de Ciencias Astron\'{o}micas y Geof\'{i}sicas, Universidad
Nacional de La Plata, Paseo del Bosque S/N, B1900FWA
La Plata, Argentina
\and
Department of Physics, Florida State University, 77 Chieftain Way, Tallahassee, FL, 32306, USA
  \and
  N. Copernicus Astronomical Center, ul. Bartycka 18, 00-716, Warszawa, Poland
\and
Leiden Observatory, Leiden University, PO Box 9513, NL-2300 RA Leiden, The Netherlands
\and
SOAR Telescope, Casilla 603, La Serena, Chile
\and
Department of Astronomy \& Astrophysics, University of
Chicago, 5640 South Ellis Avenue, Chicago, IL 60637, USA
\and
Departamento de Astronomia, Universidad de Chile, Casilla 36D, Santiago, Chile
\and
George P. and Cynthia Woods Mitchell Institute for Fundamental Physics and Astronomy, Department of Physics and Astronomy, Texas A\&M University, College Station, TX 77843, USA
 \and
 Infrared Processing and Analysis Center, Caltech/Jet Propulsion Laboratory, Pasadena, CA 91125, USA
 \and
The Oskar Klein Centre, Department of Astronomy, Stockholm University, AlbaNova, 10691 Stockholm, Sweden
\and
Department of Astronomy, University of California, Berkeley, CA 94720-3411, USA
\and
Deceased 12 December 2011
\and
Senior Miller Fellow, Miller Institute for Basic Research in Science, University of California, Berkeley, CA 94720, USA
}

\date{Received 22 March 2017 / Accepted XX August 2017}
 
\abstract{
The first phase of the {\em Carnegie Supernova Project} (CSP-I) was a dedicated supernova follow-up program based at the Las Campanas Observatory that collected science data of young, low-redshift supernovae between  2004 and 2009.
Presented in this paper is the CSP-I photometric data release of low-redshift stripped-envelope core-collapse supernovae.
The data consist of optical ($uBgVri$) photometry of \nosne\ objects, with a subset of \nosnenir\ having  near-infrared ($YJH$) photometry. 
Twenty objects have optical pre-maximum coverage with a subset of 12 beginning at least five days prior to the epoch of $B$-band maximum brightness.
In the near-infrared,  17 objects have pre-maximum observations with
a subset of 14 beginning at least five days prior to the epoch of $J$-band
maximum brightness.
 Analysis of this photometric data release is  presented in  companion papers focusing on techniques to estimate host-galaxy extinction  (Stritzinger et al., submitted)  and the light-curve and progenitor star properties  of the sample (Taddia et al., submitted). The analysis of  an accompanying visual-wavelength spectroscopy sample of $\sim 150$ spectra will be the subject of a future paper.}


\maketitle
\section{Introduction}
\label{sec:intro}

The first phase of the {\em Carnegie Supernova Project}  (CSP-I; \citealt{hamuy06}) was
an  observational supernova (SN) follow-up program
that collected data on most varieties of supernovae (SNe) between 2004 to 2009,
mainly using facilities located at the Las Campanas Observatory (LCO).
A key goal of the CSP-I was to build a well-observed sample of
stripped-envelope (SE) core-collapse SNe in a
stable and well-characterized photometric system.
SE~SNe (which we define as Type~IIb, Type~Ib, Type~Ic, and Type Ic-BL 
SNe, collectively)
 are thought to be linked to the death of massive stars which experience significant mass loss over their evolutionary lifetimes.
Contemporary estimates  from a volume-limited sample obtained during
the Lick Observatory Supernova Search 
\citep[LOSS;][]{filippenko01,filippenko05} with the 
Katzman Automatic Imaging Telescope 
suggest that SE~SNe account for
 $\sim 26$\% of the overall SN rate in the local Universe \citep{li11}.

The presence of residual hydrogen and helium maintained by the progenitor at the time of explosion, and how it manifests itself in the optical spectrum of the SN, provides the basis for today's spectral classification scheme of SE~SNe \citep[see, e.g.,][]{filippenko97, galyam16, liu16, modjaz16, shivvers17, prentice17}.
Based mostly on  historical observations, the current classification system
has become somewhat  ambiguous  with the advent of detailed spectroscopic
follow-up campaigns of low-redshift objects.
SE~SNe are grouped into subtypes  based on spectroscopic signatures ranging 
 from helium rich with traces of hydrogen (Type IIb), to helium rich with no traces of hydrogen (Type Ib), to helium and  hydrogen deficient (Type Ic). 
 Early-time spectra of SNe~IIb exhibit  prevalent
P-Cygni Balmer lines akin to normal  hydrogen-rich SN~II; however, by several weeks past explosion their spectrum experiences a metamorphosis and subsequently resembles that of a classical SN~Ib.
Given these unique characteristics, such transitional objects are
designated SNe~IIb, of which
SN 1987K was the first known example \citep{filippenko88} and
SN~1993J is the prototypical example \citep[e.g.,][]{filippenko93,swartz93}.
In both SNe~IIb and SNe~Ib,  \ion{He}{i} features tend to   dominate the spectrum around a month past explosion \citep{harkness87,matheson01}.
The  current classification scheme  therefore reflects to first order
the  mass-loss history of the progenitors with increased stripping driving
the SN~IIb$\rightarrow$SN~Ib$\rightarrow$SN~Ic spectroscopic sequence.

Initially, SNe~Ib and Ic garnered relatively little interest within the astronomical community.
 This  changed in 1998, when the gamma-ray
 burst (GRB) 980425  was demonstrated to be linked to the
 peculiar Type~Ic SN 1998bw \citep{galama98,iwamoto98}, which showed very broad spectral lines.
  Five years later the GRB--SN connection was firmly established
  \citep{hjorth03,matheson03,stanek03,mazzali06}.
   In each instance of a GRB--SN association, the underlying ``relativistic" SN exhibited
  broad absorption features indicative of expansion velocities in excess of
  $\sim$ 30,000 km s$^{-1}$, and explosion energies of $\gtrsim$~$10^{52}$ erg.
  These events are thought to be powered by a central engine or a relativistic outflow, and constitute only $\sim 1$\% of the total SE~SN population.

For comparison, the vast majority of {\em normal}
   SE~SNe typically exhibit expansion velocities of $<$~15,000~km~s$^{-1}$, and
 kinetic energies of  $\sim 10^{51}$~erg.
 Interestingly, an additional subtype has been identified that shows
enhanced expansion velocities ($\sim$ 20,000~km~s$^{-1}$), but whose kinetic energies are equivalent to those of normal SE~SNe. Given their spectral characteristics these
objects are commonly referred to as BL (broad-lined).
SN~2002ap serves as the prototypical BL SN~Ic \citep{mazzali02,foley03}, while SN~2003bg is the first identified BL SN~IIb \citep{hamuy09,mazzali09}.
 Finally, a handful of peculiar events has been studied that do not follow the light curve and/or spectral templates garnered from the normal SE~SN  population.
 These objects may, as in the case of SN~2005bf  (see light curves below),  exhibit bizarre double-humped light curves \citep[see also][]{anupama05,tominaga05,folatelli06}, or in the instances of SN~2001co, SN~2005E, and SN~2005cz, enhanced calcium abundances \citep{filippenko03,perets10,kawabata10}.

Although the identity of SE~SN progenitors  has long remained rather inconclusive, the early-time light-curve evolution of the Type~IIb SN~1993J exhibited an initial declining phase right after its explosion, followed by a rise to peak luminosity driven by the decay of $^{56}$Ni within several weeks \citep[e.g.,][]{vandriel93,richmond94}.
{This light-curve evolution  is indicative  of a massive progenitor that explodes via core collapse.
A similar early-time bump was also detected in the  Type~Ib SN~1999ex \citep{stritzinger02}, and in recent years early bumps have been used to constrain the radius of the progenitor star in a number of cases, such as SN~2008D \citep{chevalier08,bersten13}, SN~2011dh  \citep{arcavi11,soderberg12,bersten12},  SN~2011fu \citep{kumar13}, SN~2011hs \citep{bufano14}, SN~2013df \citep{morales14}, and 
SN~2016gkg \citep{tartaglia17,arcavi17}.
 
Efforts to model SE~SNe usually assume
progenitor systems  encompassing either
 single massive ($>20$ M$_{\sun}$) Wolf-Rayet stars,
or moderately massive (8--25 M$_{\sun}$) stars in binary systems
\citep[see, e.g., ][and references therein]{gaskell86,shigeyama90,podsiadlowski93,nomoto93,woosely93,iwamoto94,utrobin94,woosely95,thielemann96,heger03,fryer07,georgy09,dessart11,benvenuto13,bersten14,dessart16, sukhbold16}.
Stars within the former  scenario are thought to shed the majority of their hydrogen-rich envelopes via line-driven winds, while stars within the latter scenario  are thought to lose their hydrogen-rich envelopes through Roche-lobe overflow (RLOF) to their companions and/or during common-envelope evolution.
 Finally, rotational induced mixing has been suggested to play a role in growing the core mass at the expense of the envelope mass \citep[see, e.g.,][]{langer12}.
In the near future, robust constraints on the expelled gas will be possible by studying  spectral features associated with circumstellar material flash-ionized by ultraviolet light from the SN shock breakout in very early spectra, within one day or a few days of explosion \citep[e.g.,][]{galyam14, shivvers15, yaron17}.


Despite the heroic efforts to model SE~SNe, the evolutionary scenarios of
 the various spectroscopically defined subtypes such as SN~IIb, Ib, Ic, 
and Ic-BL (broad-lined) are not completely understood.
Ideally, the  putative detection of the progenitor  star in
pre-explosion images \citep{smartt09,eldridge13} provides
the gold standard; however, these  opportunities are rare, often controversial, and to date exist for only a handful of events including the Type~IIb
SN~1993J \citep{aldering94,cohen95},
SN~2008ax \citep{li08,crockett08},
SN~2011dh \citep{maund11,vandyk11,ergon14}, 
 iPTF~13bvn \citep{cao13, groh13}, and
SN~2013df \citep{vandyk14}.

An alternative avenue is to constrain the nature of SE~SN progenitors by observing the site of the explosion with the {\em Hubble Space Telescope (HST)} hundreds of days to a few years after the SN exploded.
For example,  deep {\it HST} imaging of iPTF~13bvn indicates a late-time ($+$740 days) SN brightness below that measured for the candidate progenitor star in pre-explosion {\it HST} images \citep{folatelli16, eldridge16}. 
 \citet{folatelli16} assert that with these observations, the characterization of the progenitor star is currently inconclusive, and the progenitor's companion was either a late-O star  or suffers from significant extinction produced by newly formed dust. On the other hand, \citet{eldridge16} suggest (from the same data) that the  companion star had a  mass of  $\sim 10$~M$_{\sun}$ or less, and that the progenitor itself was possibly a 10--12 M$_{\sun}$ helium giant star.

Similarly, {\it HST} images of the  Type~IIb SN~2011dh obtained $\sim 1160$ days after explosion
reveal a blue source that may have been the companion star \citep{folatelli14} to a yellow supergiant prior to its death \citep{bersten12,benvenuto13}, though this interpretation has been challenged \citep{maund15}. 
Of course, these cases follow in the footsteps of the well-studied Type~IIb SN~1993J, whose massive companion was identified in late-phase {\it HST} imaging \citep{maund04,fox14}.

Taken together, the current literature on SE~SNe suggests that  at least a portion of these cosmic explosions arise  from binary-star systems.
Contemporary studies  of hot and cool, single, massive
stars indicate mass-loss rates  $\sim 2$--10 times less than those typically adopted
in  late-stage stellar evolutionary models \citep[][and references therein]{puls08}.
When considering revised  mass-loss rates, in combination with the observed
SN rates,  one obtains  rather strong (albeit indirect) constraints that the majority of
SE~SN progenitors are moderately massive stars in binary systems \citep{smith11}.
 Moreover, the implied lower mass-loss rates suggest that single  massive stars
 require more efficient mass-stripping mechanisms in order to
 shed their outer  hydrogen and helium layers prior to exploding.
Potential mechanisms that have been suggested to drive enhanced mass loss include  pulsation-driven superwinds \citep[e.g.,][]{yoon10} and episodic luminous blue variable (LBV)  outbursts during the pre-SN evolution \citep[e.g.,][]{smith06,smith11,vink11}.

 Given the lack of constraints on SE~SN progenitor properties from direct detection methods, investigations of their local environments  have  been pursued.
\citet{anderson12} found that SNe~Ic are projected on regions of higher star-formation intensity within host galaxies than both SNe~Ib and SNe~II, and suggested  this implies higher-mass progenitors for the SNe~Ic (with an only marginal difference of environment found between the SNe~Ib and SNe~IIb).
Host \ion{H}{ii} region metallicity studies
of SE~SNe have attempted to probe progenitor metallicity differences;
however, results have thus far been inconclusive as to whether
metallicity inferred from line diagnostics \citep[e.g.,][]{pettini04} significantly affects the final outcome of progenitor stars through the role of line-driven winds; see \citet{anderson10}, \citet{modjaz11}, \citet{leloudas11}, and
\cite{sanders12}. 
 A  compilation of all literature measurements is presented in  \citet{galbany16}. 
Most recently, \citet{graur17a} revisited the LOSS sample in order to search for correlations between SN rates and galaxy properties, while  \citet{graur17b} compare the relative rates of different SN types to the masses of the host galaxies.

Until recently, the  literature sample of SE SNe with multiband light curves was rather limited, with only a handful of objects  having been well studied.
However, with an enhanced interest in this SN class and the advent of
monitoring programs, detailed studies of nearby objects are  more common.
For example, \citet{drout11} presented $V$- and $R$-band light curves of 25 SE SNe. Combined with the best-observed objects in the literature, they were able to study a statistically significant sample.
Most recently,  \citet{bianco14} and \citet{taddia15} have presented expanded  sets of multiband SE~SN light curves obtained by the CfA Supernova Group and the  SDSS-II SN survey, respectively, while detailed studies of literature-based samples have been presented by \citet{cano13}, \citet{lyman16}, and \citet{prentice16}. These papers are discussed in a companion paper \citep{taddia17}, which also contains a detailed study of the CSP-I SE~SN light-curve sample. 
It includes the estimation of key explosion parameters by comparing best-fit analytical and hydrodynamical modeled bolometric light curves to bolometric light curves constructed from the broad-band observations presented here.
In short, consistent with previously studied SE~SN samples, we estimate the $^{56}$Ni abundance, the ejecta mass ($M_{\rm ej}$), and the explosion energy ($E_{K}$). 
Quantitatively, for the majority of normal SE~SNe, these parameters range from $\sim 0.03$ to 0.38 M$_{\sun}$ of radioactive $^{56}$Ni,  $\sim 1.1$ to 6.2 M$_{\sun}$ of $M_{\rm ej}$, and (excluding SNe~Ic-BL) $E_{K}$ between $0.25\times10^{51}$ and $3.0\times10^{51}$ erg \citep{taddia17}.

Continued advances in the understanding of SE~SNe will come from the detailed studies of {\em homogeneous} samples.
New datasets will enable (i) improved methods to estimate
host-galaxy dust extinction,  (ii) the construction of  accurate
ultraviolet--optical--near-infrared (UVOIR) bolometric light curves useful for model comparisons, and (iii) better knowledge of their spectroscopic properties.
Furthermore, with the advent of new, high-cadence surveys, observations of SE~SNe discovered just after explosion combined with improved models \citep[e.g.,][]{bersten12, nakar14, piro15, sapir17, piro17} will provide better constraints on the progenitors and their pre-SN evolution.

In this paper, we  present  photometry of the CSP-I sample of
low-redshift SE~SNe.
 Our dataset consists of optical  light curves
of  $\nosne$ SE~SNe, with a  subset of \nosnenir\ objects also
having  accompanying near-infrared (NIR) light curves.
Three of these objects ---
SN~2005bf \citep{folatelli06}, SN~2007Y \citep{stritzinger09}, and SN~2009bb \citep{pignata11} --- have been the subject of individual case studies by the CSP-I.
 The photometry reported here of these objects supersedes what has been previously published, and is based on definitive color and extinction coefficient terms characterizing the CSP-I natural photometric system as presented by \citet{krisciunas17}. These updates have minimal differences compared to the originally published photometry and in no way affect any results previously presented.

This   paper is accompanied by three  companion papers.
In  \citet{stritzinger17b}, our SE~SN photometry   is used to devise robust methods to estimate host-galaxy extinction.  \citet{taddia17} use the photometry to construct UVOIR bolometric light curves which are modeled to estimate key explosion parameters.
Finally, \citet{holmbo17} present a detailed study of the accompanying sample of visual-wavelength spectroscopy.


\section{The Sample}
\label{sample}

The CSP-I performed follow-up observations on all varieties of southern and equatorial SNe discovered by both professional and amateur-led search programs.
The general restrictions considered to place an object in  the follow-up queue were its temporal phase and brightness.
Selected targets were  thought to have been discovered prior to or just after maximum luminosity and were estimated to reach peak apparent brightness no fainter than 18 mag.
Owing to these observational guidelines and the methods used by the  search programs that discovered the SNe, the majority of the sample is located rather nearby (redshift $z \leq 0.03$).
This is demonstrated in the redshift distribution presented as a stacked histogram in Figure~\ref{redshift}, where only two objects within the sample are located at $z > 0.035$.
Given that the majority of the CSP-I SE~SNe were discovered by targeted surveys, the sample is biased, as the targeted search strategy leads to an underrepresentation of segments of the population, particularly objects located in faint, metal-poor galaxies \citep{arcavi10}.

General properties of each of the objects in the CSP-I sample are summarized in Table~\ref{table1}, including (i) their J2000.0 coordinates, (ii) host-galaxy type and morphology as provided by NED, (iii) the heliocentric redshift as listed in NED or  as determined by our own spectral analysis, (iv) a reference to the reported discovery,  (v) the group or individual(s) credited with the discovery,
(vi) the spectral classification of each object, (vii) the phase range of the obtained spectra,  (viii) and an estimate of the epoch of  $B$-band maximum brightness (see below).

 Spectral classification was made through the careful inspection of all available spectra \citep[see][]{holmbo17}.
 The decision to classify a particular object as  a SN~IIb or  a SN~Ib was based on
 the strength of H$\alpha$ and the presence and strength of any conspicuous \ion{He}{i} absorption features.
 If H$\alpha$ was prominent while  \ion{He}{i} was either weak or absent, a SN~IIb
 classification was preferred, while weak H$\alpha$ and  conspicuous \ion{He}{i}
absorption yielded a SN~Ib subtyping.
However, these  criteria can be misleading, as they are highly dependent
on the temporal phase of the earliest spectrum used to make the classification
\citep{stritzinger09,chornock11,milisavljevic13,shivvers17}.
For example,  it is possible that an object classified as a SN~Ib based on a spectrum taken at maximum brightness or even two weeks earlier could  have appeared as a SN~IIb soon after explosion. 
The temporal phases of the spectroscopic sample listed in Table~\ref{table1} indicate there are a few such cases: SN~2004ew, SN~2006ep, and SN~2008gc.
 Similarly, attempting to make a distinction between some SNe~Ib and SNe~Ic was not always clear, and (as previously mentioned) highly dependent on the number of spectra and the breadth of the temporal phase covered. This includes SN~2005em, SN~2007ag, and SN~2009dt, whose spectroscopic coverage does not extend sufficiently far past maximum brightness to ensure the lack of \ion{He}{i} features.
 With these caveats in mind, Table~\ref{table1}  includes our best
 spectral classifications for each object within the sample, and we note that these classifications agree with best-fit matches made with SNID (SuperNova IDentification; \citealt{blondin07}) using both standard and expanded template sets \citep[][]{liu14}.
 The final tally is as follows:  10 SNe~IIb, 11 SNe~Ib, 11 SNe~Ic, and 2 SNe~Ic-BL.

Figure~\ref{FC} contains a single $V$-band image for  each of the SE~SNe in the CSP-I obtained with the Henrietta Swope 1~m telescope at Las Campanas
Observatory (LCO).
Each image is oriented such that the SE~SN, marked with a blue circle, is located at the center of the field.
 In addition,  visible local sequence stars are indicated with red squares. 

\section{Optical and NIR Passbands}
\label{csppassbands}

 The majority  of imaging of the CSP-I SE SN sample was performed by the LCO Swope 1~m telescope equipped with a direct CCD camera named after its detector ``SITe3" and a NIR imager specifically built for the CSP-I named ``RetroCam".
 A significant amount of NIR imaging was also performed by
 the LCO Ir\'en\'ee  du Pont 2.5~m telescope equipped with the Wide Field IR Camera (WIRC),  and a small amount of optical imaging was performed with the facility Tek5 CCD camera using the same filter set used with the Swope$+$SITe3 telescope. 
 Readers are  referred to  \citet{hamuy06} for details on CSP-I instrumental setups and observing procedures.

The CSP-I carried out a  detailed photometric calibration program in order to accurately measure the Swope$+$SITe3 and Swope$+$RetroCam and the du Pont$+$WIRC system (telescope$+$instrument$+$filters) response functions.
Unfortunately no scans were made for the du Pont$+$Tek5, however, after some detailed experimentation we have found that the du Pont$+$Tek5 system response functions are in complete agreement with the  Swope$+$SIT3 system response functions \citep[see][]{krisciunas17}. 

A detailed account of  the experimental setup  used to conduct the measurements of the system response functions is presented by \citet{rheault10},  and the  optical ($uBgVri$ band) system response functions are presented by \citet{stritzinger11}.
 NIR  ($YJH$ band) system response functions 
 are presented here, and in a companion paper presenting  the final CSP-I Type~Ia supernova light-curve data release \citep[][]{krisciunas17}\footnote{The final CSP-I SNe~Ia data release paper  contains complete details  concerning our natural photometric systems including robust measurements of  extinction-term coefficients,  color-term coefficients,  and photometric zero-points.}.
Here we provide a brief overview of the instrumental calibration project and present definitive  system response functions used to observe the CSP-I SE SN sample.\footnote{Optical/NIR system response functions are available in ASCII format on the CSP-I webpage located here: \href{http://csp.obs.carnegiescience.edu/}{http://csp.obs.carnegiescience.edu/}}  

The experimental setup consists of an appropriate light source connected to a monochromator that allows one to step through wavelength in a narrow bandwidth of light \citep[see][for detailed specifics]{stritzinger11}. 
The light is fed from the monochromator through a fiber optic bundle and projected onto a highly reflective flat-field screen. 
Light entering the telescope  is measured by a system of calibrated germanium (optical: 2500--10,000~\AA) and InGaAs (NIR: 8000--16,500~\AA) photodiodes placed behind the secondary mirror. 
 Exposures are  taken with the facility instrument and the resulting signal  is compared to the incoming signal measured by the photodiodes. This is a relative measurement that yields  the system response function. The process is  performed in a  scanning manner as the monochromator is used to step through wavelength.
 
 Scanning was performed over the course of two calibration runs and  each passband was scanned a minimum of  two times per  run. 
The repeatability of the measured optical system response function indicates the relative throughputs were measured with a  precision of  $1\%$ or less \citep[see][]{stritzinger11}, while repeatable scanning of the NIR system response functions indicate relative throughputs were measured with a precision of 2-3\%. 

Plotted in Figure~\ref{passbands} are the scanned optical (top panel) and   NIR (bottom panel) system response functions used to obtain data of the CSP-I SE SN sample. 
The scanned system response functions have been multiplied by an atmospheric transmission function for an airmass of 1.2 and a telluric absorption spectrum appropriate to LCO.  The optical response functions  correspond to the filter set used with the Swope ($+$SITe3) telescope, and as described in \citet{stritzinger11}, the original $V$-band filter (``$LV-3014$") broke on 14 January 2006 UT and was replaced with a similar $V$-band filter (``$LV-9844$") on 25 January 2006 UT. 
From our photometric calibration program it has been determined that the same color term applies to transform  natural  $LV-3014$ and $LV-9844$ system photometry to the standard Landolt $V$-band photometric system, and therefore any differences in photometry obtained with the passbands is well below 1\%.

Turning to the NIR, the $YJH$-band system response functions shown in Figure~\ref{passbands} correspond to filter sets installed in RetroCam and WIRC. 
Inspection of the RetroCam and WIRC system response functions reveals  $J$ and $H$ throughputs, while clear differences are apparent between the  two $Y$ system response functions \citep[see][for details]{krisciunas17}.  
The $Y$-band was originally introduced by  \citet{hillenbrand02} and is calibrated relative to $Y$-band magnitudes of the \citet{persson98} standard stars.
The $Y$ magnitudes of the   \citeauthor{persson98} standards are computed 
 using  $(Y-K_s)$ and $(J-K_s)$ color relations derived from a grid of Kurucz synthetic stellar atmosphere models with the added requirement  that 
 $(Y-K_s) = 0$  when $(J-K_s) = 0$ \citealt[see][]{krisciunas17}. 
 This stipulation is an agreement with the primary standard star Vega  being characterized by definition with zero magnitude in the $JHK_s$ bands \citep{elias82}.

Finally, inspection of Figure~\ref{passbands} also reveals that two $J$-band filters were used with RetroCam.
As discussed in \citeauthor{krisciunas17}, the  $J_{RC1}$ filter experienced contamination issues in early 2009 and was replaced on 15 January 2009 with a new filter called,  $J_{RC2}$, which was used throughout the remainder of the CSP-I (mid-2009). 
From our photometric calibration program we  found the color term required to transform $J_{RC1}$ and $J_{RC2}$ magnitudes to the \citeauthor{persson98} standard photometric system are  discrepant enough to warrant two different $J$-band Swope ($+$ RetroCam) natural systems. The difference in the color term computed for  the $J_{RC1}$ and $J_{RC1}$ system response functions  implies  differences in SN photometry  of 2\% or less.

\section{Observations and Data Reduction}
\label{observations}

A detailed description of the observing  methodology and facilities at LCO used by the CSP-I   is provided by  \citet{hamuy06}, while a thorough  account of
 the data-reduction techniques including (i) the full processing of science and host-galaxy template images, (ii)  subtraction of the host templates from each science image, (iii) construction of local sequences, and (iv) subsequent computation of definitive photometry is presented by \citet{contreras10} and \citet{stritzinger11}.

 All science images, host-galaxy templates, and standard-star fields were reduced in the conventional manner as described by \citet[][and references therein]{stritzinger11}. Optical host-galaxy template images were obtained with the same filter set as the science images, in most cases with the du Pont telescope equipped with either facility Tek-5 or SITe2 CCD cameras, or  occasionally  with the Swope ($+$SITe3) telescope.
All  NIR template images  were taken with the du Pont ($+$ WIRC) telescope.
Templates  deemed suitable for galaxy subtraction were those in which the
SN had sufficiently faded  such that it was below our detection threshold, and were of excellent image quality, i.e., their seeing conditions are equal or superior to the science images.
Once template subtraction was performed on each of the science images, 
photometry of the local sequence and SN was computed.

 SN photometry was computed differentially with respect to a sequence of local standards calibrated with respect to standard star fields typically observed over a minimum of three photometric nights \citep[see][for details]{krisciunas17}.
Definitive photometry of the local sequences in the standard 
 \citet{smith02} ($ugri$), \citet{landolt92} ($BV$), and \citet{persson98} ($YJH$)  photometric systems is listed in 
 Table~\ref{table2} and Table~\ref{table3}, respectively. 
 The quoted uncertainties accompanying the magnitudes of each of the local sequence stars correspond to the weighted average of the instrumental errors computed from measurements obtained over the course of multiple photometric nights.
 
 Although the local sequences are reported in the standard systems as a service to the community, the CSP-I has a long tradition of publishing SN photometry in the natural photometric system. Photometry in the natural system is the purest form of the data, and allows us to  avoid the notorious problem associated with using color terms derived from standard stars to transform SN photometry to the standard photometric system \citep{suntzeff00}. 
 To compute natural system SN photometry therefore requires us to construct catalogs of each of the local sequences that correspond to our various natural systems (i.e., Swope$+$SITe3, Swope$+$RetroCam, and du Pont$+$WIRC). 
 To do this the standard system photometry of the local sequences is placed on the 
 natural system using  transformation equations and color terms provided in \citet{krisciunas17}. As mentioned above we have reason to believe the du Pont$+$Tek5 natural system is equivalent to the Swope$+$Site3 natural system. 

Armed with local sequence photometry on the natural system, we proceeded to compute PSF photometry of the CSP-I SE~SNe. 
Definitive natural system optical photometry of the  SN sample obtained with the Swope and du Pont telescopes is listed in Table~\ref{table4} and  Table~\ref{table5}, respectively.
 Definitive NIR photometry of the SN sample in the Swope natural system is given in Table~\ref{table6} and Table~\ref{table7}, where the former table contains  $J_{RC1}$ photometry and the later table contains $J_{RC2}$ photometry.
 Finally, natural system NIR photometry obtained with the du Pont ($+$ WIRC) telescope is listed in Table~\ref{table8}.

Reported along with each photometric measurement is a photometric error obtained by adding in quadrature the instrumental
error of each SN photometry measurement with the nightly zero-point error.
By electing to publish photometry on the natural system, the CSP-I data can be readily and accurately transformed to any user-defined photometric system through the use of S-corrections \citep{stritzinger02}.
Finally, we note that all of the photometry presented in this paper is also available in electronic format on the CSP-I webpage.

\section{Final Light Curves}
\label{lightcurves}

 The total number of photometry points contained within the CSP-I  SE~SN sample amounts to 
2075 in the optical ($u$, 217 epochs; $g$, 357; $r$, 366; $i$, 376; $B$, 367; $V$, 391) and  468 in the NIR ($Y$, 165; $J$, 162; $H$, 141).
Definitive light curves of  the \nosne\  CSP-I SE~SNe  are plotted in Figure~\ref{fig:LCs}.
 Overplotted on each of the light curves are  Gaussian Process spline fits  (dashed lines) computed within the SNooPy environment \citep{burns11}.
The SNooPy fits provide an  estimate of the peak magnitudes and the time of maximum brightness, along with robust uncertainties estimated via Monte Carlo simulations.
For those objects with observations beginning past maximum, an estimate of the time of $B$-band maximum was obtained using  the relation of Figure~3 from \citet{taddia17}.  The relation correlates the time of $B$-band maximum to the time of maximum of the other passbands, all of which except the $u$ band peak at phases after the time of $B$-band maximum.
In these cases, the peak of the light curve was covered in at least one of the redder bands.
Epochs of $B$-band maximum are listed in Table~\ref{table1}.
Twenty of the \nosne\ objects are found to have pre-maximum $B$-band
light-curve coverage, with 12 of these having observations beginning at least five days prior to the time of $B$-band maximum.
Furthermore, of the  \nosnenir\ objects with NIR follow-up observations, 17 have photometric measurements obtained before
$J$-band maximum, and a subset of  14 have NIR coverage beginning at least 5 days earlier.

\section{Prospects for the Future}
\label{summary}
We have presented the definitive optical and NIR photometry of an expanded sample of well-observed SE~SNe.
The high-quality and dense photometric coverage for many of the objects within  this data release, extending over eight filters, is particularly well-suited for tackling the problem of estimating host-galaxy extinction.
In a companion paper \citep{stritzinger17b}, the present observations are used to devise improved methods to estimate host-galaxy reddening.
This is accomplished through the construction of intrinsic  color-curve templates allowing for the inference of optical and NIR color excesses, the host reddening, $A_V^{\rm host}$, and (in some instances) the reddening-law parameter $R_V^{\rm host}$.

Armed with accurate estimates of host reddening, the  CSP-I photometry sample is studied in detail in an additional companion paper by  \citet{taddia17}.
This includes the analysis of the light-curve properties and luminosities, the construction of light-curve templates, and UVOIR bolometric light curves. The bolometric light curves are  modeled with both semi-analytical and hydrodynamical modeling methods in order to  constrain key explosion parameters (e.g., the $^{56}$Ni abundance, the ejecta mass, and the explosion energy).
Finally, the accompanying  visual-wavelength spectroscopy of this sample is  presented and studied in detail in a third companion paper by \citet{holmbo17}.

Since the completion of the CSP-I follow-up program, we performed a second stage named the {\em Carnegie Supernova Project II} (CSP-II; 2011--2015).
The  main goal of CSP-II was to achieve the most accurate  rest-frame NIR Hubble diagram  by pushing the accuracy of Type~Ia SN distances to the 1--2\% level.
However, observations of many young SNe of various types were also obtained, including more than a dozen SE SNe  discovered prior to  maximum brightness.
The CSP-II SE SN sample is distinguished from others by our concerted efforts to build a large dataset of multi-epoch NIR spectroscopy.
These data will offer a new avenue to study helium and  hydrogen signatures in SE SNe, and ultimately will enable us to better understand the nature of such cosmic explosions.

We end with a few words on the future study of young SE~SNe. In light of current and upcoming surveys such as ASAS-SN (All-Sky Automated Survey for Supernovae), ATLAS, DLT40 (Distance Less Than 40 Mpc), Pan-Starrs, and the Zwicky Transient Facility, progressively more young SE~SNe will be discovered within hours of explosion.
These early discoveries, coupled with high-cadence spectroscopy at medium or high resolutions, offer the potential to provide new constraints on SE~SN progenitors, their pre-SN history, and the reddening of light produced by circumstellar dust.

\begin{acknowledgements}
 We thank the referee for their thorough review of this manuscript and their constructive comments.
Special thanks to the technical staff and telescope operators of Las Campanas Observatory for their professional  assistance over the course of the CSP-I, and James Hughes for providing invaluable maintenance for our computer network.
M.~D. Stritzinger, C.~Contreras, and E.~Hsiao  acknowledge support provided by the Danish Agency for Science and Technology and Innovation realized through a Sapere Aude Level 2 grant and the  Instrument-center for Danish Astrophysics (IDA).
M.~D.~Stritzinger also acknowledges support by a research grant (13261) from the VILLUM FONDEN.
M.~D.~Stritzinger conducted part of this research at the Aspen Center for Physics, which is supported by NSF grant PHY-1066293. 
M.~Hamuy acknowledges support provided by the Millennium Center for Supernova Science through grant P10-064-F (funded by Programa Iniciativa Cientifica Milenio del Ministerio de Economia, Fomento y Turismo de Chile).
This material is based upon work supported by the US National Science
Foundation (NSF) under
grants AST--0306969, AST--0607438, AST--0908886, AST--1008343, AST--1211916, AST--1613426, AST--1613455, and AST--1613472.
A.~V. Filippenko is also grateful for financial assistance from
TABASGO Foundation, the Christopher R. Redlich Fund, and the
Miller Institute for Basic Research in Science.
 The work of Filippenko was conducted in part at the 
Aspen Center for Physics, which is supported by NSF grant PHY--1607611; he thanks the 
Center for its hospitality during the neutron stars workshop in June and July 2017.
 This research has made use of the NASA/IPAC Extragalactic Database (NED), which is operated by the Jet Propulsion Laboratory, California Institute of Technology, under contract with the National Aeronautics and Space Administration.
\end{acknowledgements}

\bibliographystyle{aa}

\begin{thebibliography}{100}

\bibitem[Aldering et al.(1995)]{aldering94}
Aldering, G., Humphreys, R. M., Odewahn, S., Thurmes, P. 1994, \aj, 107, 662,

\bibitem[Anderson et al.(2010)]{anderson10}
Anderson, J. P., Covarrubias, R. A. , James, P. A., et al. 2010, \mnras, 407, 2660

\bibitem[Anderson et al.(2012)]{anderson12}
Anderson, J. P., Habergham, S. M., Jampes, P. A., et al. 2012, \mnras, 424, 1372

\bibitem[Anupama et al.(2005)]{anupama05}
Anupama, N., et al. 2005, \apj, 631L, 125

\bibitem[Arcavi et al.(2010)]{arcavi10}
Arcavi, I., Gal-Yam, A., Kasliwal, M. M., et al. 2010, \apj, 721, 777

\bibitem[Arcavi et al.(2011)]{arcavi11}
Arcavi, I., Gal-Yam, A., Yaron, O., et al. 2011, \apj, 742L, 18

\bibitem[Arcavi et al.(2017)]{arcavi17}
Arcavi, I., Hosseinzadeh, G., Brown, P. J., et al. 2017, \apj, 837L, 2

\bibitem[Benvenuto et al.(2013)]{benvenuto13}
Benvenuto, O. G., Bersten, M. C., \& Nomoto, K. 2013, \apj, 762, 74

\bibitem[Bersten et al.(2012)]{bersten12}
Bersten, M. C., Benvenuto, O., \& Nomoto, K. 2012, \apj, 767, 143

\bibitem[Bersten et al.(2013)]{bersten13}
Bersten, M., Tanaka, M., Tominaga, N., et al. 2013, \apj, 767, 143

\bibitem[Bersten et al.(2014)]{bersten14}
Bersten, M., Benvenuto, O. G., Folatelli, G., et al. 2014, \aj, 148, 68

\bibitem[Bianco et al.(2014)]{bianco14}
Bianco, F. B., Modjaz, M., Hicken, M., et al. 2014, \apjs, 213, 19

\bibitem[Blanco et al.(1987)]{blanco87}
Blanco, V. M.,Gregory, B., Hamuy, M., et al. 1987, \apj, 320, 589

\bibitem[Blondin \& Tonry(2007)]{blondin07}
Blondin, S., \& Tonry, J. L., 2007, \apj, 666, 1024

\bibitem[Bufano et al.(2014)]{bufano14}
Bufano, F., Pignata, G., Bersten, M., et al. 2014, \mnras, 439, 1807

\bibitem[Burns et al.(2011)]{burns11}
Burns, C. R., Stritzinger, M. D., Phillips, M. M., et al. 2011, \aj, 141, 19

\bibitem[Cano(2013)]{cano13}
Cano, Z. 2013, \mnras, 434, 1098

\bibitem[Cao et al.(2013)]{cao13}
Cao, Y., Kasliwal, M. M., Arcavi, I., et al. 2013, \apj, 775, L7

\bibitem[Chevalier \& Fransson(2008)]{chevalier08}
Chevalier, R. A., \& Fransson, C. 2008, \apj, 683L, 135

\bibitem[Chornock et al.(2011)]{chornock11}
Chornock, R., Filippenko, A. V., Li., W., et al. 2011, \apj, 739, 41

\bibitem[Cohen et al.(1995)]{cohen95}
Cohen, J. G., Darling, J., \& Porter, A. 1995, \aj, 110, 308

\bibitem[Contreras et al.(2010)]{contreras10}
Contreras, C., Hamuy, M., Phillips, M. M., et al. 2007, \aj, 139, 519

\bibitem[Crockett et al.(2008)]{crockett08}
Crockett, R. M., et al. 2008, \mnras, 391, L5

\bibitem[Dessart et al.(2011)]{dessart11}
Dessart, L., et al. 2011, \mnras, 414, 2985

\bibitem[Dessart et al.(2016)]{dessart16}
Dessart, L., Hillier, D. J., Woosley, S., et al. 2016, \mnras, 458, 1618

\bibitem[Drout et al.(2011)]{drout11}
Drout, M. R., Soderberg, A. M., Avishay, G. Y., et al. 2011, \apj, 741, 97

\bibitem[Eldridge et al.(2013)]{eldridge13}
Eldridge, J. J., Morgan, F., Smartt, S. J. et al. 2013, \mnras, 436, 774

\bibitem[Eldridge \& Maund(2016)]{eldridge16}
Eldridge, J. J., \& Maund, J. R. 2016, \mnras, 461L, 117

\bibitem[Elias et al.(1982)]{elias82}
Elias, J. H., Frogel, J. A., Matthews, K., \& Neugebauer, G. 1982, \aj, 87, 1029

\bibitem[Ergon et al.(2014)]{ergon14}
Ergon, M., et al. 2014, \aap, 562, A17

\bibitem[Filippenko(1988)]{filippenko88}
Filippenko, A. V. 1988, AJ, 96, 1941                             

\bibitem[Filippenko(1997)]{filippenko97}
Filippenko, A. V. 1997, ARA\&A, 35, 309

\bibitem[Filippenko(2005)]{filippenko05}
Filippenko, A. 2005, in The Fate of the Most Massive Stars, ed. R. Humphreys 
\& K. Stanek (San Francisco: ASP), p. 33

\bibitem[Filippenko et al.(2003)]{filippenko03}
Filippenko, A. V., Chornock, R., Swift, B., Modjaz, M., Simcoe, R., 
\& Rauch, M. 2003, IAU Circ. No. 8159, 2

\bibitem[Filippenko et al.(2001)]{filippenko01}
Filippenko, A. V., Li, W. D., Treffers, R. R., \& Modjaz, M. 2001,  
in Small-Telescope Astronomy on Global Scales, ed.                   
W. P. Chen, C. Lemme, \& B. Paczy\'{n}ski (San Francisco:
ASP Conf. Ser. Vol. 246), 121   

\bibitem[Filippenko, Matheson, \& Ho(1993)]{filippenko93}
Filippenko, A. V., Matheson, T., Ho, L. C. 1993, \apj, 415L, 103

\bibitem[Folatelli et al.(2006)]{folatelli06}
Folatelli,~G., et al. 2006,   \apj, 641, 1039

\bibitem[Folatelli et al.(2014)]{folatelli14}
Folatelli,~G., Bersten, M. C., Benvenuto, O. G., et al. 2014, \apj, 793L, 22

\bibitem[Folatelli et al.(2016)]{folatelli16}
Folatelli,~G.,  Van Dyk, S. D., Kuncarayakti, H., et al. 2016, \apj, 825L, 22

\bibitem[Foley et al.(2003)]{foley03}
 Foley, R. J., et al., 2003, \pasp, 115, 1220

\bibitem[Fox et al.(2014)]{fox14}
Fox, O. D., Bostroem, K. A., Van Dyk, S. D., et al. 2014, \apj, 790, 17

\bibitem[Frieman et al.(2008)]{frieman08}
Frieman, J., et al. 2008, \aj, 135, 338

\bibitem[Fryer et al.(2007)]{fryer07}
Fryer, C. L., et al. 2007, \pasp, 119, 1211

\bibitem[Galama et al.(1998)]{galama98}
Galama, T. J., et al. 1998, Nature, 395, 670

\bibitem[Galbany et al.(2016)]{galbany16}
Galbany, L., et al. 2016, \aap, 591, 48

\bibitem[Gal-Yam (2017)]{galyam16}
Gal-Yam, A. 2016, in press (arXiv:1611.09353)

\bibitem[Gal-Yam et al.(2014)]{galyam14}
Gal-Yam, A., Arcavi, I., Ofek, E. O., et al. 2014, \nat, 509, 471

\bibitem[Gaskell et al.(1986)]{gaskell86}
Gaskell, C. M., et al. 1986, \apj, 306L, 77

\bibitem[Georgy et al.(2009)]{georgy09}
Georgy, C., et al. 2009, \aap, 502, 611

\bibitem[Graur et al.(2017a)]{graur17a}
Graur, O., Bianco, F. B., Huang, S., et al. 2017, \apj, 837, 120

\bibitem[Graur et al.(2017b)]{graur17b}
Graur, O., Bianco, F. B., Modjaz, M., et al. 2017, \apj, 837, 121

\bibitem[Groh et al.(2013)]{groh13}
Groh, J.~H., Georgy, C., \& Ekstr{\"o}m, S. 2013, \aap, 558L, 1

\bibitem[Hamuy et al.(2006)]{hamuy06}
Hamuy,~M., et al. 2006, \pasp, 118, 2

\bibitem[Hamuy et al.(2009)]{hamuy09}
Hamuy,~M., et al. 2009, \apj, 703, 1612

\bibitem[Harkness et al.(1987)]{harkness87}
Harkness, R. P., et al. 1987, \apj, 317, 355

\bibitem[Heger et al.(2003)]{heger03}
Heger, A., et al. 2003, \apj, 591, 288

\bibitem[Hillenbrand et al.(2002)]{hillenbrand02}
Hillenbrand, L. A., Foster, J. B., Persson, S. E., \& Matthews, K. 2002,
\pasp, 114, 708

\bibitem[Hjorth et al.(2003)]{hjorth03}
Hjorth, J., et al. 2003, Nature, 423, 847

\bibitem[Holmbo et al., in preparation]{holmbo17}
Holmbo, S., Stritzinger, M. D., Hsiao, E. Y., et al., in preparation

\bibitem[Iwamoto et al.(1994)]{iwamoto94}
Iwamoto, K., et al. 1994, \apj, 437, L115

\bibitem[Iwamoto et al.(1998)]{iwamoto98}
Iwamoto, K., Mazzali, P. A., Nomoto, K., et al. 1998, \nat, 395, 672

\bibitem[Kawabata et al.(2010)]{kawabata10}
Kawabata, K. S., et al. 2010, Nature, 465, 326

\bibitem[Kirshner et al.(1987)]{kirshner87}
Kirshner, R. P., Sonneborn, G., Crenshaw, D. M., et al. 1987, \apj, 320, 602

\bibitem[Krisciunas et al., in perparation]{krisciunas17}
Krisciunas, K., Contreras, C., Burns, C. R. et al., in perparation

\bibitem[Kumar et al. (2013)]{kumar13}
Kumar, B., Pandey, S. B., Sahu, D. K., et al. 2013, \mnras, 431, 308

\bibitem[Landolt(1992)]{landolt92}
Landolt,~A.~U. 1992, \aj, 104, 340

\bibitem[Langer(2012)]{langer12}
Langer, N. 2012, \araa, 50, 107

\bibitem[Leloudas et al.(2011)]{leloudas11}
Leloudas, G., Gallazzi, A., Sollerman, J., et al. 2011, \aap, 530, 95

\bibitem[Li(2008)]{li08}
Li, W. 2008, ATel, 1433, 1

\bibitem[Li et al.(2011)]{li11}
Li, W., Leaman, J., Chornock, R., et al. 2011, \mnras, 412, 1441

\bibitem[Liu et al.(2014)]{liu14}
Liu, Y. Q., \& Modjaz, M., 2014, arXiv:1405:1437

\bibitem[Liu et al.(2016)]{liu16}
Liu, Y. Q., Modjaz, M., \& Bianco, F. B., 2016, 1612.07321

\bibitem[Lyman et al.(2016)]{lyman16}
Lyman, J. D., Bersier, D., James, P. A., et al. 2016, \mnras, 457, 328

\bibitem[Malesani et al.(2009)]{malesani09}
Malesani, D., et al. 2009, \apj, 692, 84

\bibitem[Matheson et al.(2001)]{matheson01}
Matheson, T., et al. 2001, \aj, 121, 1648

\bibitem[Matheson et al.(2003)]{matheson03}
Matheson, T., et al. 2003, \apj, 599, 394

\bibitem[Maund et al.(2004)]{maund04}
Maund, J. R., Smartt, S. J., Kudritzki, R. P., et al. 2004, \nat, 427, 129

\bibitem[Maund et al.(2011)]{maund11}
Maund, J. R., et al. 2011, \apj, 739, L37

\bibitem[Maund et al.(2015)]{maund15}
Maund, J. R., Arcavi, I., Ergon, M., et al. 2015, \mnras, 454, 2580

\bibitem[Mazzali et al.(2002)]{mazzali02}
Mazzali, P. A., et al. 2002, \apj, 572, L61

\bibitem[Mazzali et al.(2006)]{mazzali06}
Mazzali, P. A., Deng, J., Pian, E., et al. 2006, \apj, 645, 1323

\bibitem[Mazzali et al.(2009)]{mazzali09}
Mazzali, P. A., et al. 2009, \apj, 703, 1624

\bibitem[Milisavljevic et al.(2013)]{milisavljevic13}
Milisavljevic, D., Margutti, R., Soderberg, A. M., et al. 2012, \apj, 767, 71

\bibitem[Modjaz et al.(2016)]{modjaz16}
Modjaz, M., Liu, Y. Q., \& Bianco, F. B., 2016, \apj, 832, 108

\bibitem[Modjaz et al.(2011)]{modjaz11}
Modjaz, M., Kewley, L., Bloom, J. S., et al. 2011, \apj, 731L, 4

\bibitem[Morales-Garoffolo et al.(2014)]{morales14}
Morales-Garoffolo, A., Elias-Rosa, N., Benetti, S., et al. 2014, \mnras, 445, 1647

\bibitem[Nakar \& Piro(2014)]{nakar14}
Nakar, E., \& Piro, A. L. 2014, \apj, 788, 193

\bibitem[Nomoto et al.(1993)]{nomoto93}
Nomoto, K., et al. 1993, Nature, 364, 507


\bibitem[Patat et al.(2007)]{patat07}
Patat, F., Chandra, P., Chevalier, R., et al. 2007, Science, 317, 924

\bibitem[Perets et al.(2010)]{perets10}
Perets, H. B. et al., 2010, Nature, 465, 322

\bibitem[Persson et al.(1998)]{persson98}
Persson,~S.~E., Murphy,~D.~C.,  Krzeminski,~W., Roth,~M., \& Rieke,~M.~J. 1998, \aj, 116, 2475

\bibitem[Pettini \& Pagel(2004)]{pettini04}
Pettini, M., \& Pagel, B. E. J. 2004, \mnras, 348, 56

\bibitem[Pignata et al.(2011)]{pignata11}
Pingata, G., et al. 2011, \apj, 728, 14

\bibitem[Piro(2015)]{piro15}
Piro, A. L. 2015, \apj, 808L, 51

\bibitem[Piro et al.(2017)]{piro17}
Piro, A. L., Muhleisen, M. E., Arcavi, I., et al., submitted to \apj, arXiv:1703.00913

\bibitem[Podsiadlowski et al.(1993)]{podsiadlowski93}
Podsidlowski, P., Hsu, J. J. L., Joss, P. C., Ross, R. R. 1993, Nature, 364, 509

\bibitem[Prentice et al.(2016)]{prentice16}
Prentice, S. J., Mazzali, P. A., Pian, E., et al. 2016, \mnras, 458, 2973

\bibitem[Prentice \& Mazzali(2017)]{prentice17}
Prentice, S. J., \& Mazzali, P. A. 2017, \mnras, in press, arXiv:1704.06635

\bibitem[Puls, Vink, \& Najarro(2008)]{puls08}
Puls, J., Vink, J. S., \& Najarro, F. 2008, A\&ARv, 16, 209

\bibitem[Rheault et al.(2010)]{rheault10}
Rheault, J. P., et al. 2010, Proc. SPIE, 7735, 773564

\bibitem[Richmond et. al.(1994)]{richmond94}
Richmond, M., W., Treffers, R. R., Filippenko, A. V., et. al. 1994, \aj, 107, 1022

\bibitem[Sanders et al.(2012)]{sanders12}
Sanders, N. E., Caldwell, N., McDowell, J., Harding P. 2012, \apj, 758, 133

\bibitem[Sapir \& Waxman(2017)]{sapir17}
Sapir, N., \& Waxman, E. 2017, \apj, 838, 130


\bibitem[Shigeyama et al.(1990)]{shigeyama90}
Shiteyama, T., Nomoto, K., Tsujimoto, T., Hashimoto, M. A. 1990, \apj, 361, L23

\bibitem[Shivvers et al.(2015)]{shivvers15} 
Shivvers, I., Groh, J., Mauerhan, J. C., et al. 2015, ApJ, 806, 213

\bibitem[Shivvers et al.(2017)]{shivvers17} 
Shivvers, I., et al. 2017, PASP, 129, 054201

\bibitem[Smartt(2009)]{smartt09}
Smartt, S. J. 2009, ARA\&A, 47, 63

\bibitem[Smith et al.(2002)]{smith02}
Smith,~J.~A., et al. 2002, \aj,  123, 2121

\bibitem[Smith \& Owocki(2006)]{smith06}
Smith, N., \& Owocki, S. P. 2006, \apj, 645, L45

\bibitem[Smith et al.(2011)]{smith11}
Smith, N., Li, W., Filippenko, A. V., Chornock, R. 2011, \mnras, 412, 1522

\bibitem[Soderberg et al(2012)]{soderberg12}
Soderberg, A. M., Margutti, R., Zauderer, B. A., et al. 2012, \apj, 752, 78

\bibitem[Stanek et al.(2003)]{stanek03}
Stanek, K. Z., et al. 2003, \apj, 591, L17

\bibitem[Stritzinger et al.(2002)]{stritzinger02}
Stritzinger, M., et al. 2002, \aj, 124, 2100

\bibitem[Stritzinger et al.(2009)]{stritzinger09}
Stritzinger,~M.,  et al. 2009, \apj, 696, 713

\bibitem[Stritzinger et al.(2011)]{stritzinger11}
Stritzinger, M., et al. 2011, \aj, 142, 156

\bibitem[Stritzinger et al., submitted to A\&A]{stritzinger17b}
Stritzinger, M. D., Taddia, F., Bersten, M., et al. 2017, submitted to \aap

\bibitem[Sukhbold et al.(2016)]{sukhbold16}
Sukhbold, T., Ertl, T, Woosley, S. E., et al. 2016, \apj, 821, 38

\bibitem[Suntzeff(2000)]{suntzeff00}
Suntzeff, N. B. 2000, in Cosmic Explosions, ed. S. S. Holt \& W. W. Zhang
(New York: AIP), 65

\bibitem[Swartz et al.(1993)]{swartz93}
Swartz, D. A., Clocchiatti, A., Benjamin, R., Lester, D. F., Wheeler, J. C. 1993, Nature, 365, 232

\bibitem[Taddia et al.(2015)]{taddia15}
Taddia, F., Sollerman, J., Leloudas, G., et al. 2015, \aap, 574, 60

\bibitem[Taddia et al., submitted to \aap]{taddia17}
Taddia, F., Stritzinger, M., Bersten, M., et al. 2017, submitted to \aap
 
 \bibitem[Tartaglia et al.(2017)]{tartaglia17}
 Tartaglia, L., Fraser, M., Sand, D. J., et al. 2017, \apj, 836L, 12
 
 \bibitem[Thielemann, Nomoto, \& Hashimoto(1996)]{thielemann96}
 Thielemann, F. K., Nomoto, K., Hashimoto, \& M. A., 1996, \apj, 460, 408

\bibitem[Tominaga et al.(2005)]{tominaga05}
Tominaga, N., et al. 633L, 97

 \bibitem[Utrobin(1994)]{utrobin94}
Utrobin, V. 1994, \aap, 281, 89

\bibitem[van Driel et al.(1993)]{vandriel93}
van Driel, W., Yoshida, s., Nakada, Y., et al. 1993, \pasj, 45L, 59

\bibitem[Van Dyk et al.(2011)]{vandyk11}
 Van Dyk, S. D., et al. 2011, \apj, 741, L28

\bibitem[Van Dyk et al.(2014)]{vandyk14}
 Van Dyk, S. D., et al. 2014, \aj, 147, 37

\bibitem[Vink(2011)]{vink11}
 Vink, J. S. 2011, Ap\&SS, 336, 163

\bibitem[Woosely, Langer, \& Weaver(1993)]{woosely93}
 Woosley, S. E., Langer, N., \& Weaver, T. A. 1993, \apj, 411, 823

\bibitem[Woosely, Langer, \& Weaver(1995)]{woosely95}
 Woosley, S. E., Langer, N., \& Weaver, T. A. 1995, \apj, 448, 315

\bibitem[Yaron et al.(2017)]{yaron17}
Yaron, O., Perley, D. A., Gal-Yam, A., et al. 2017, Nature Physics, 13, 510

\bibitem[Yoon, \& Cantiello(2010)]{yoon10}
Yoon, S. C., \& Cantiello, M. 2010, \apj, 717, L62

\end{thebibliography}

\clearpage
\setcounter{figure}{0}
\begin{figure}[h]
\centering
\includegraphics[width=5.6in]{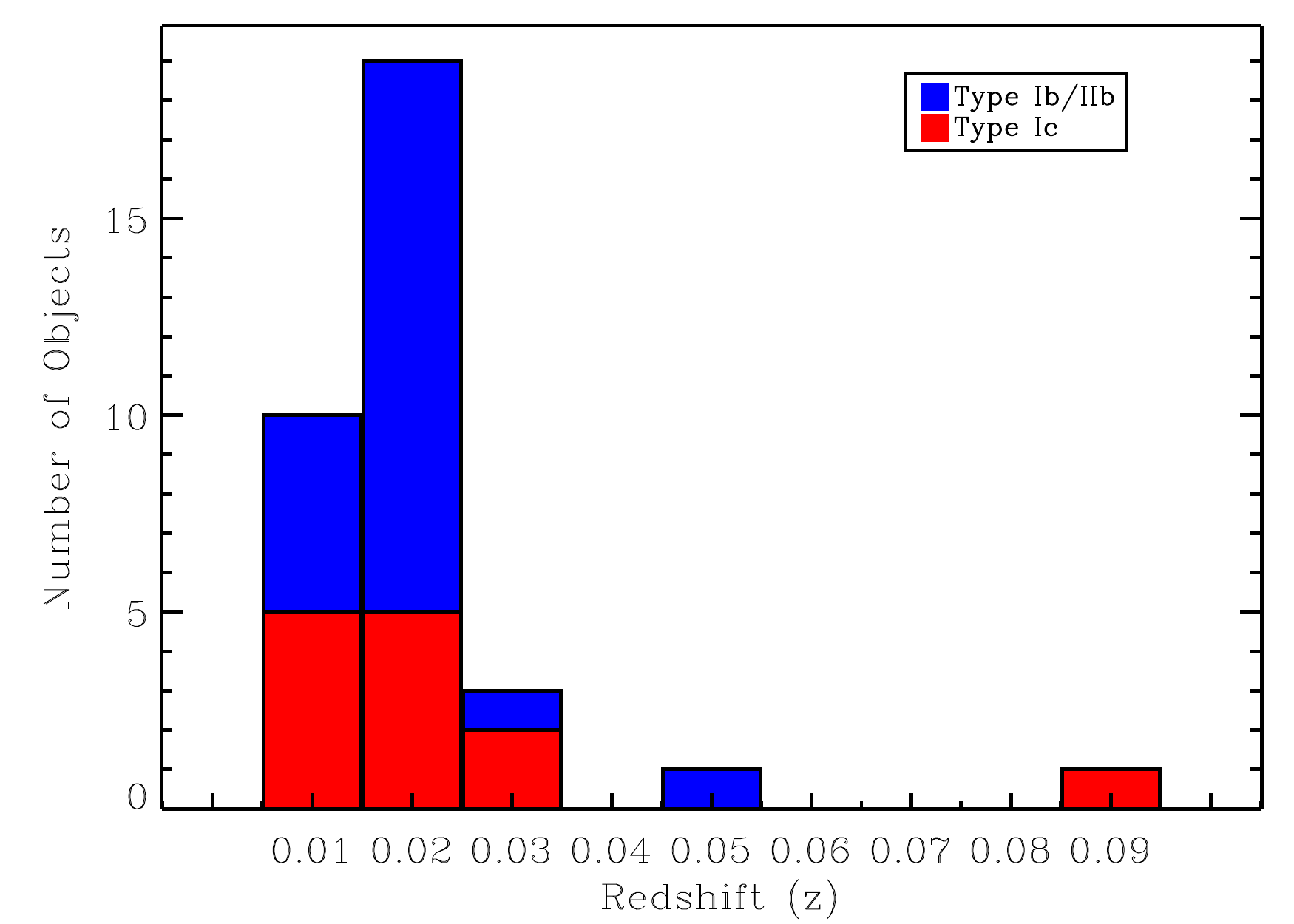}
\caption{Stacked histogram of the redshift distribution of
SN~Ib/IIb (blue) and SN~Ic (red) observed by the CSP-I.\label{redshift}}
\end{figure}

\clearpage
\setcounter{figure}{1}
\begin{figure}
\begin{center}$
\begin{array}{cc}
\includegraphics[width=3cm]{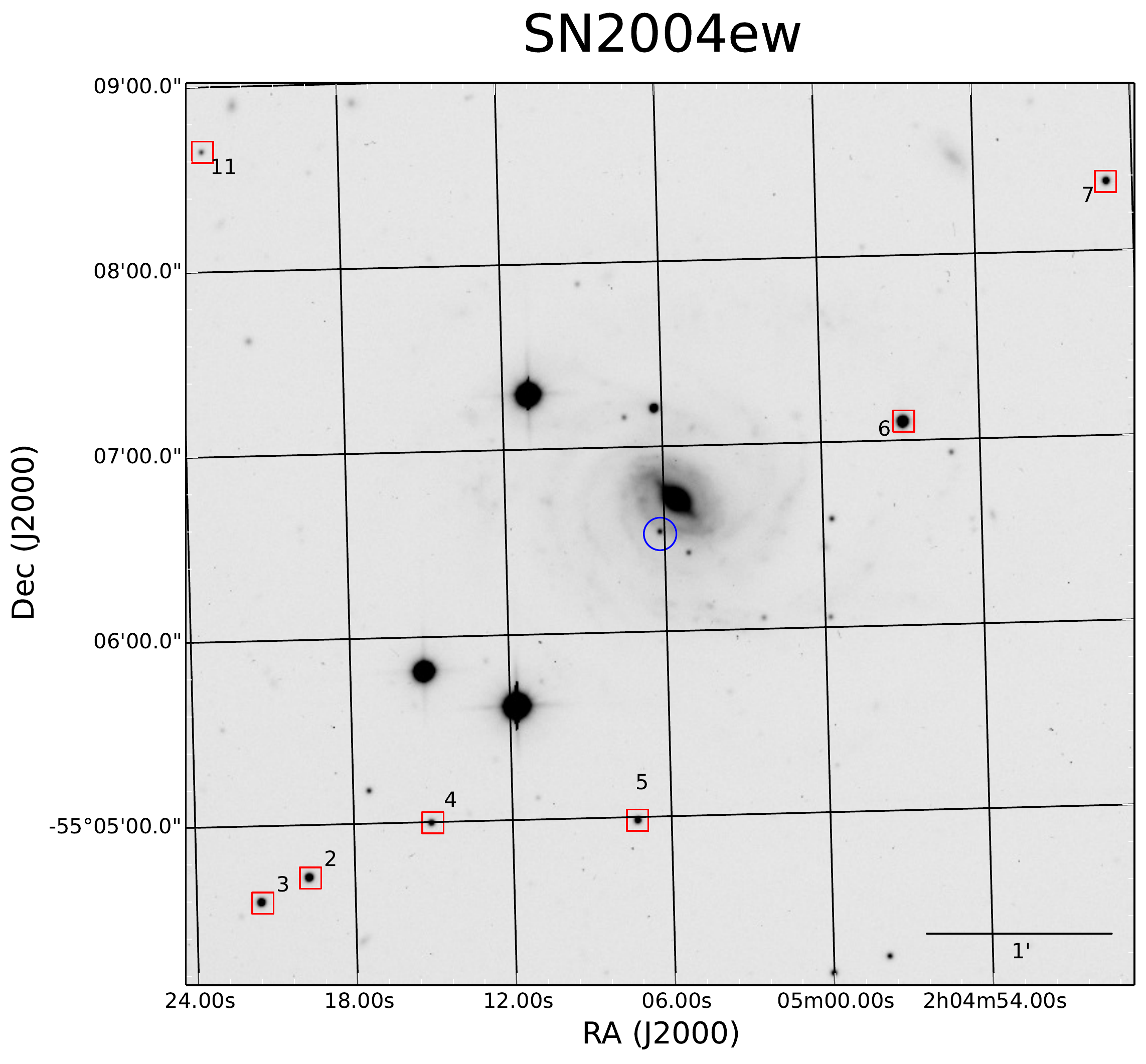}
\includegraphics[width=3cm]{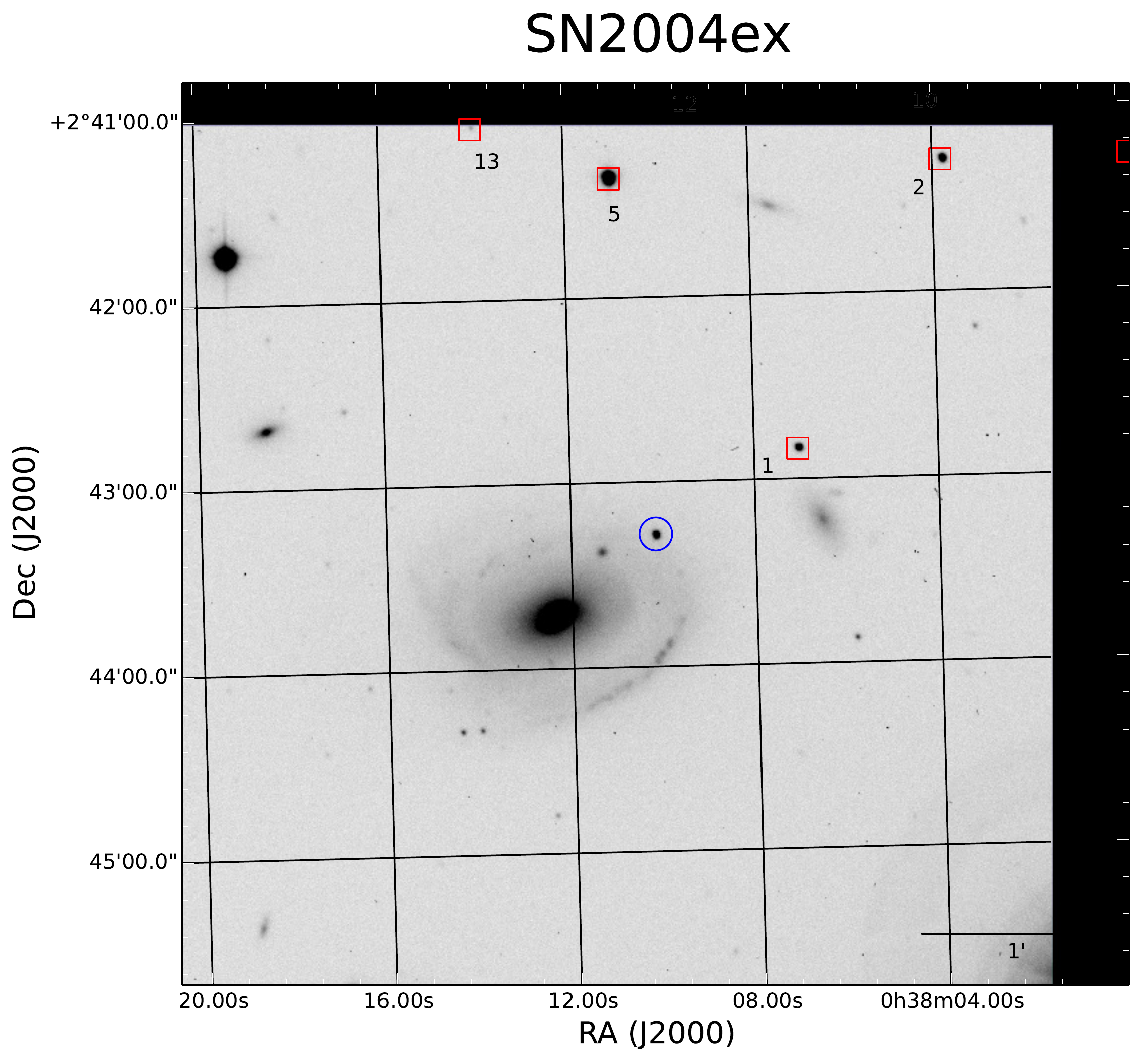}
\includegraphics[width=3cm]{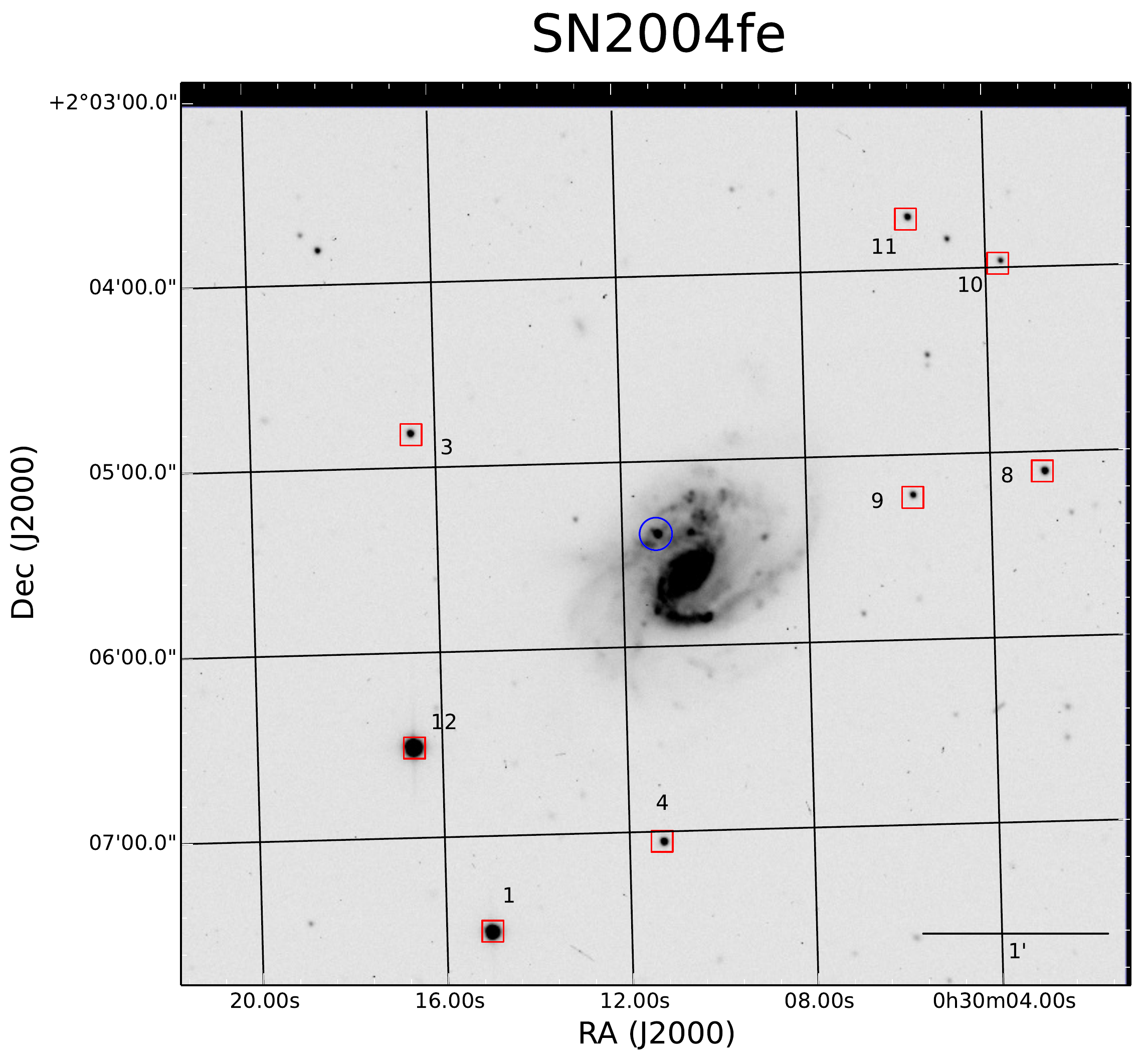}
\includegraphics[width=3cm]{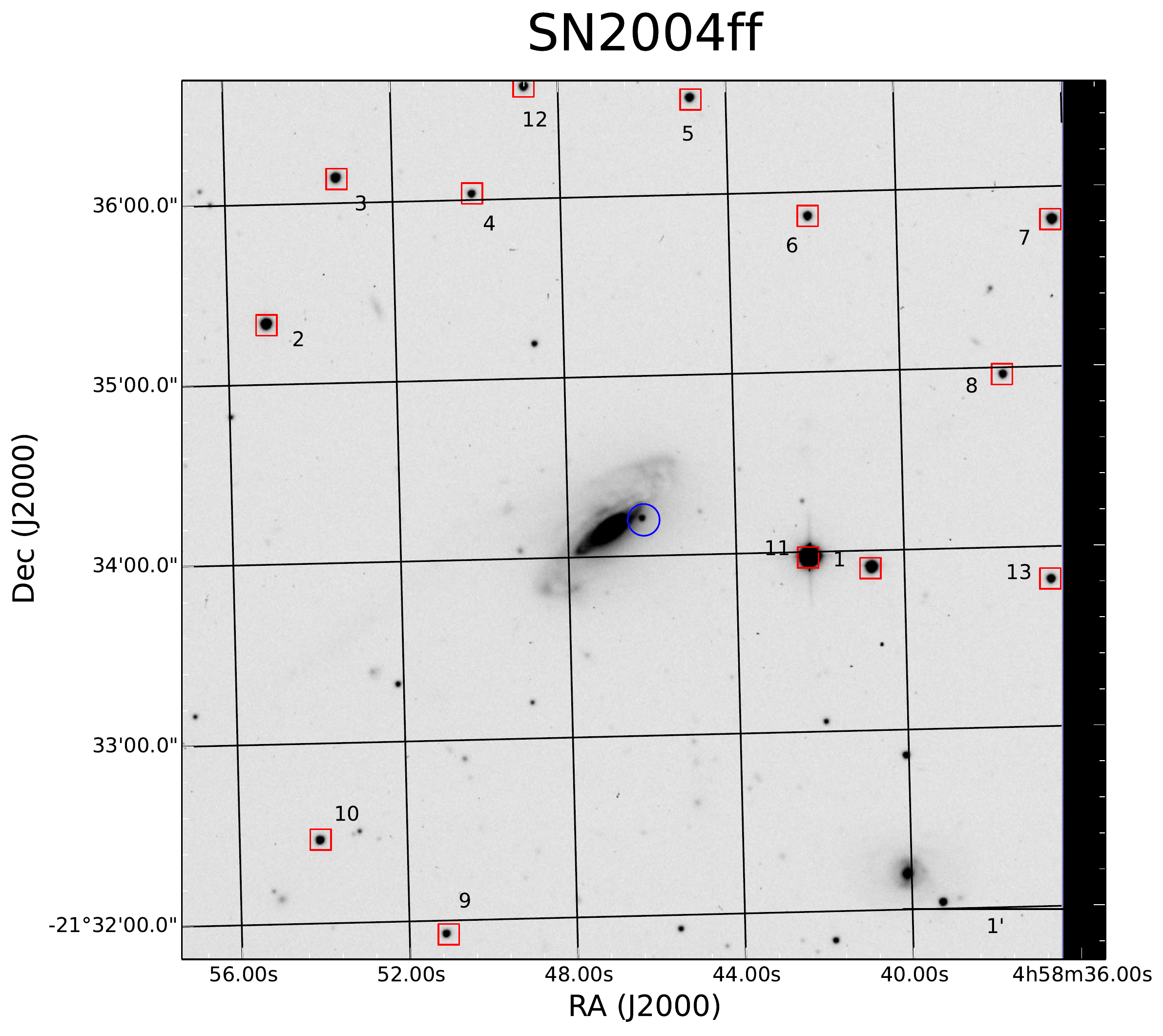}
\includegraphics[width=3cm]{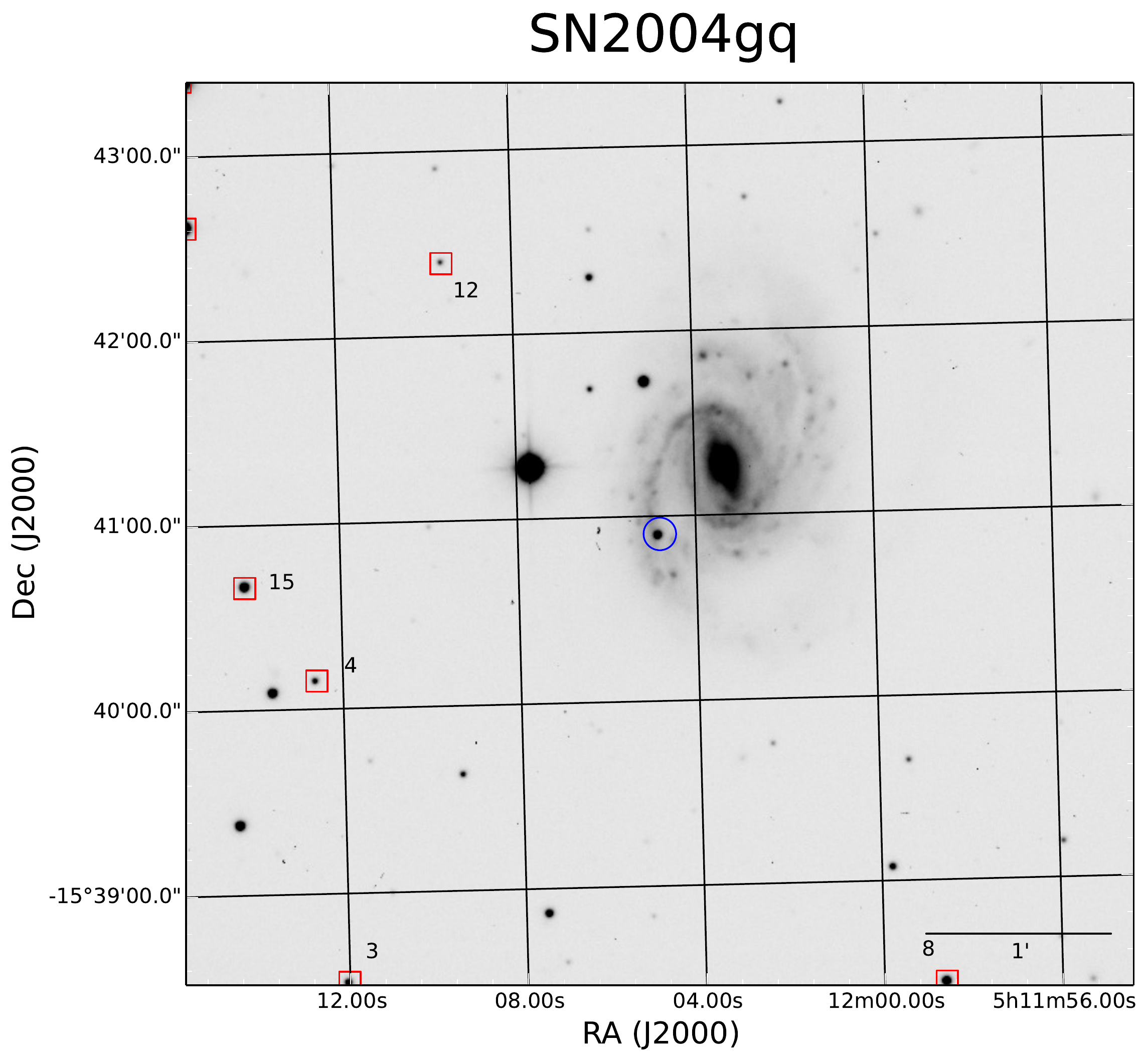}
\includegraphics[width=3cm]{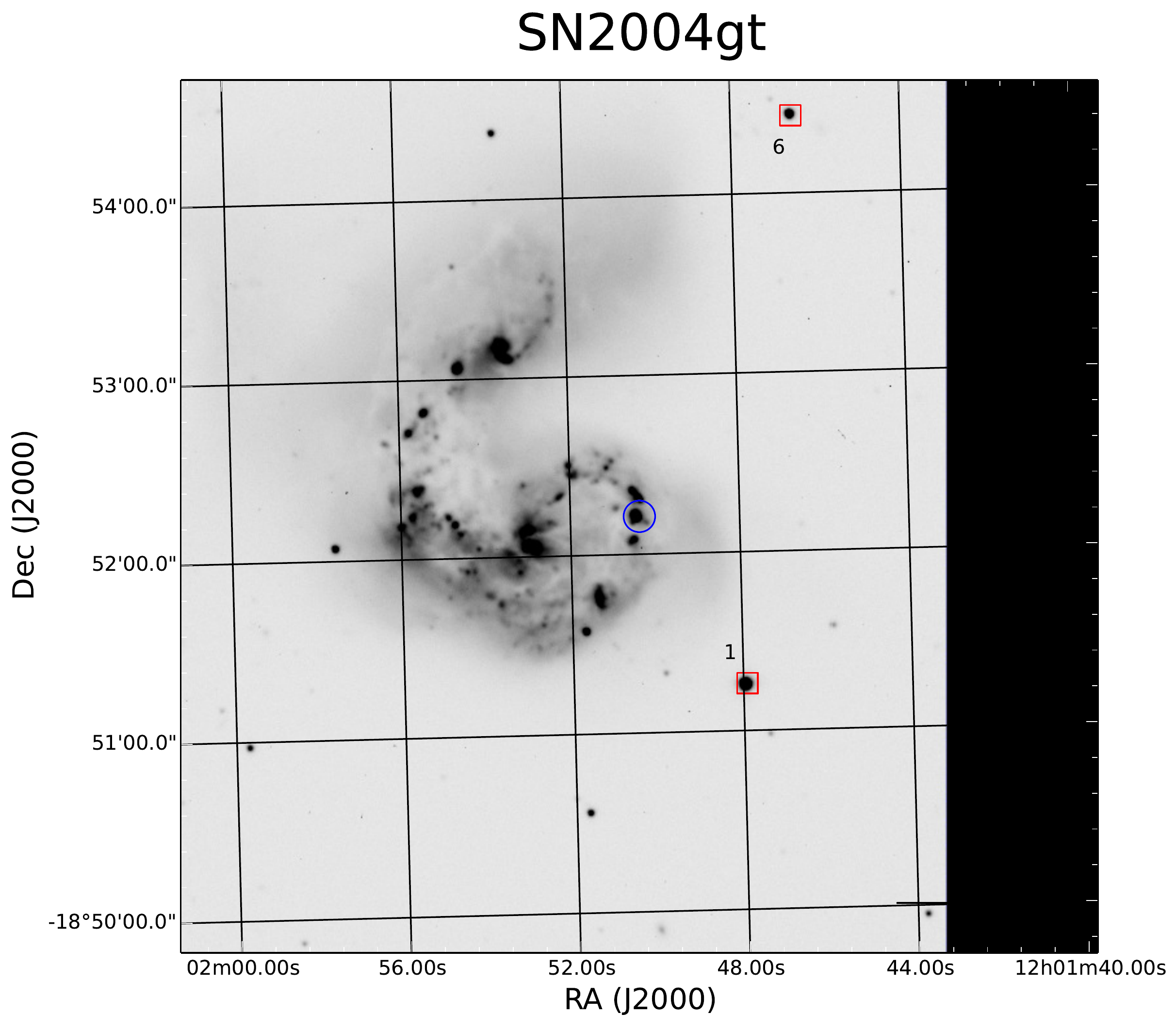}\\
\includegraphics[width=3cm]{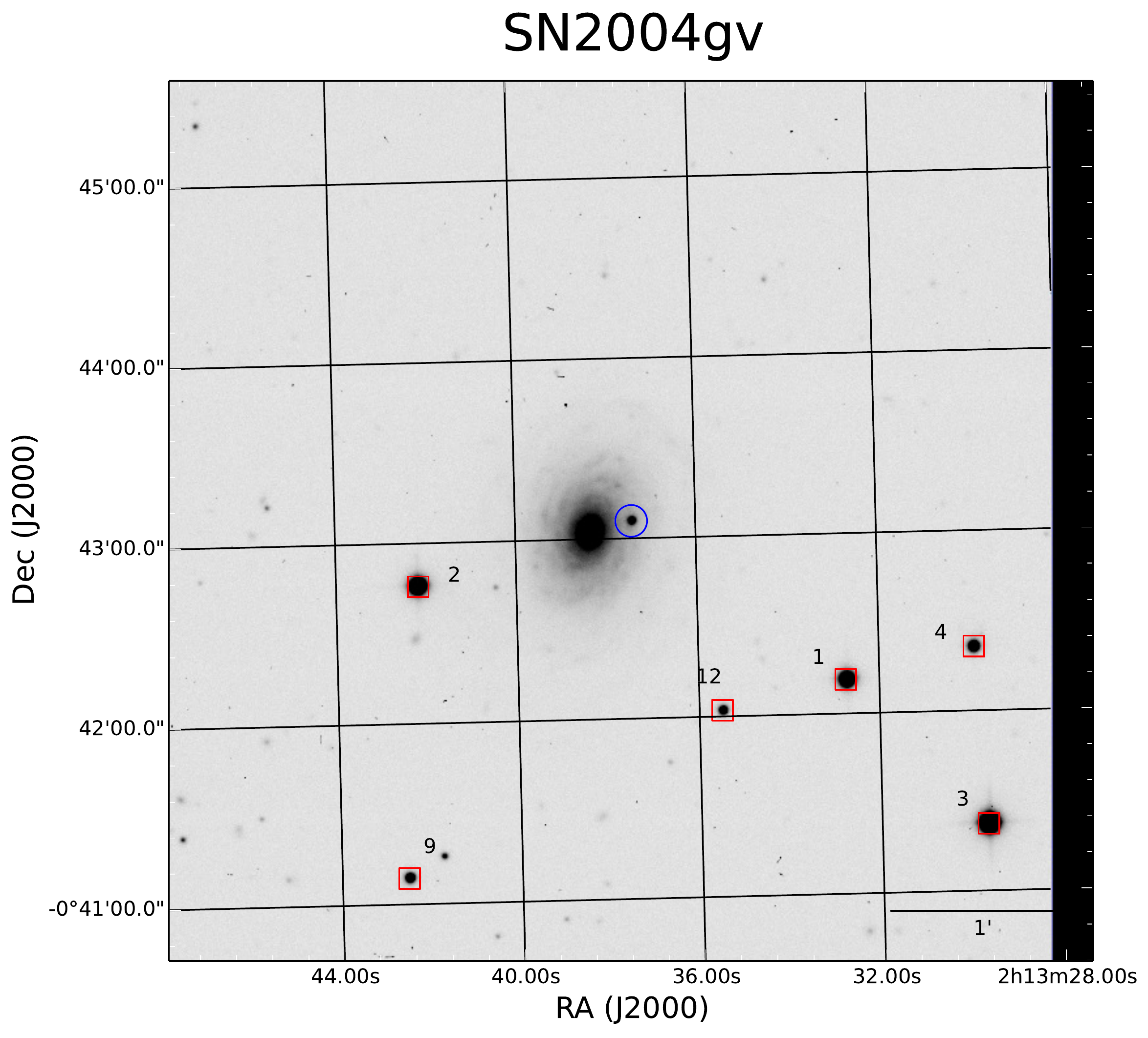}
\includegraphics[width=3cm]{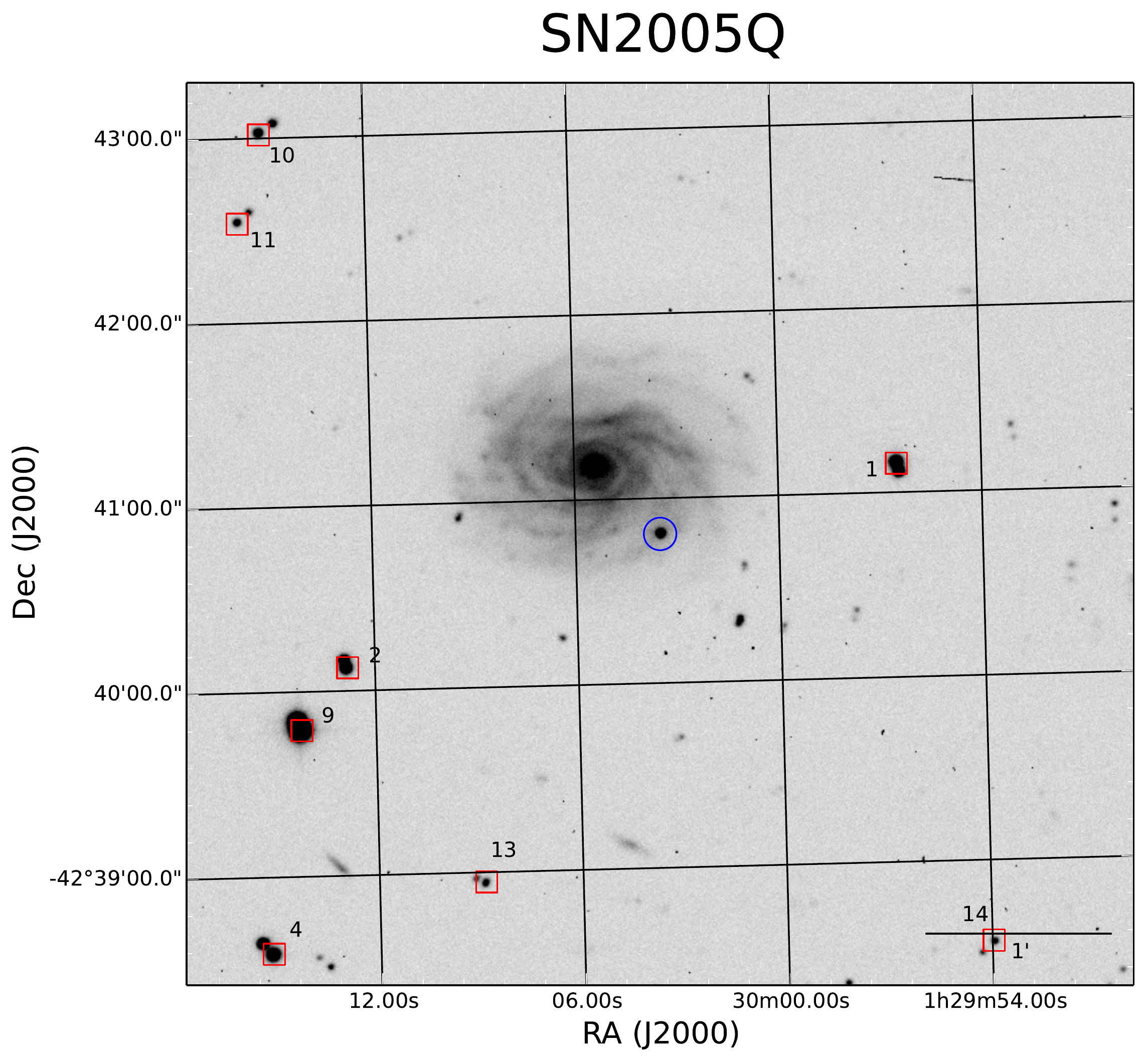}
\includegraphics[width=3cm]{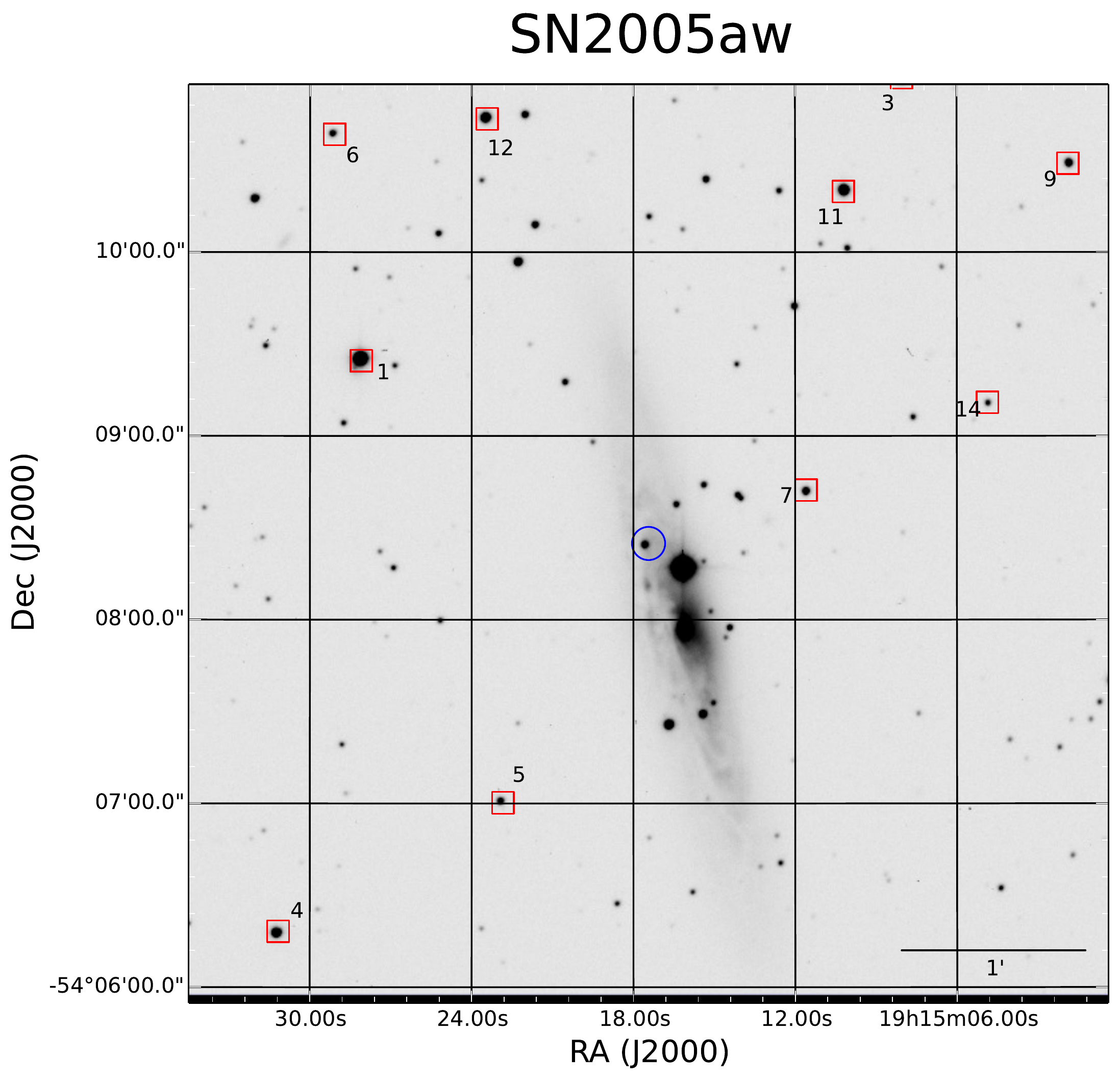}
\includegraphics[width=3cm]{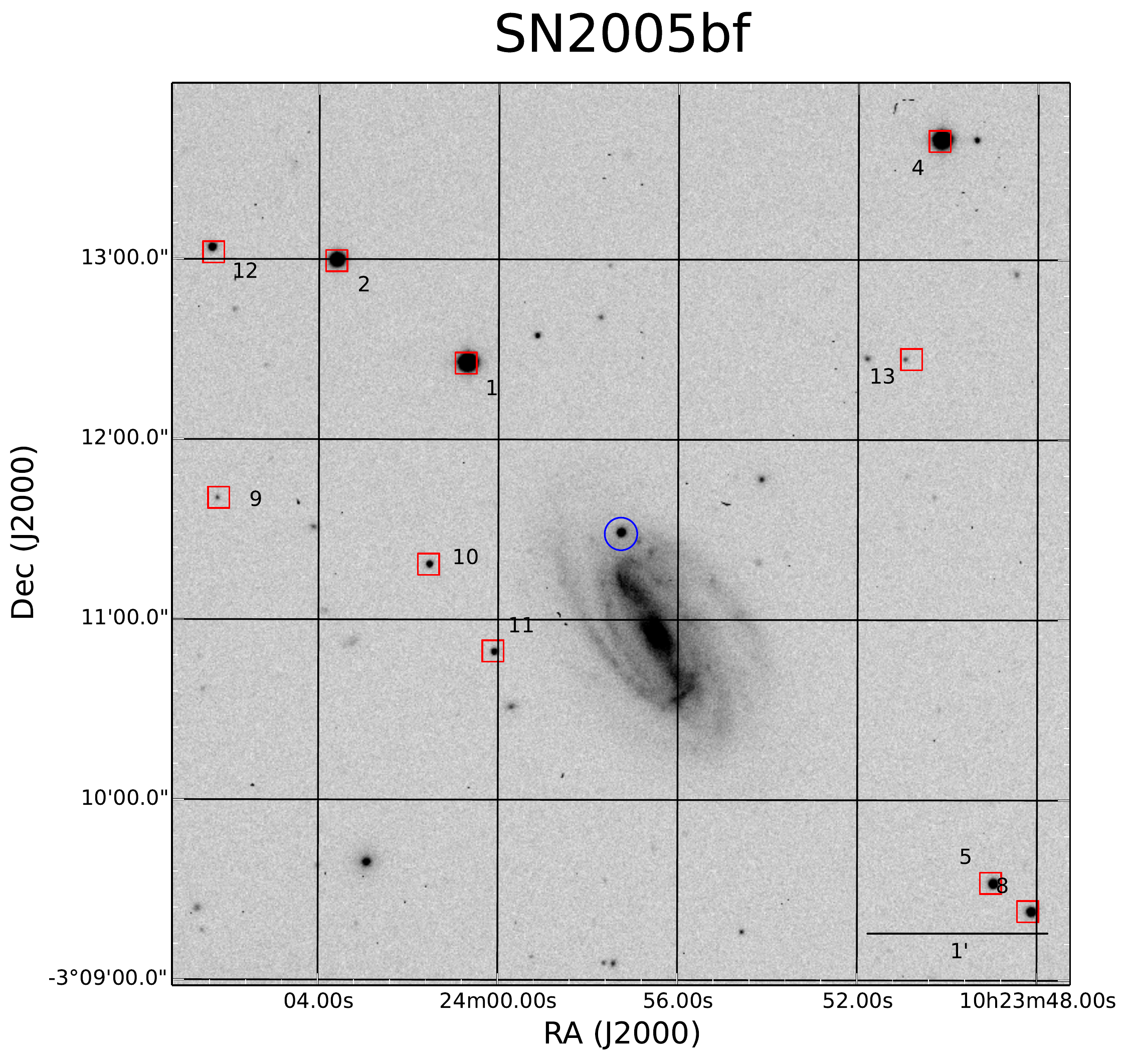}
\includegraphics[width=3cm]{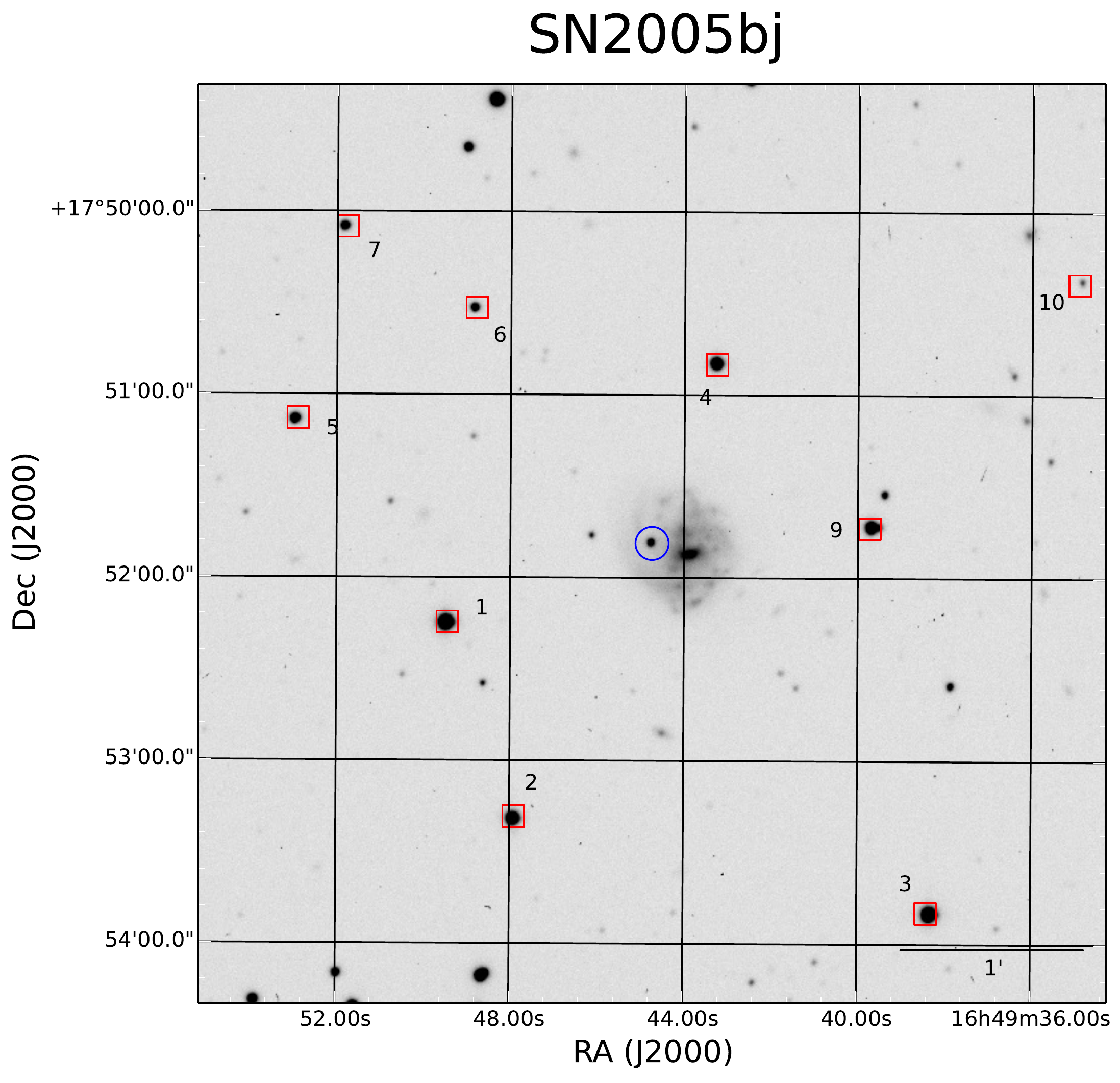}
\includegraphics[width=3cm]{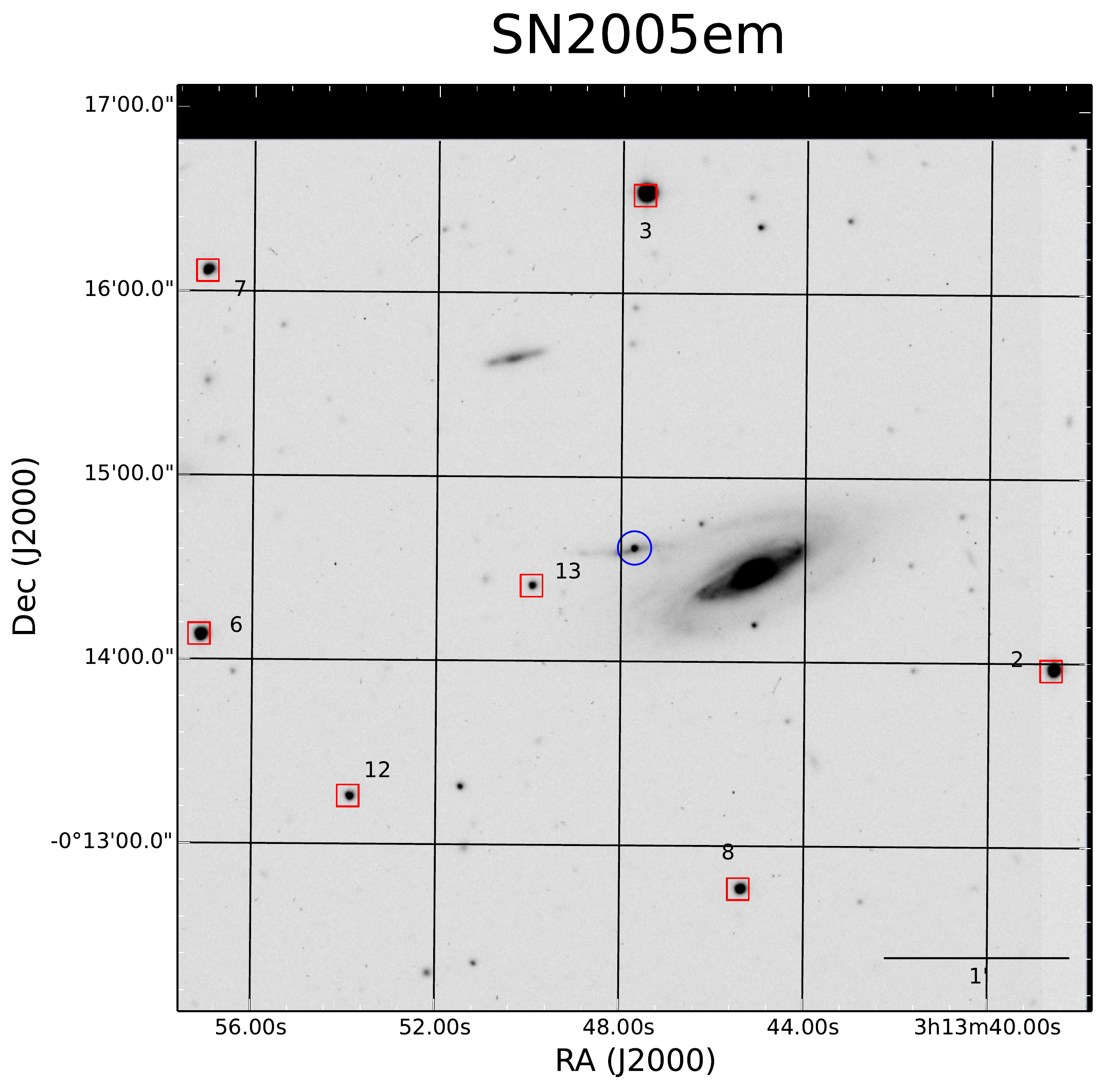}\\
\includegraphics[width=3cm]{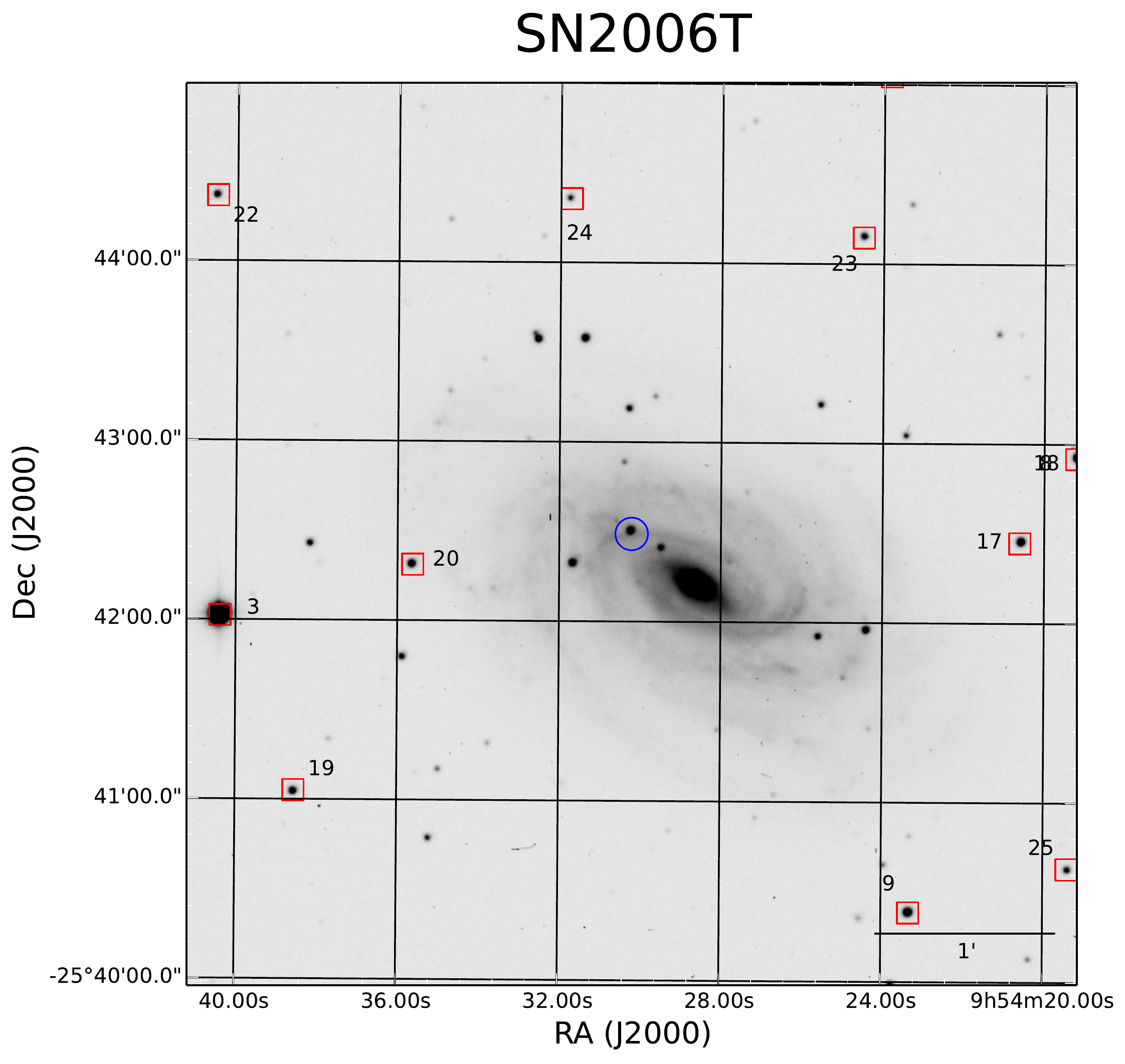}
\includegraphics[width=3cm]{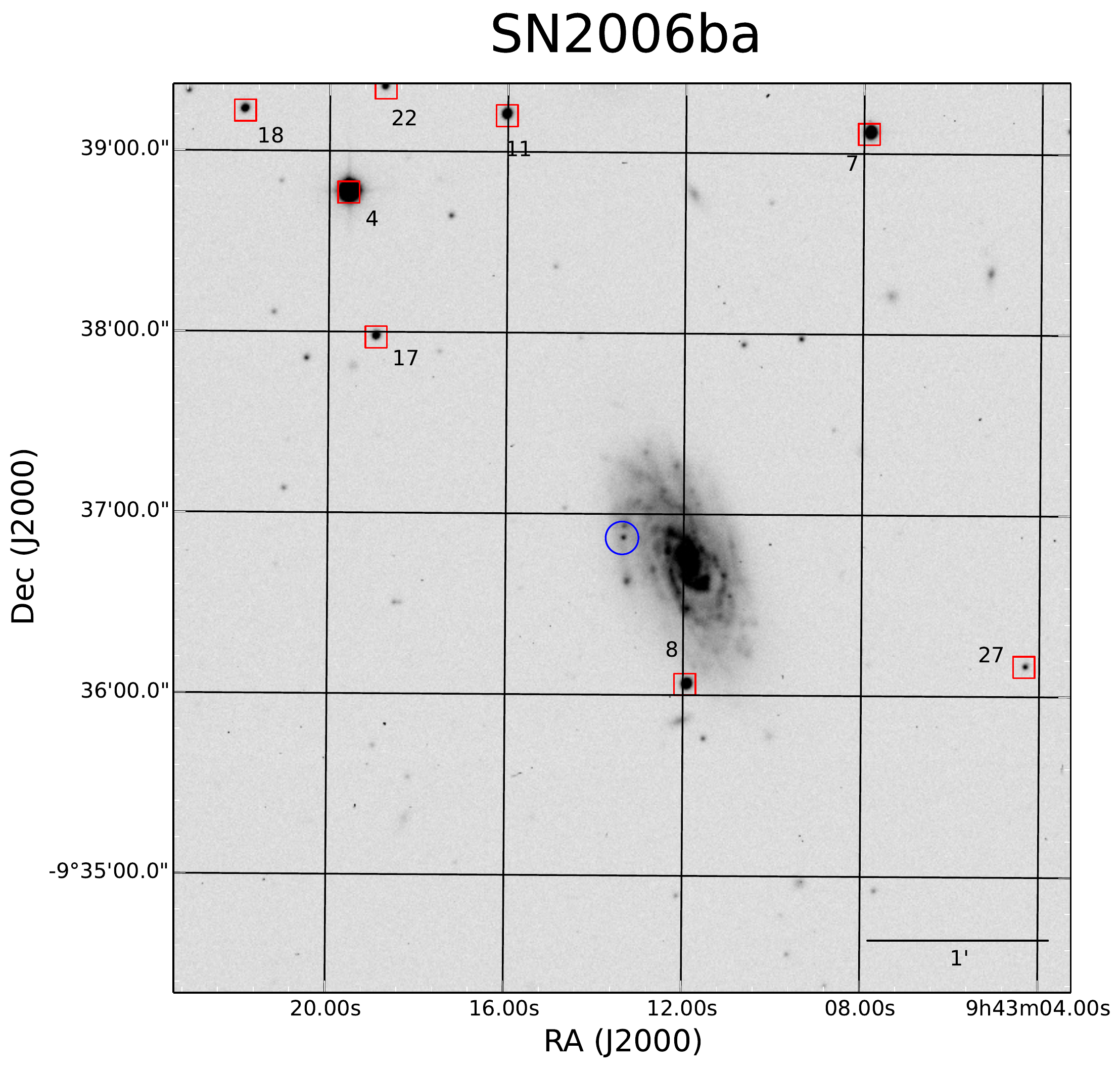}
\includegraphics[width=3cm]{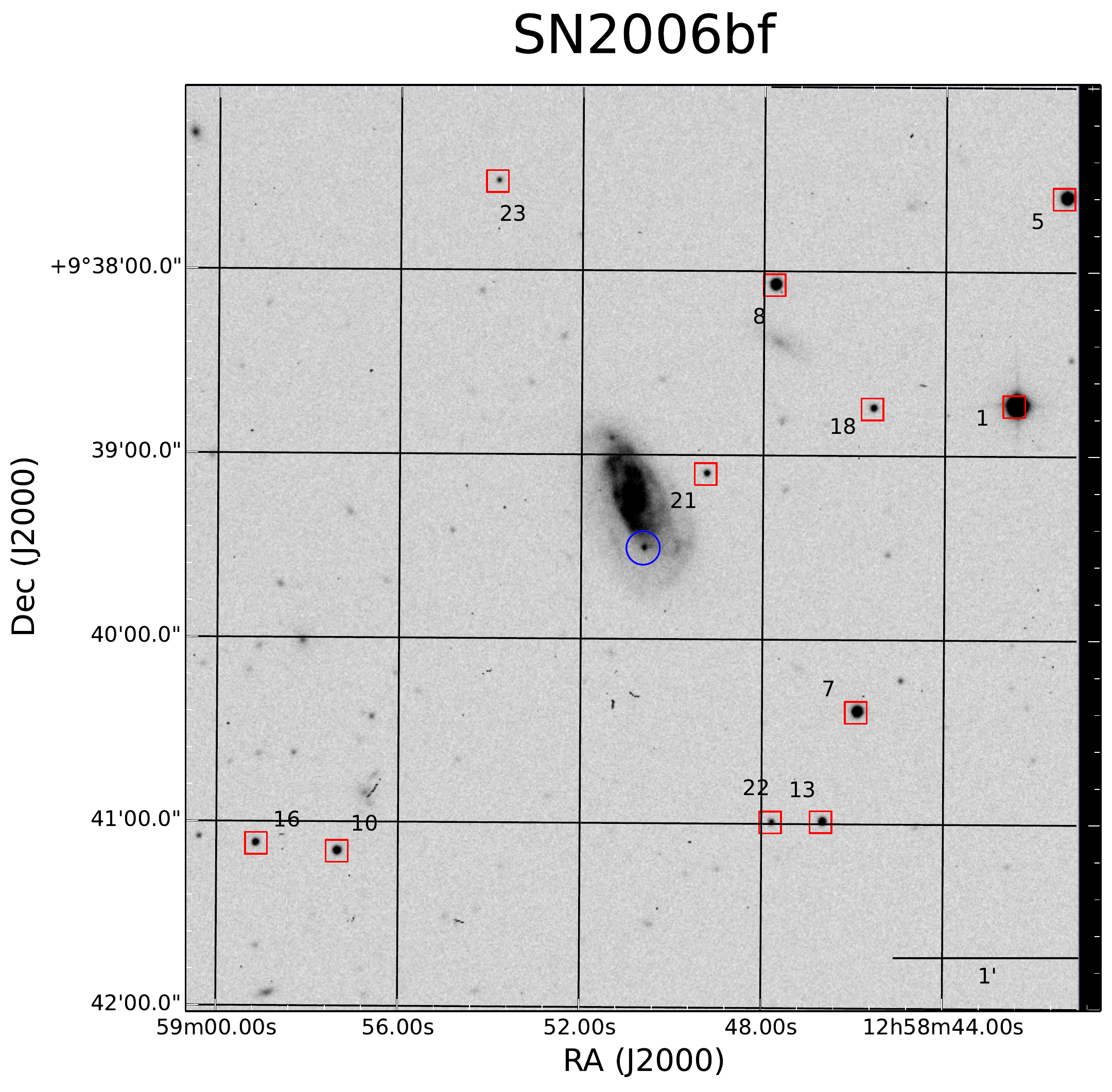}
\includegraphics[width=3cm]{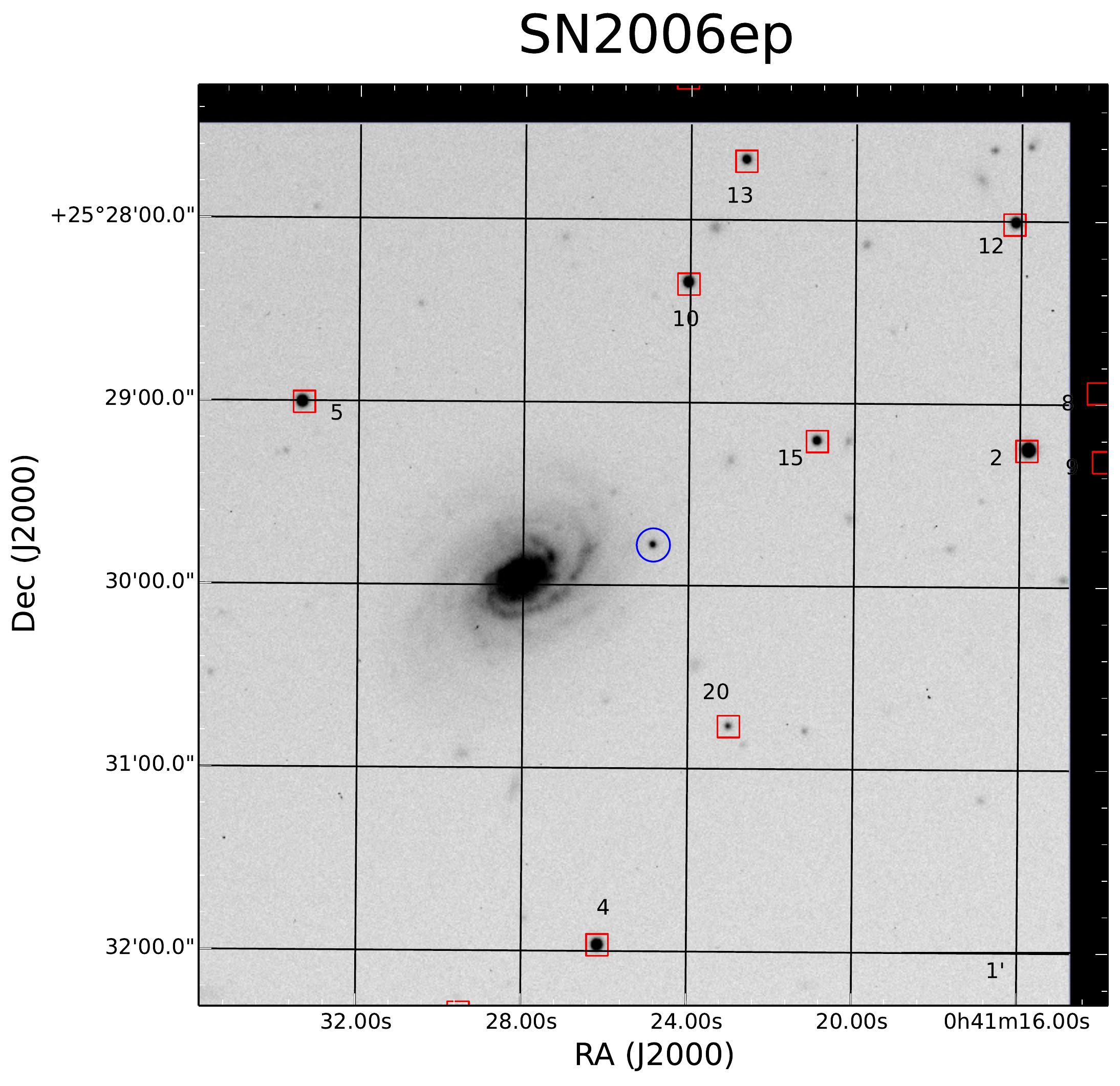}
\includegraphics[width=3cm]{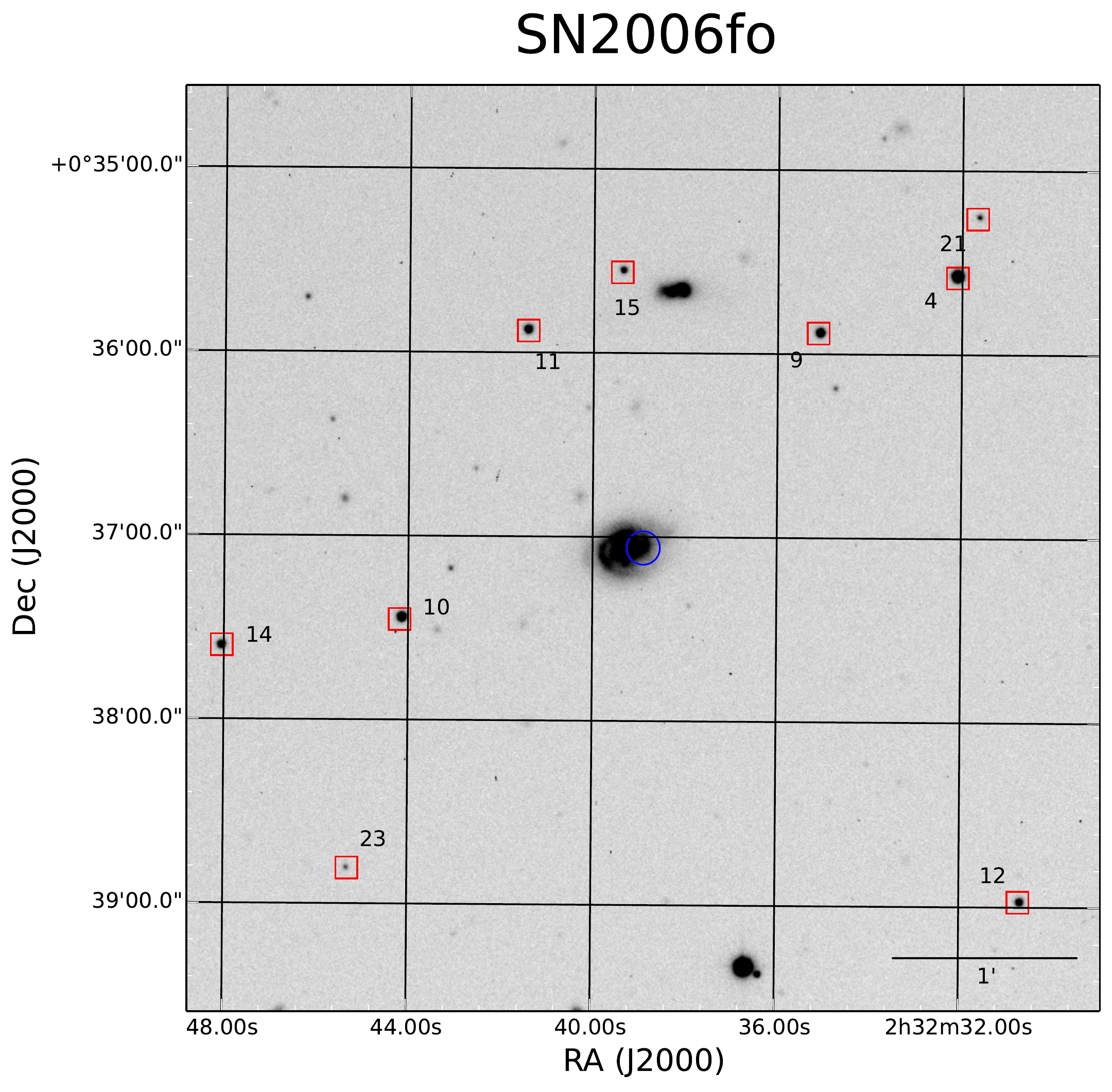}
\includegraphics[width=3cm]{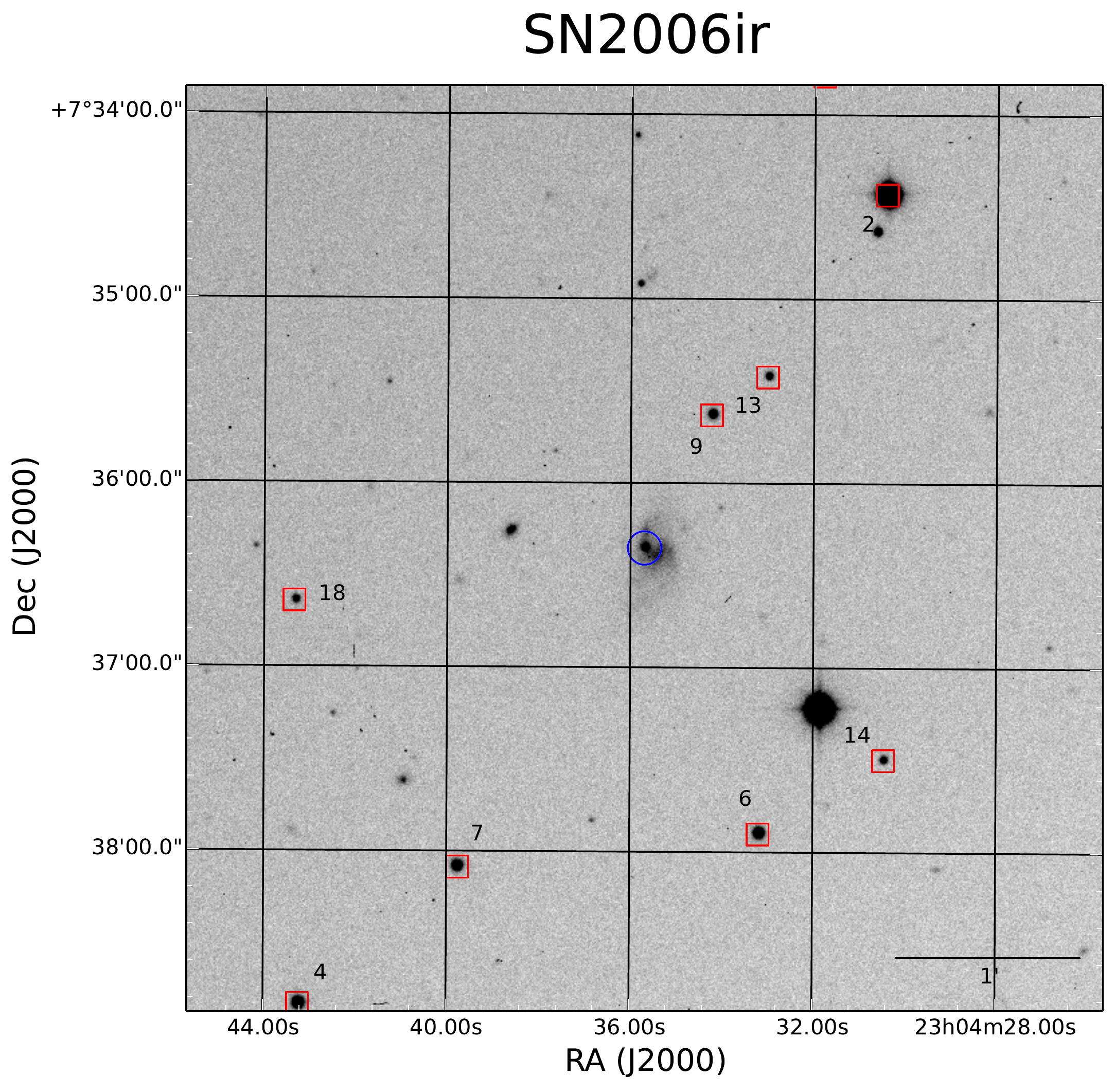}\\
\includegraphics[width=3cm]{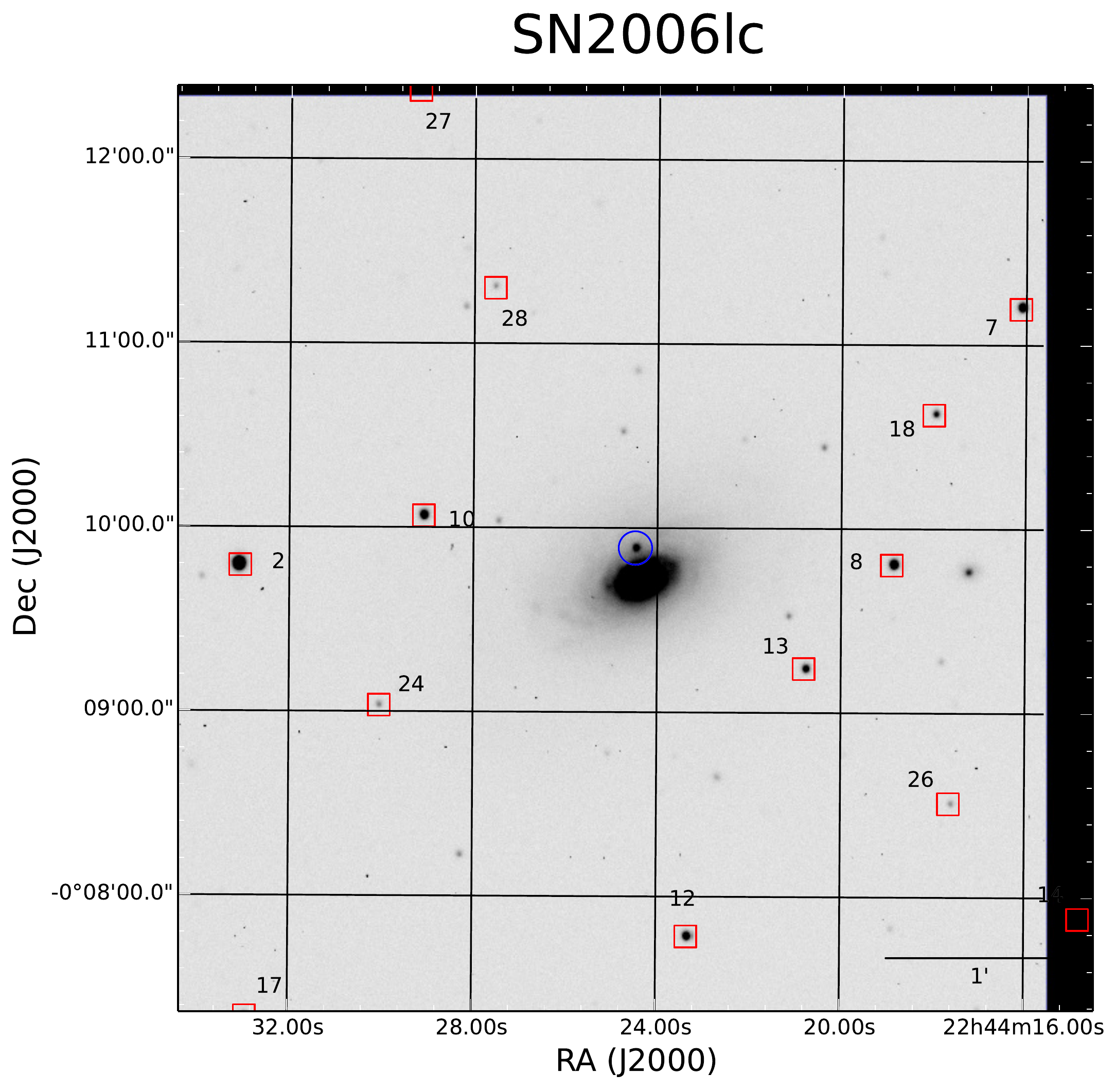}
\includegraphics[width=3cm]{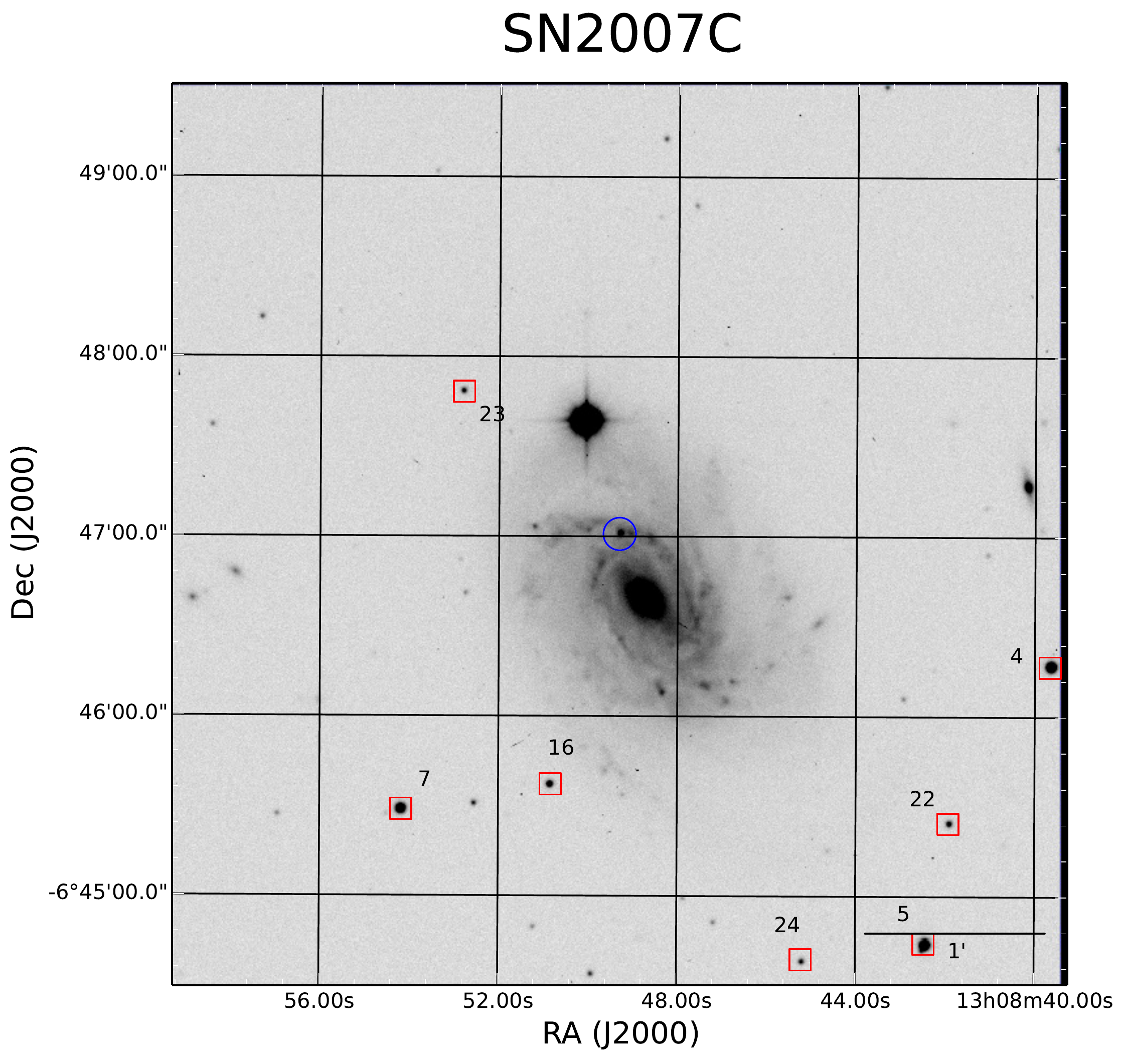}
\includegraphics[width=3cm]{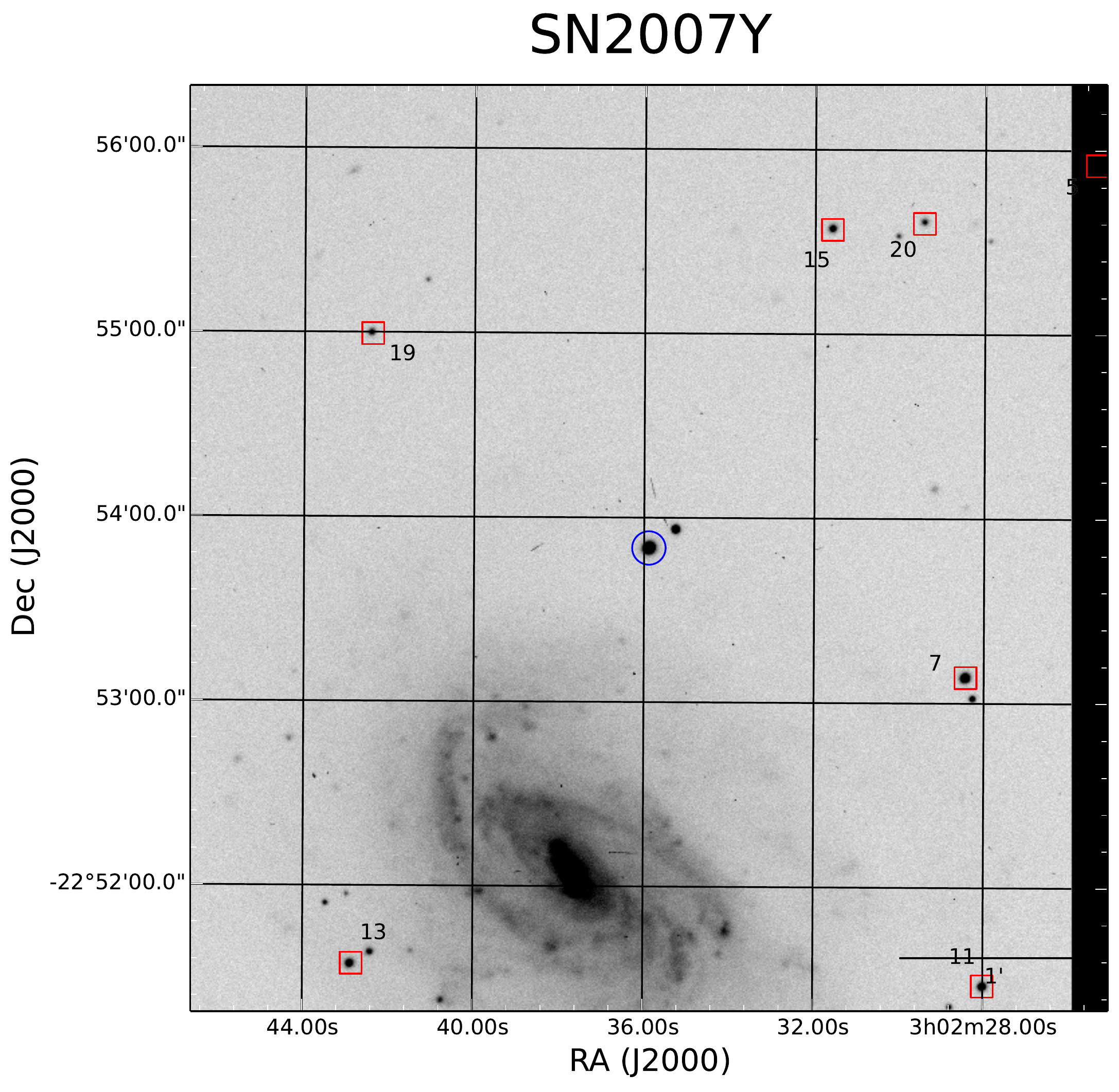}
\includegraphics[width=3cm]{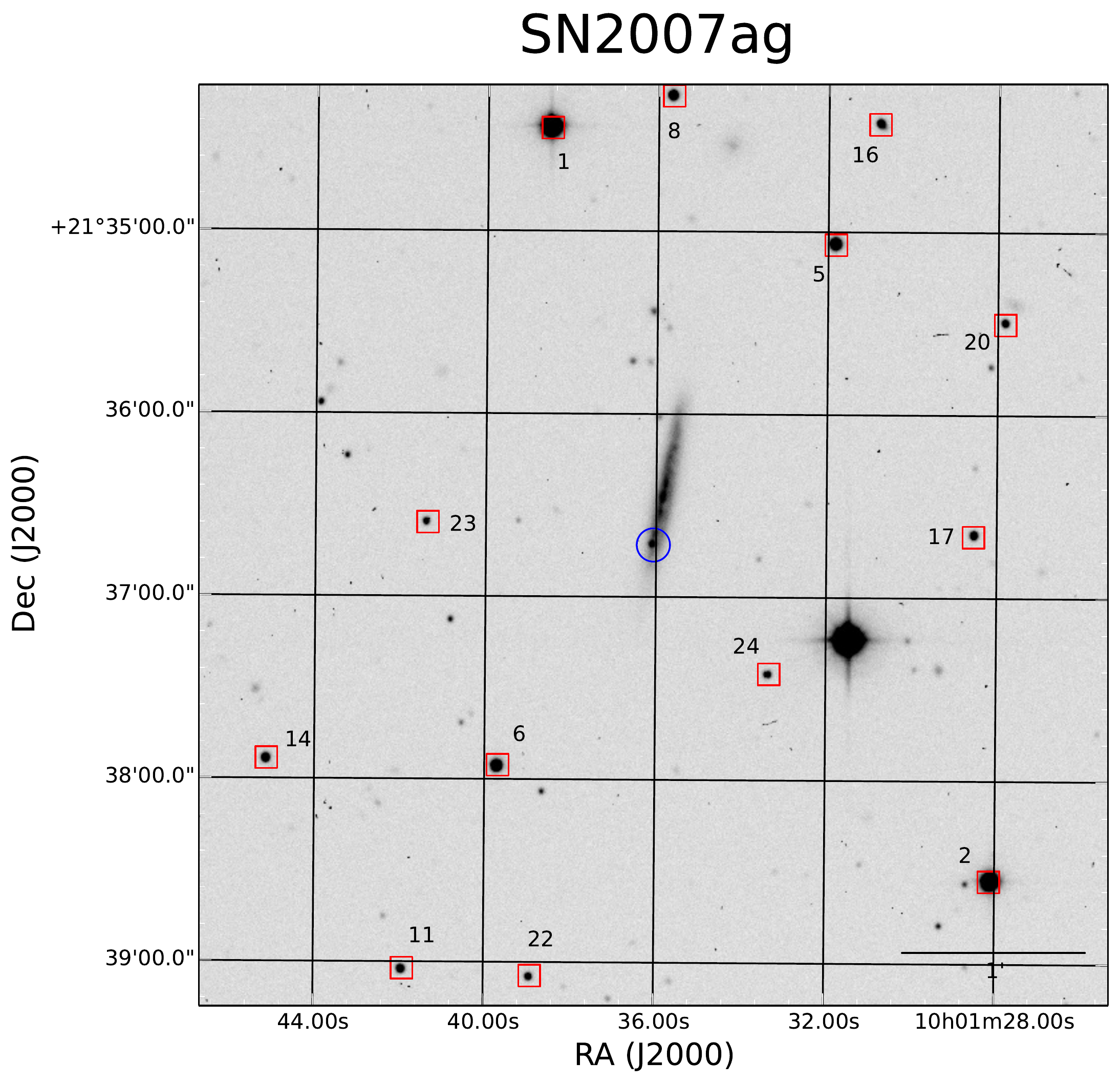}
\includegraphics[width=3cm]{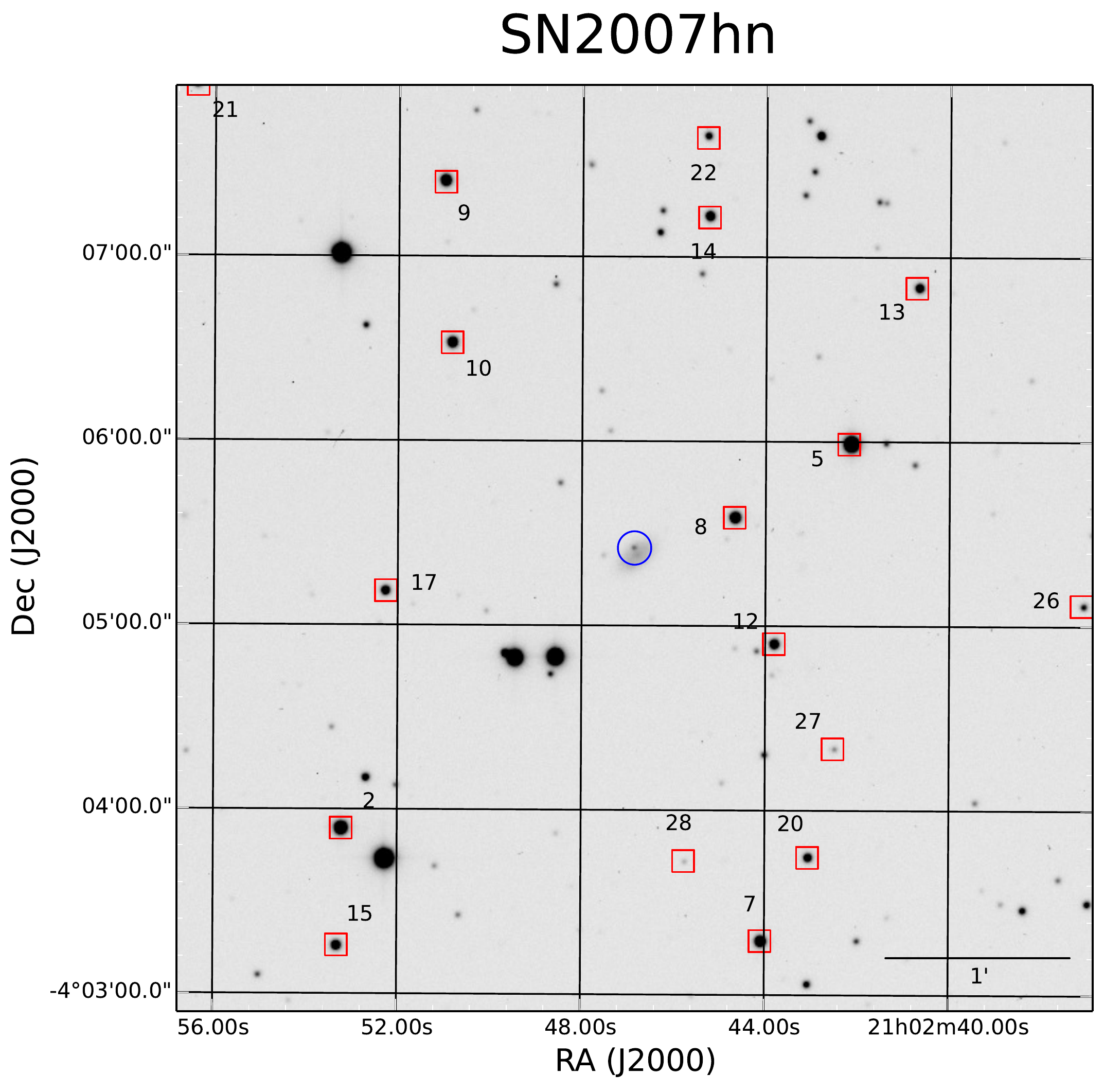}
\includegraphics[width=3cm]{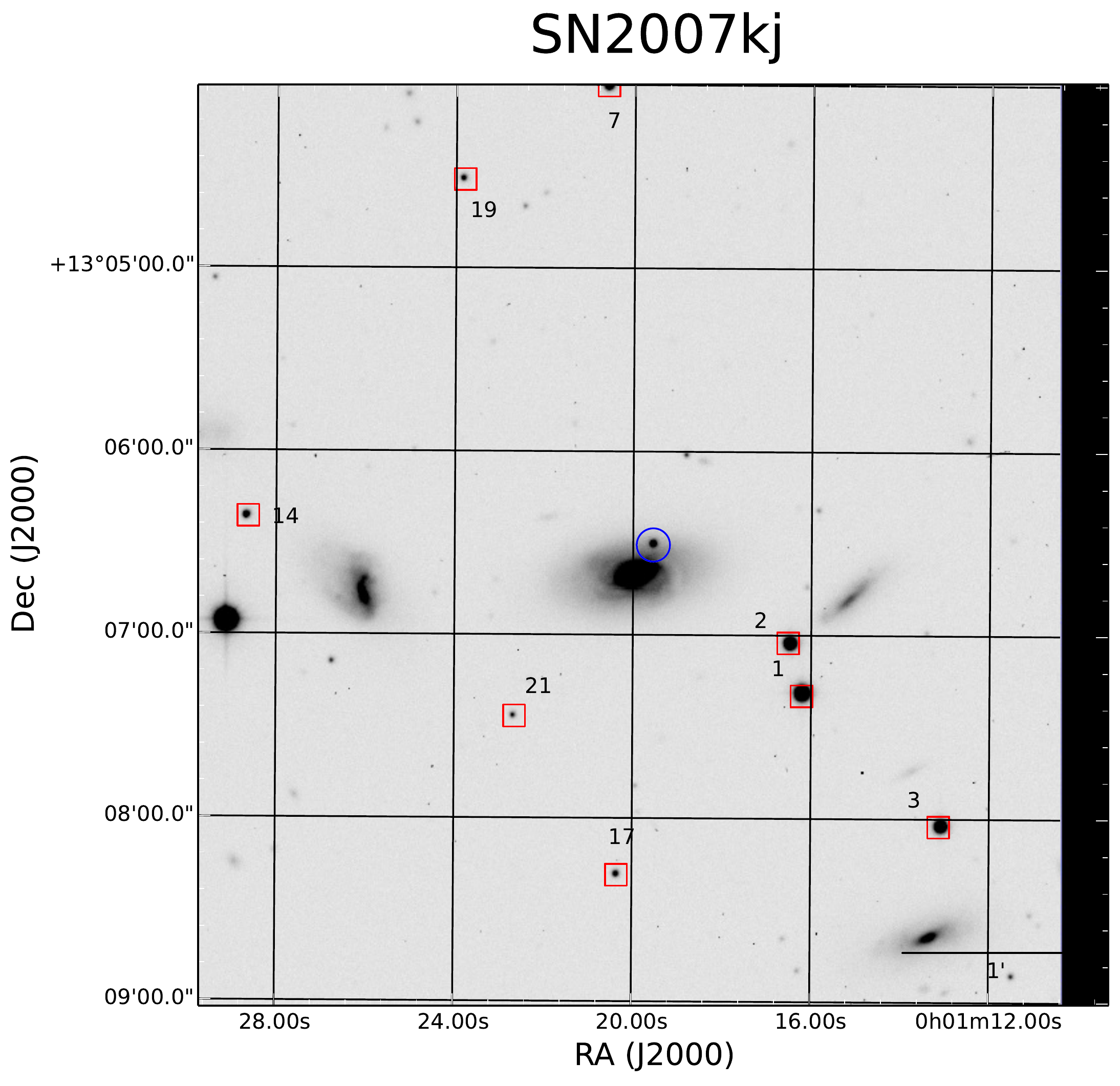}\\
\includegraphics[width=3cm]{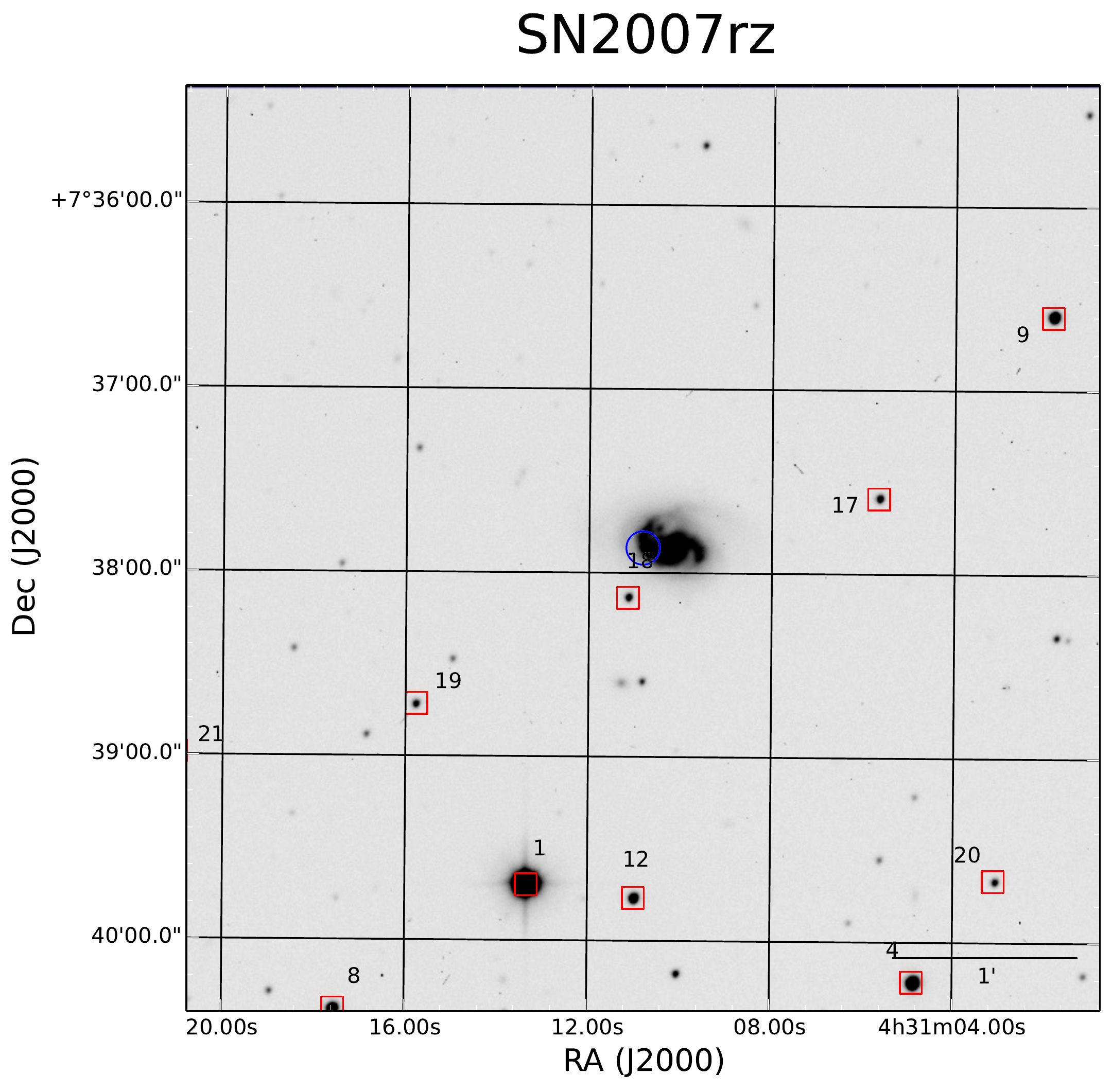}
\includegraphics[width=3cm]{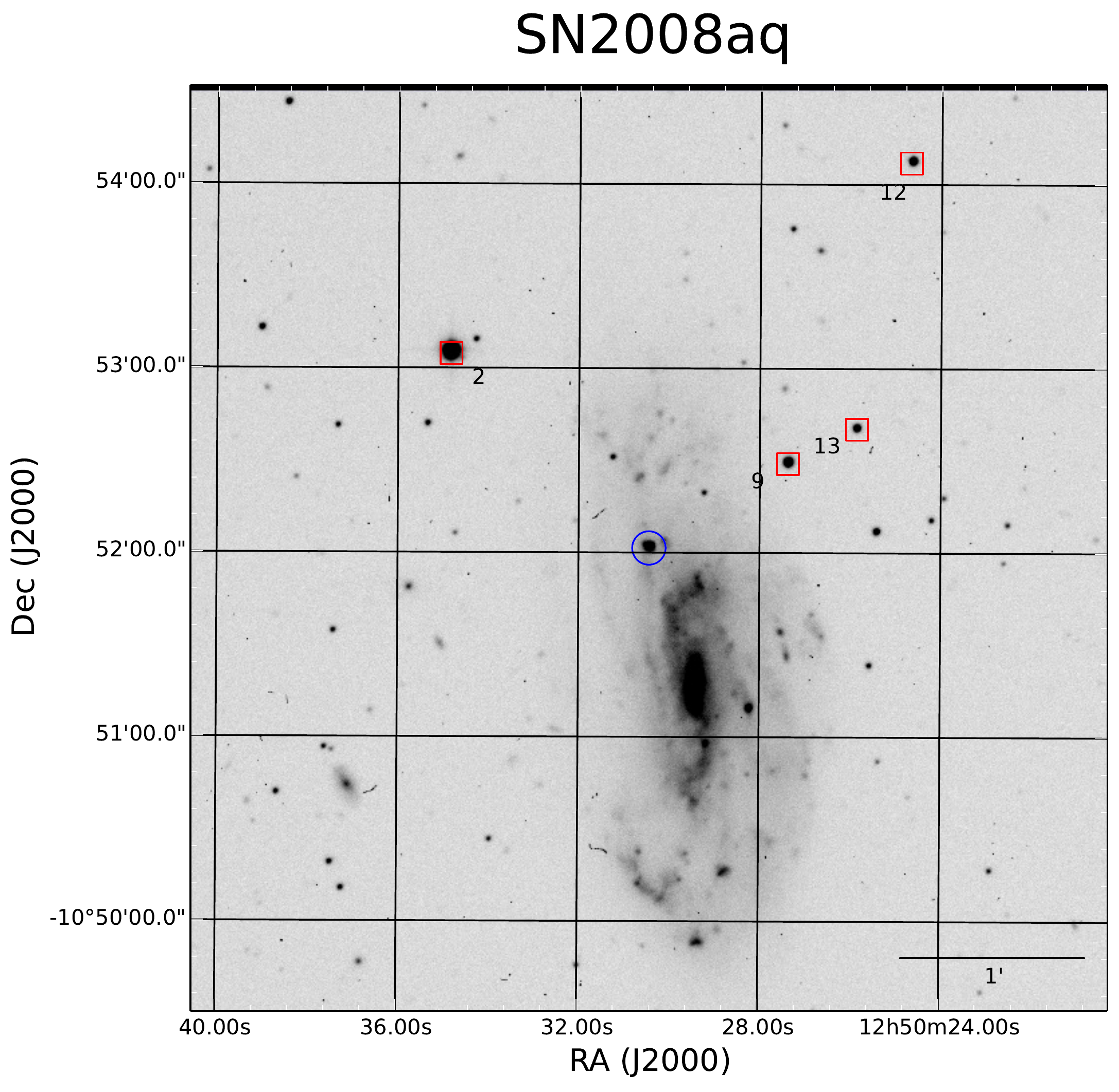}
\includegraphics[width=3cm]{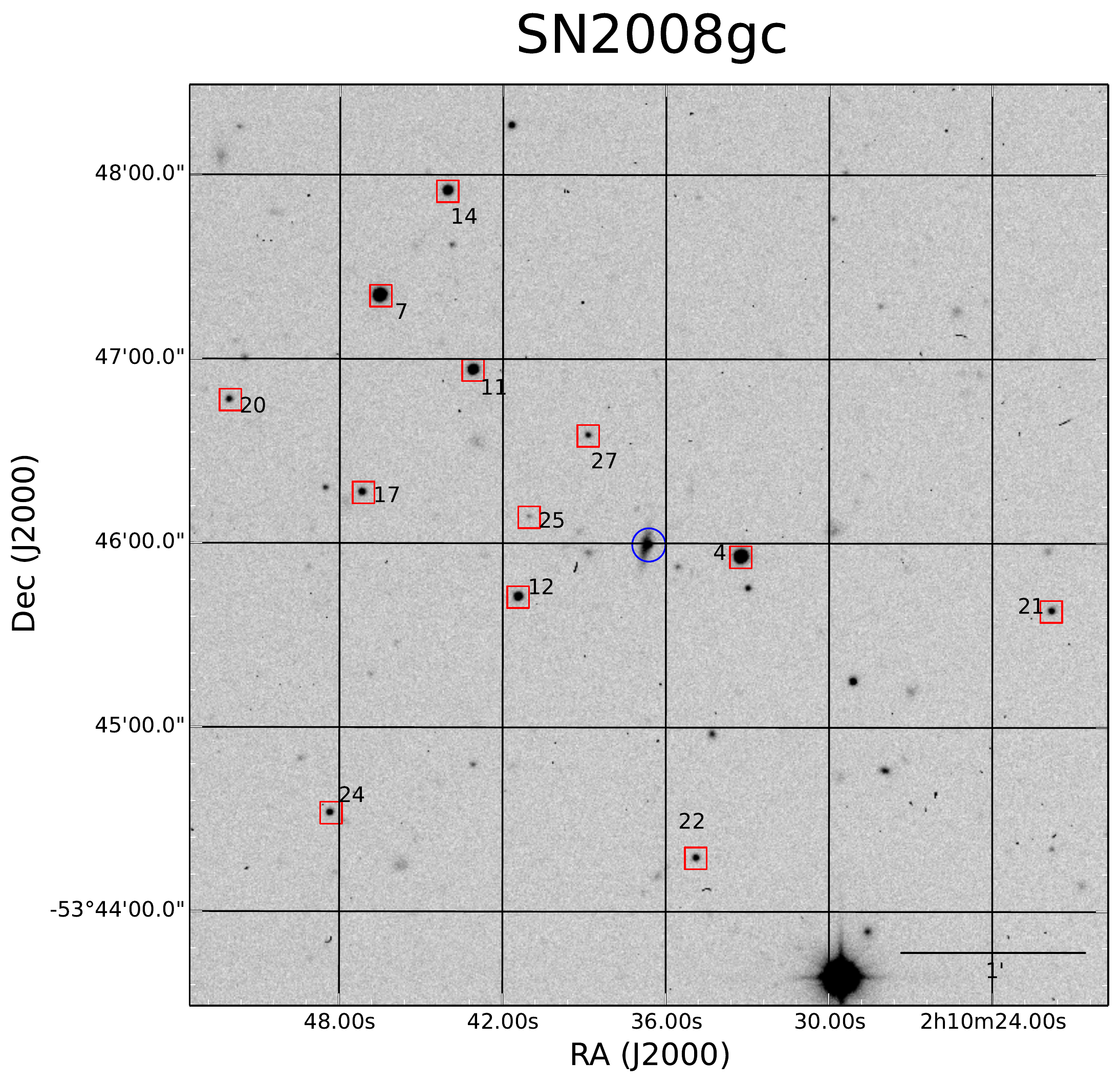}
\includegraphics[width=3cm]{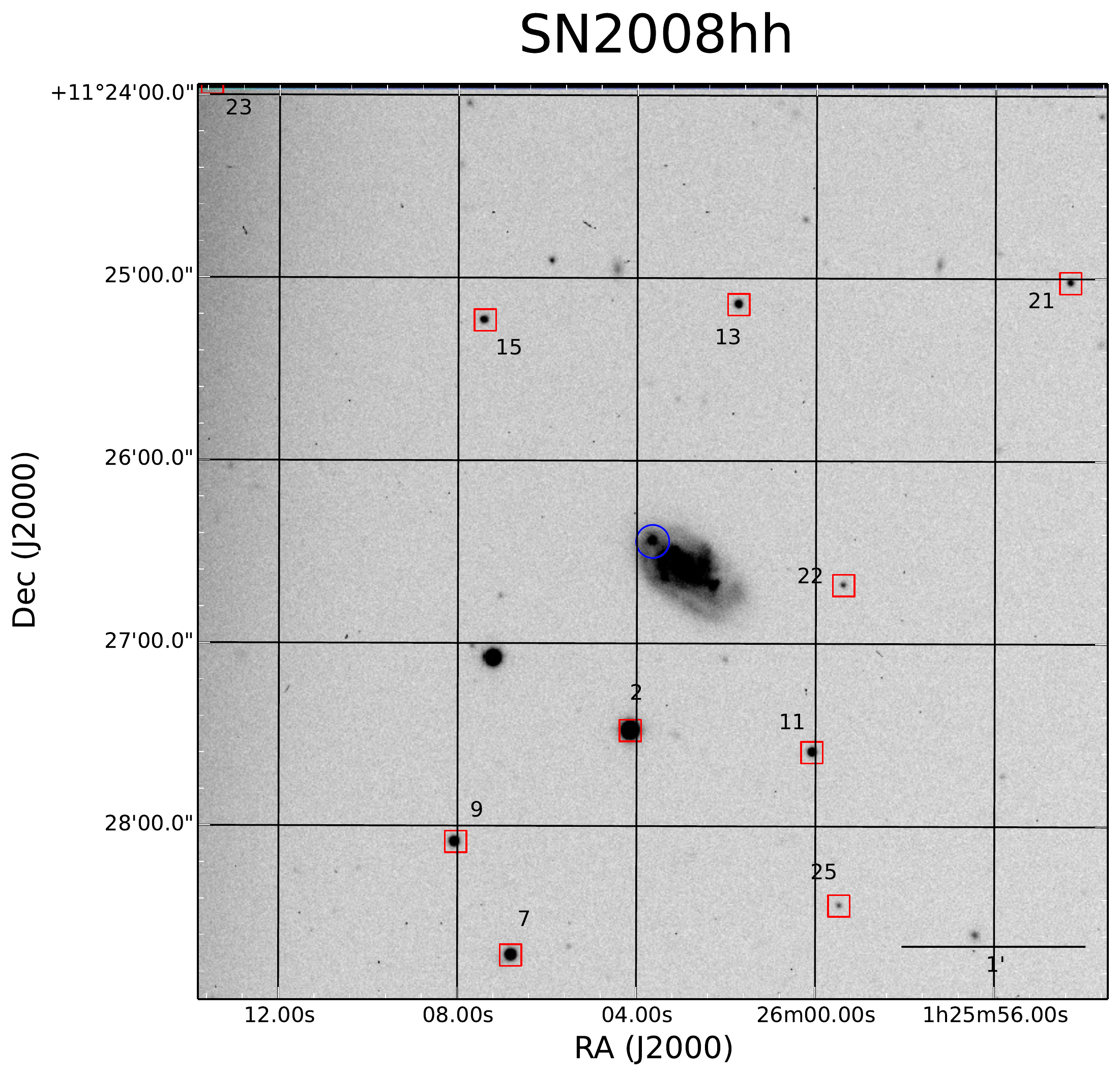}
\includegraphics[width=3cm]{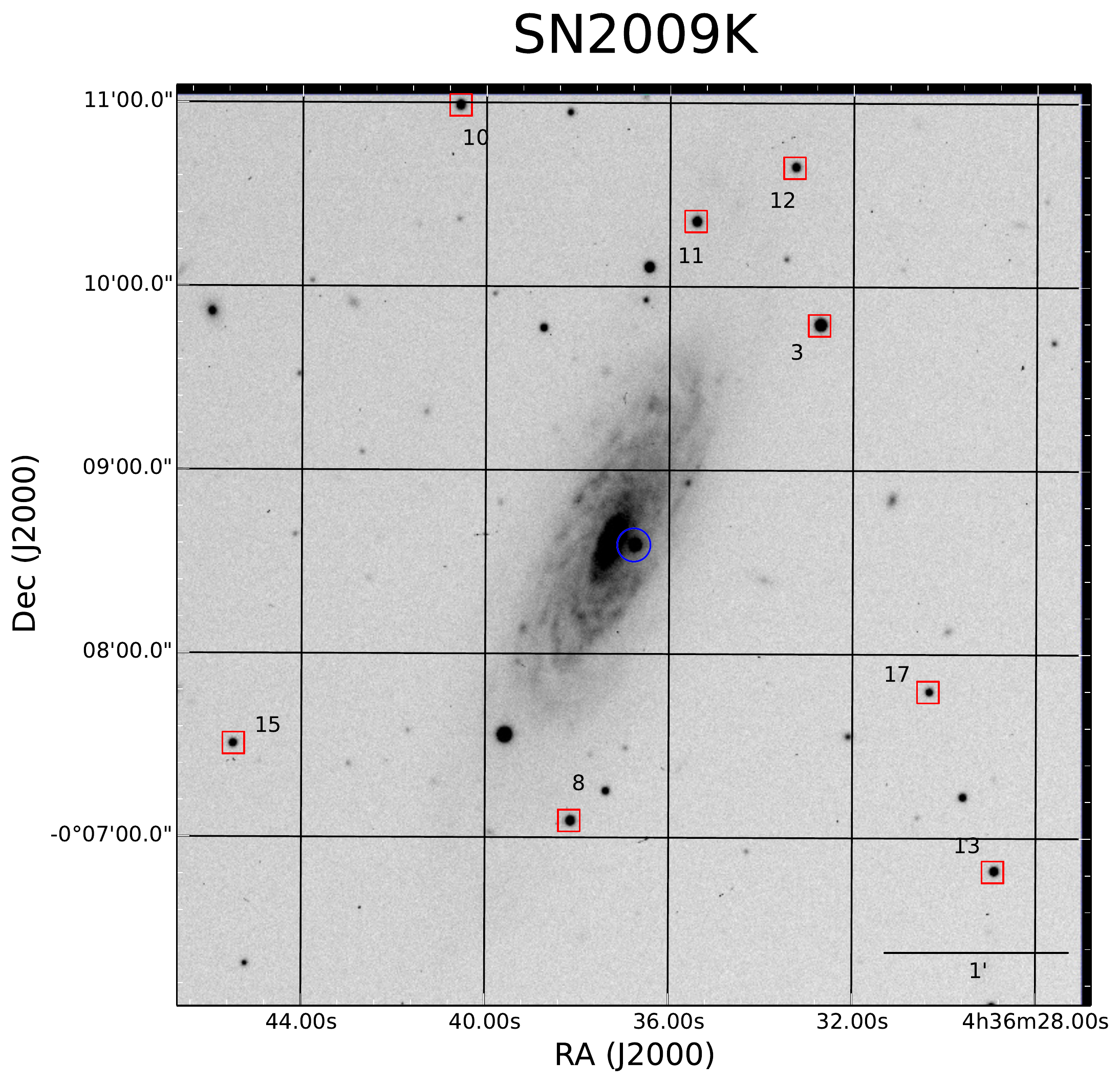}
\includegraphics[width=3cm]{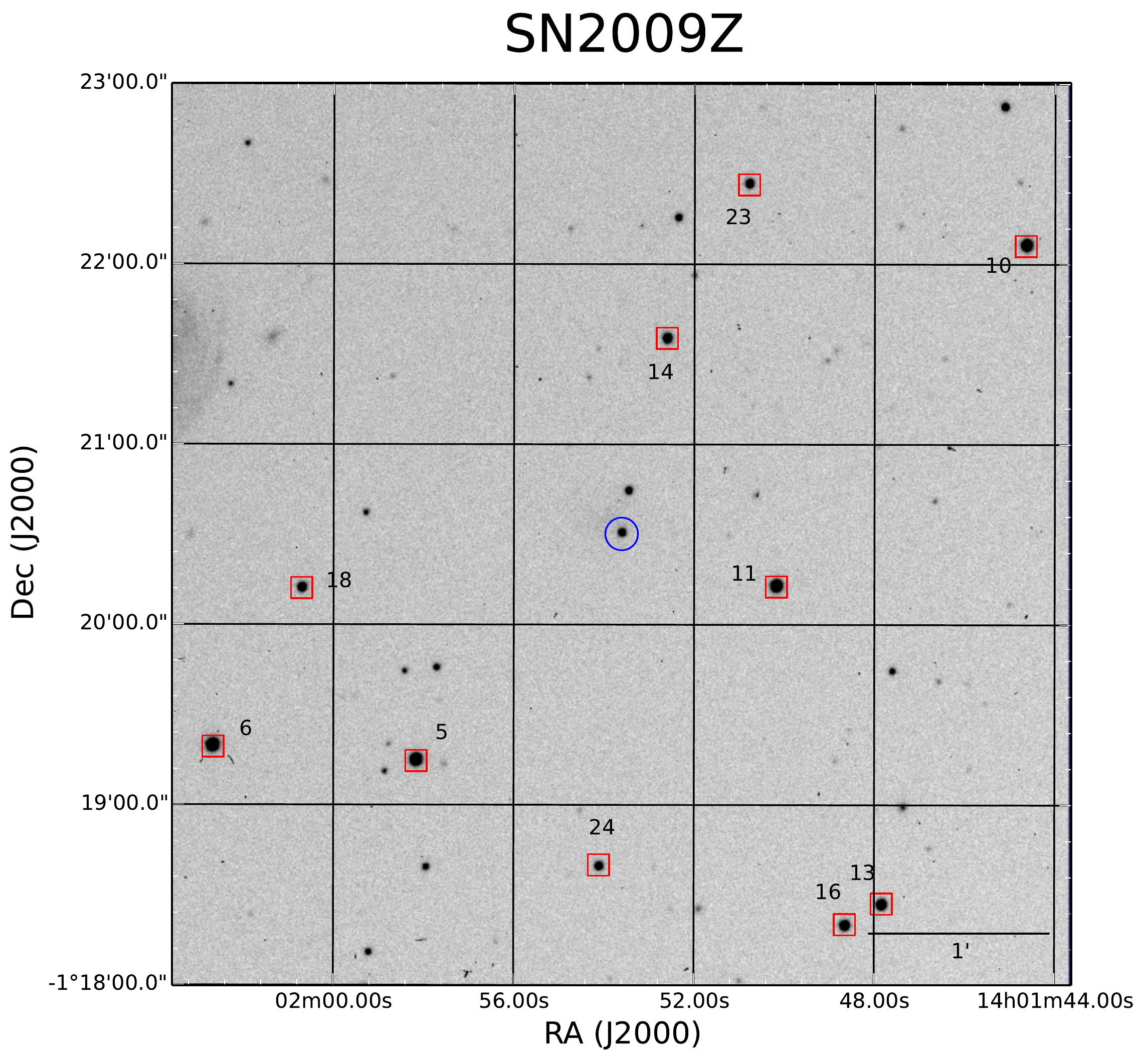}\\
\includegraphics[width=3cm]{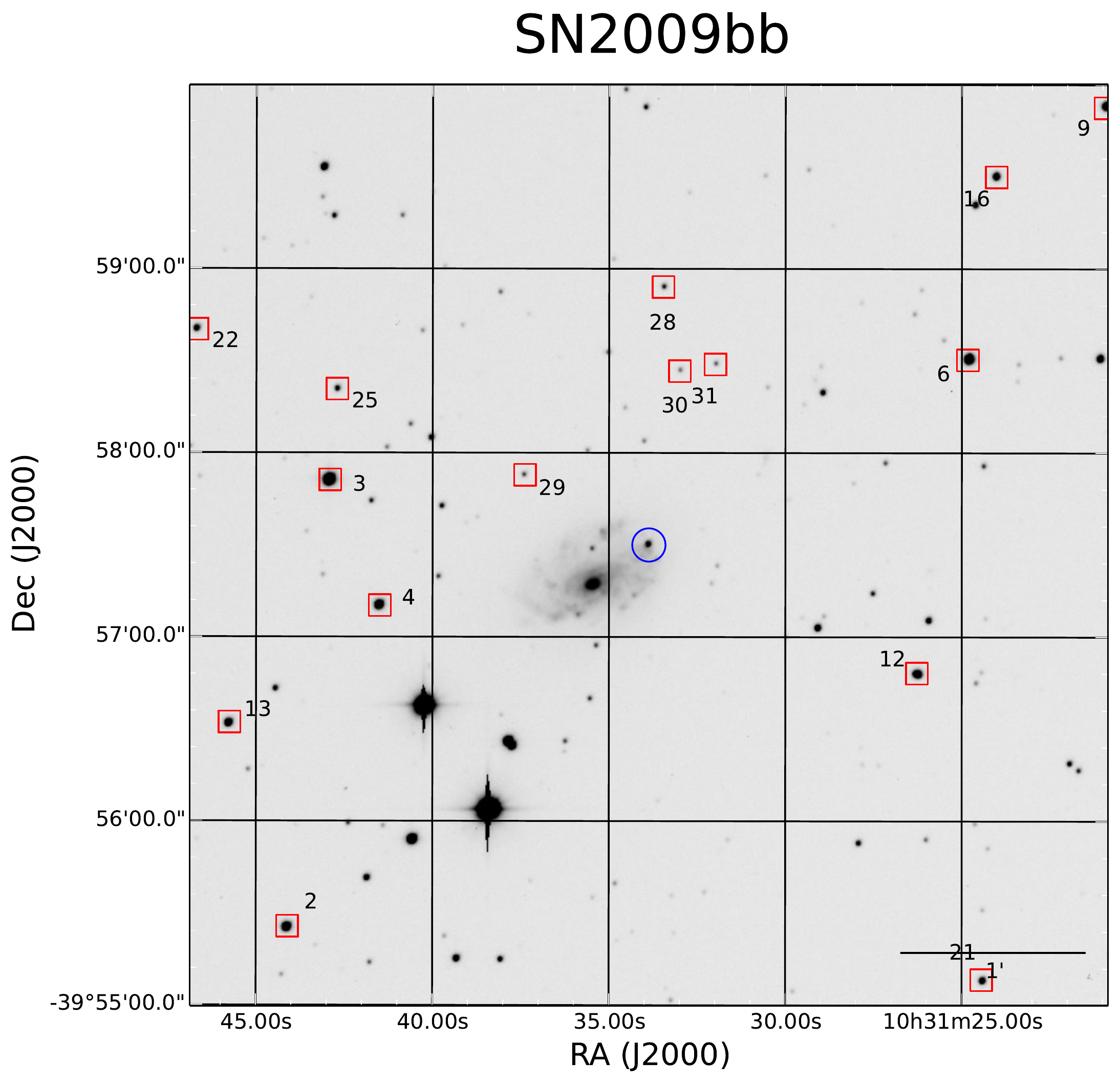}
\includegraphics[width=3cm]{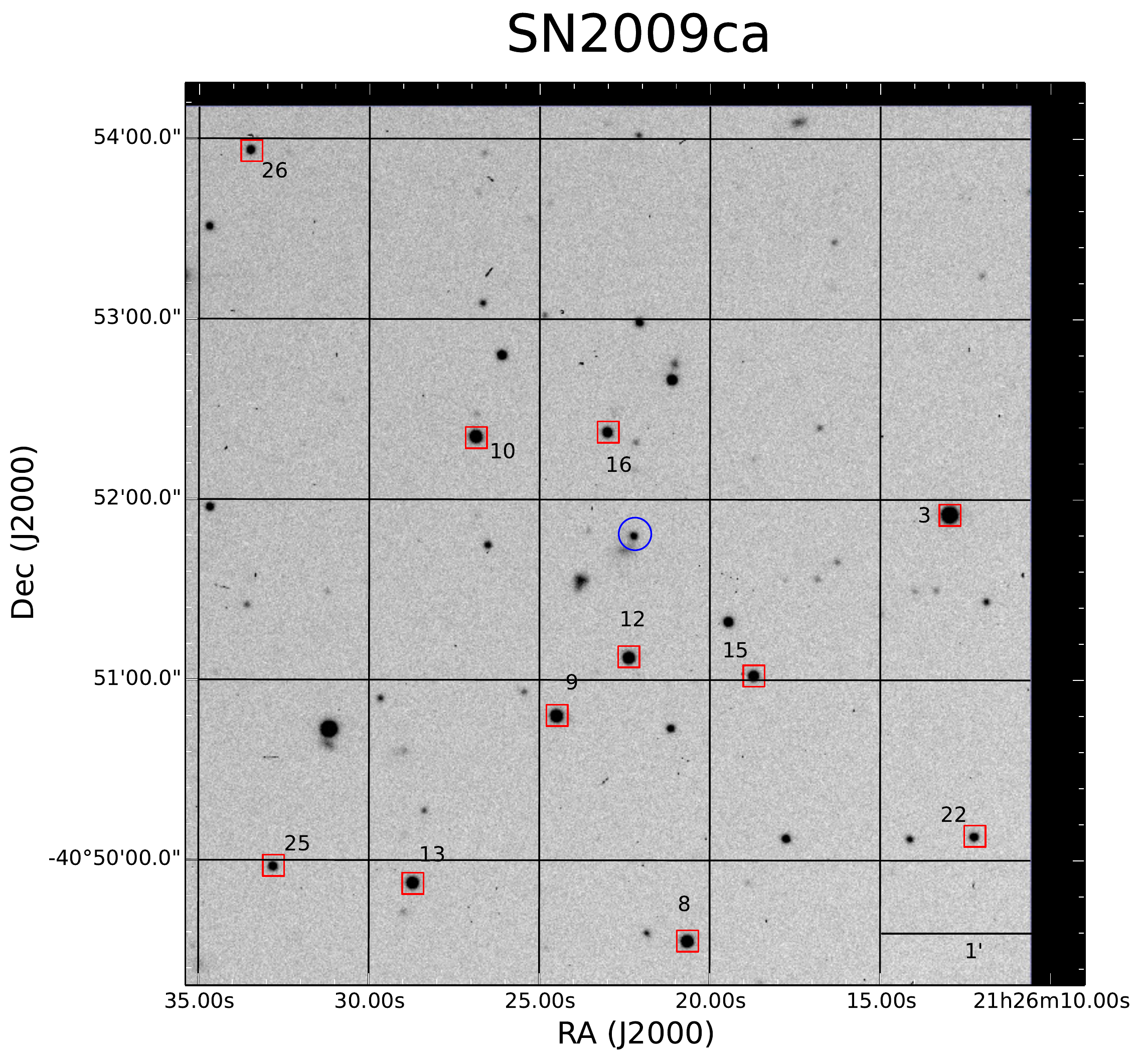}
\includegraphics[width=3cm]{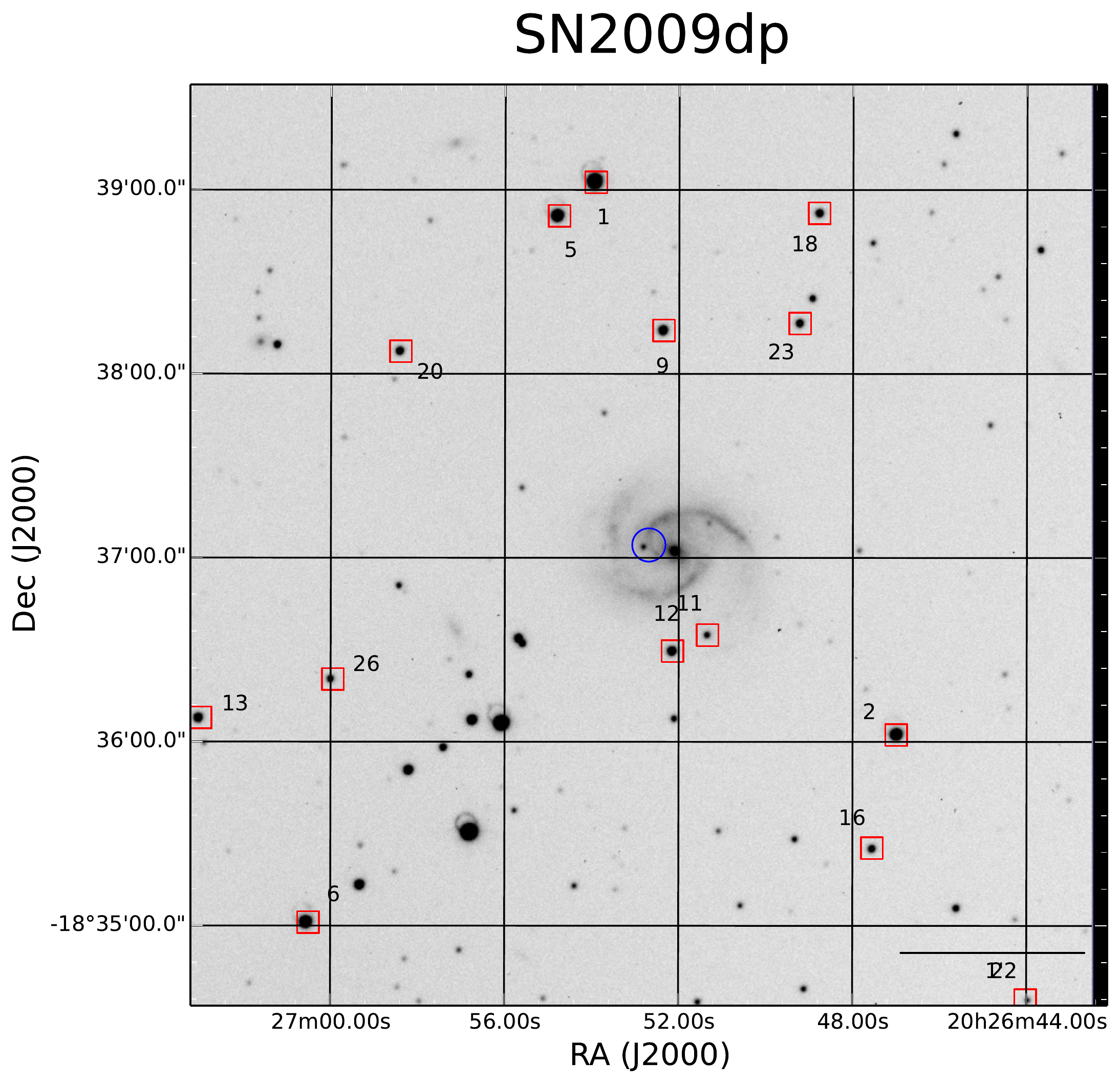}
\includegraphics[width=3cm]{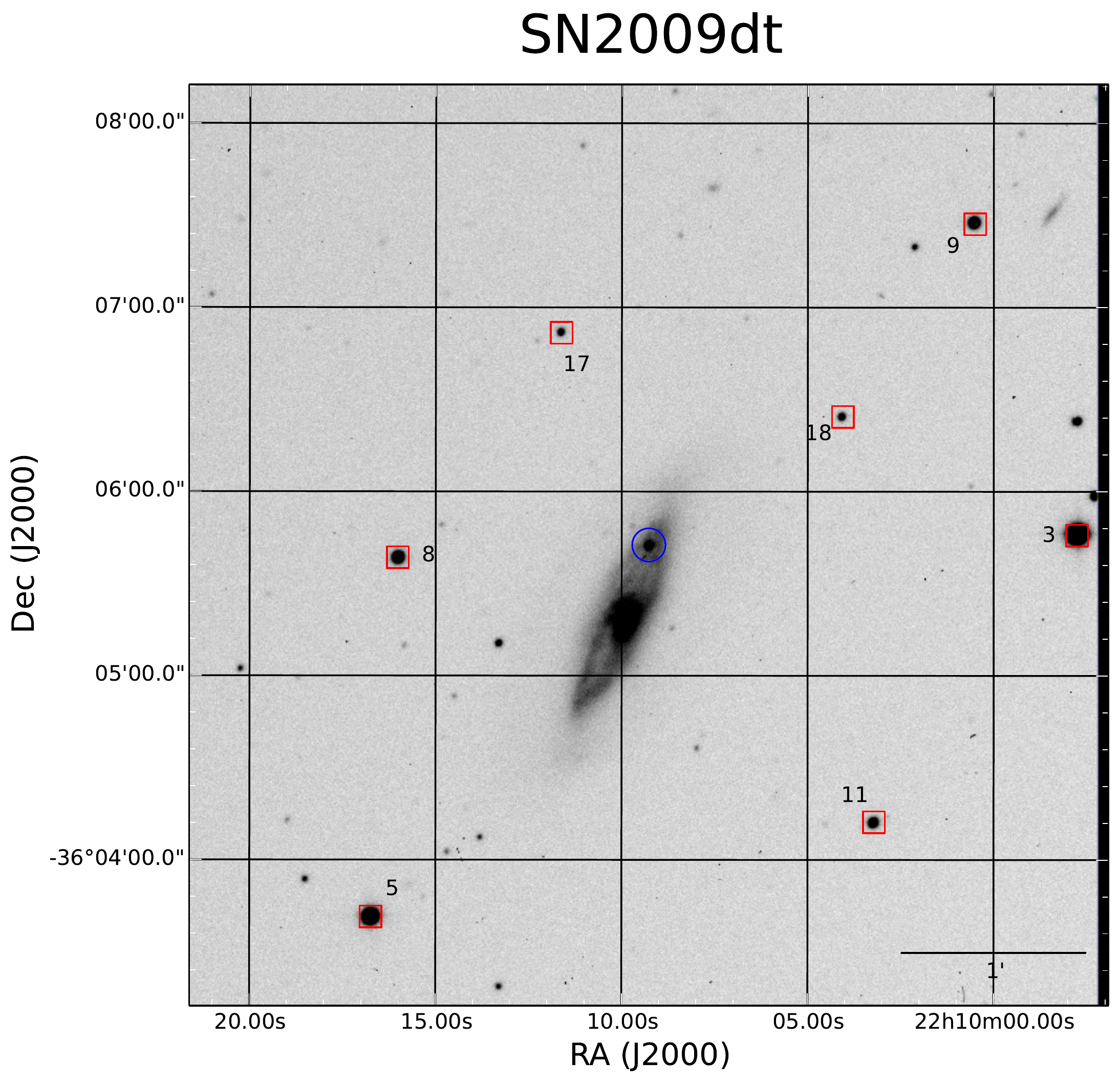}
\end{array}$
\end{center}
\caption{A mosaic of $V$-band CCD images of \nosne\ SE~SNe observed by the CSP-I. Each supernova is indicated with a blue circle, while optical local sequence stars are marked with red squares.\label{FC}}
\end{figure}

\clearpage
\setcounter{figure}{2}
\begin{figure}[h]
\centering
\includegraphics[width=5.6in]{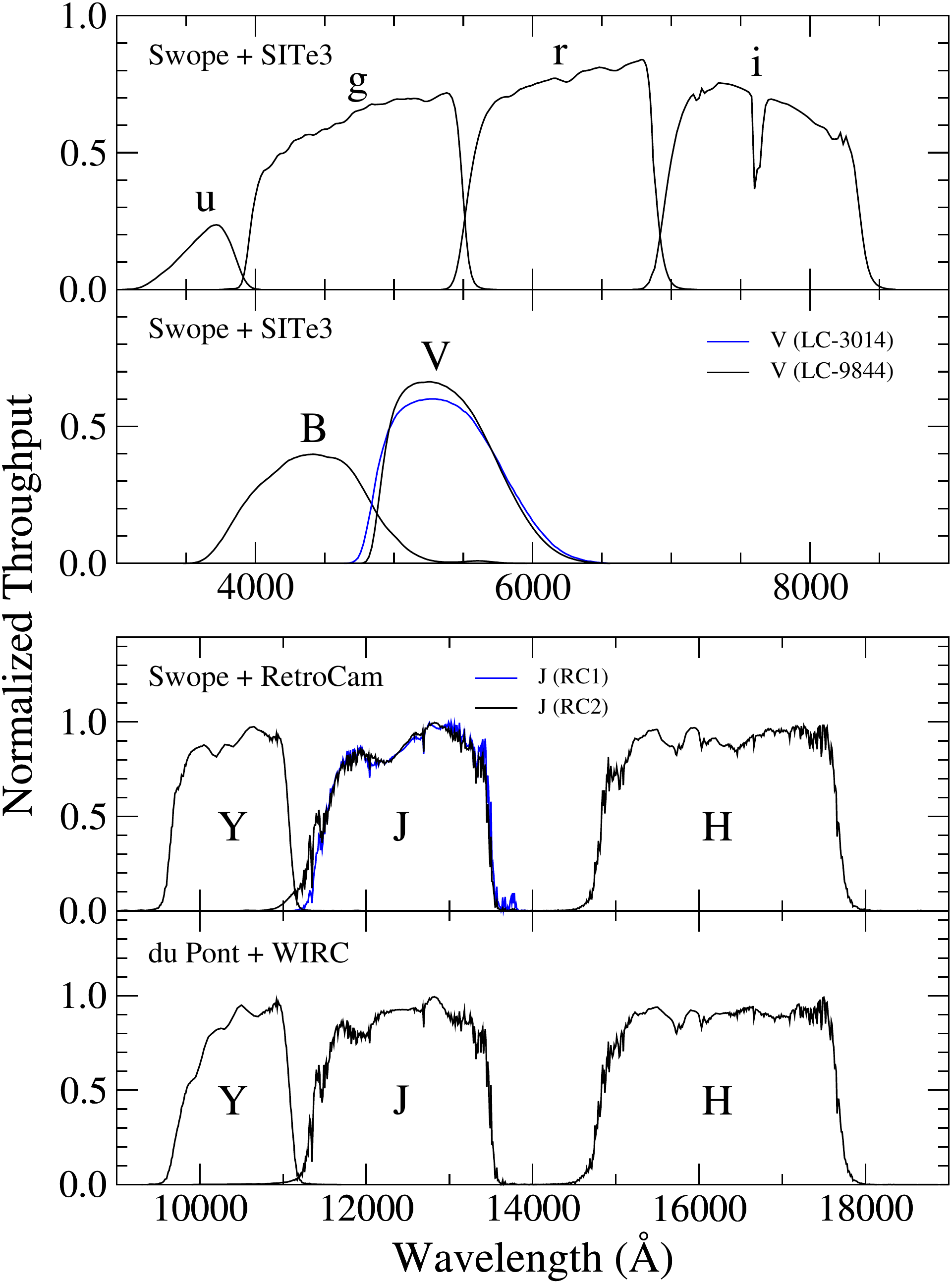}
\caption{CSP-I optical (top panel) and NIR (bottom panel) system response (telescope$+$detector$+$filters) functions for the Swope  ($+$SITe3 and $+$RetroCam)  and  du Pont ($+$WIRC) telescopes.
The normalized response functions include multiplications by an atmospheric transmission function for   an airmass of 1.2 and telluric absorption spectrum appropriate to LCO.\label{passbands}}
\end{figure}

\clearpage
\begin{figure}
\setcounter{figure}{3}
\begin{center}$
\begin{array}{cc}
\includegraphics[width=0.5\textwidth]{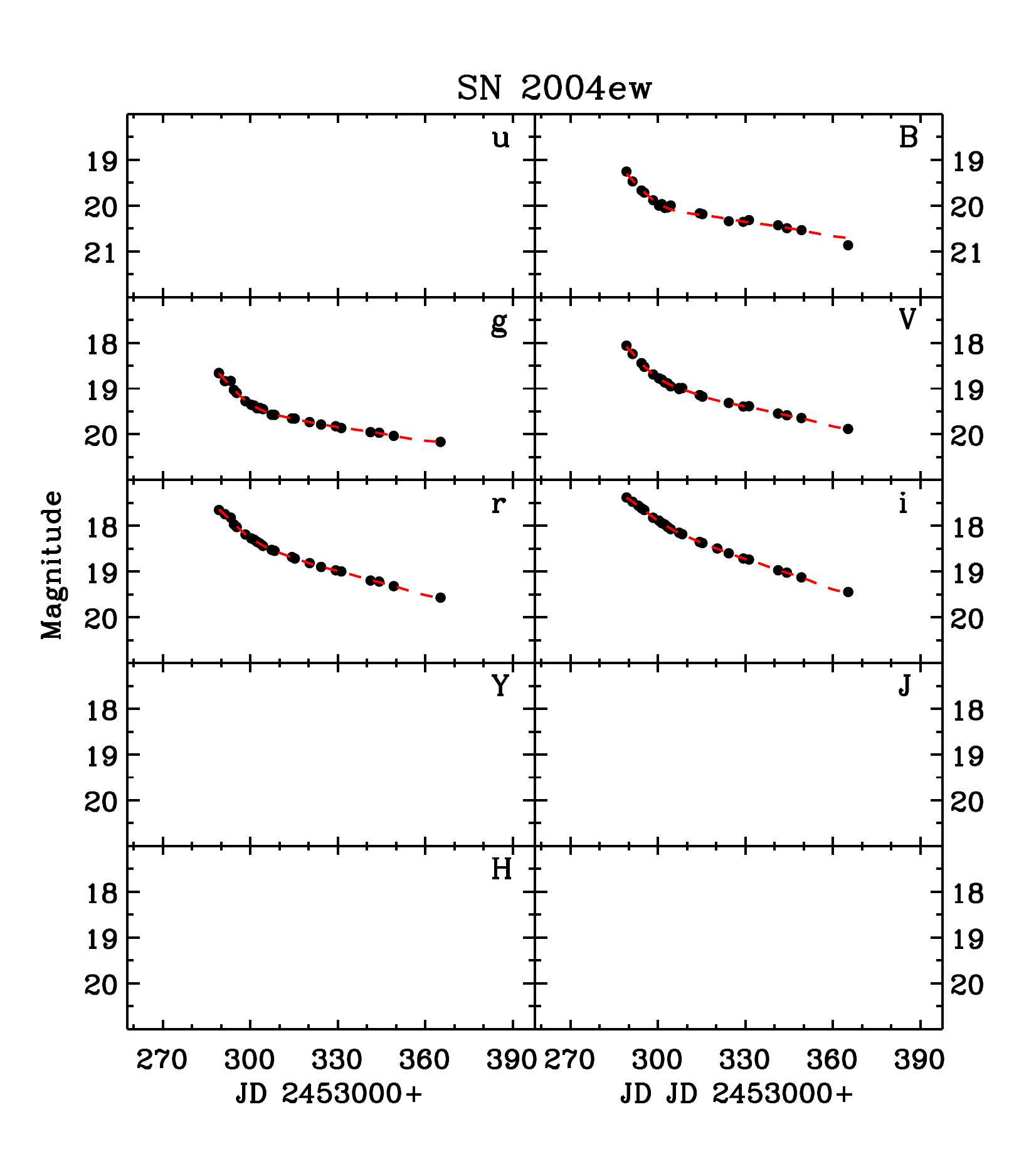}
\includegraphics[width=0.5\textwidth]{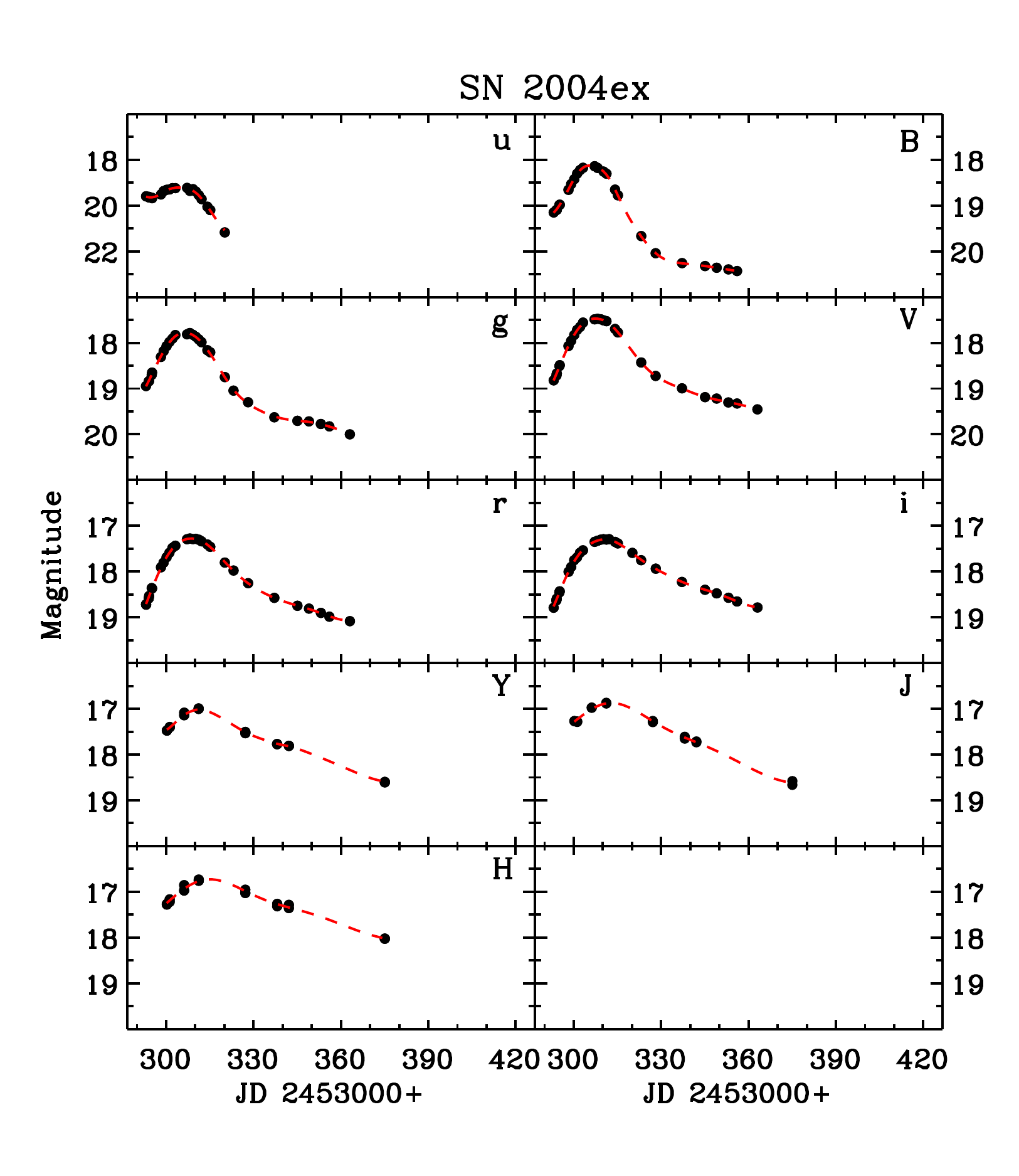}\\
\includegraphics[width=0.5\textwidth]{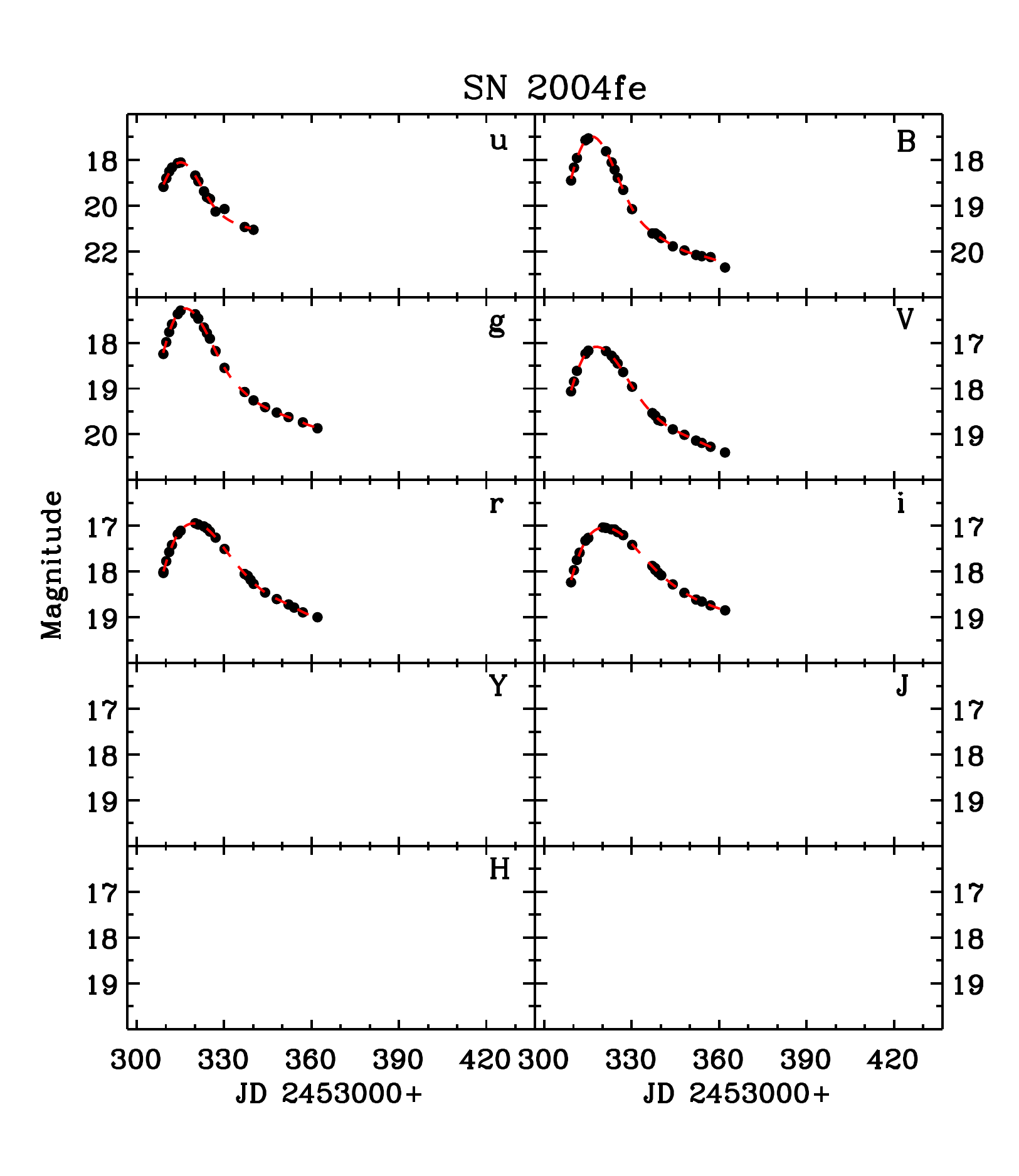}
\includegraphics[width=0.5\textwidth]{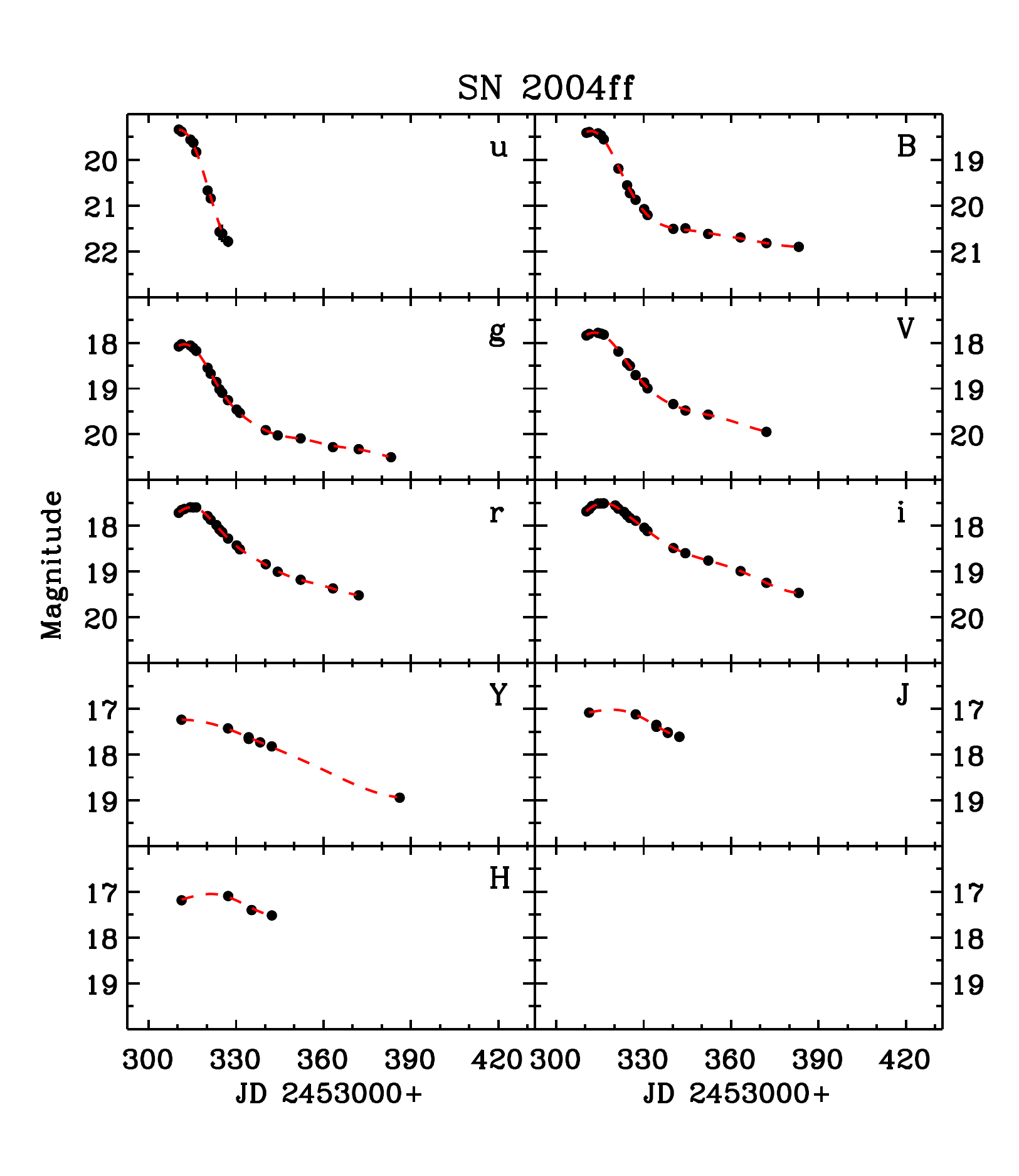}\\
\end{array}$
\end{center}
\caption{Light curves of  CSP-I SE~SNe in the 
 Swope natural photometric system.  
 Uncertainties in the photometry are smaller
  than the points, unless shown.  The smooth curves correspond to  Gaussian process spline  
  fits computed within the light-curve analysis software package {\tt SNooPy} \citep{burns11}. \label{fig:LCs}  
}
\end{figure}
\clearpage

\setcounter{figure}{3}
\begin{figure}
\begin{center}$
\begin{array}{cc}
\includegraphics[width=0.5\textwidth]{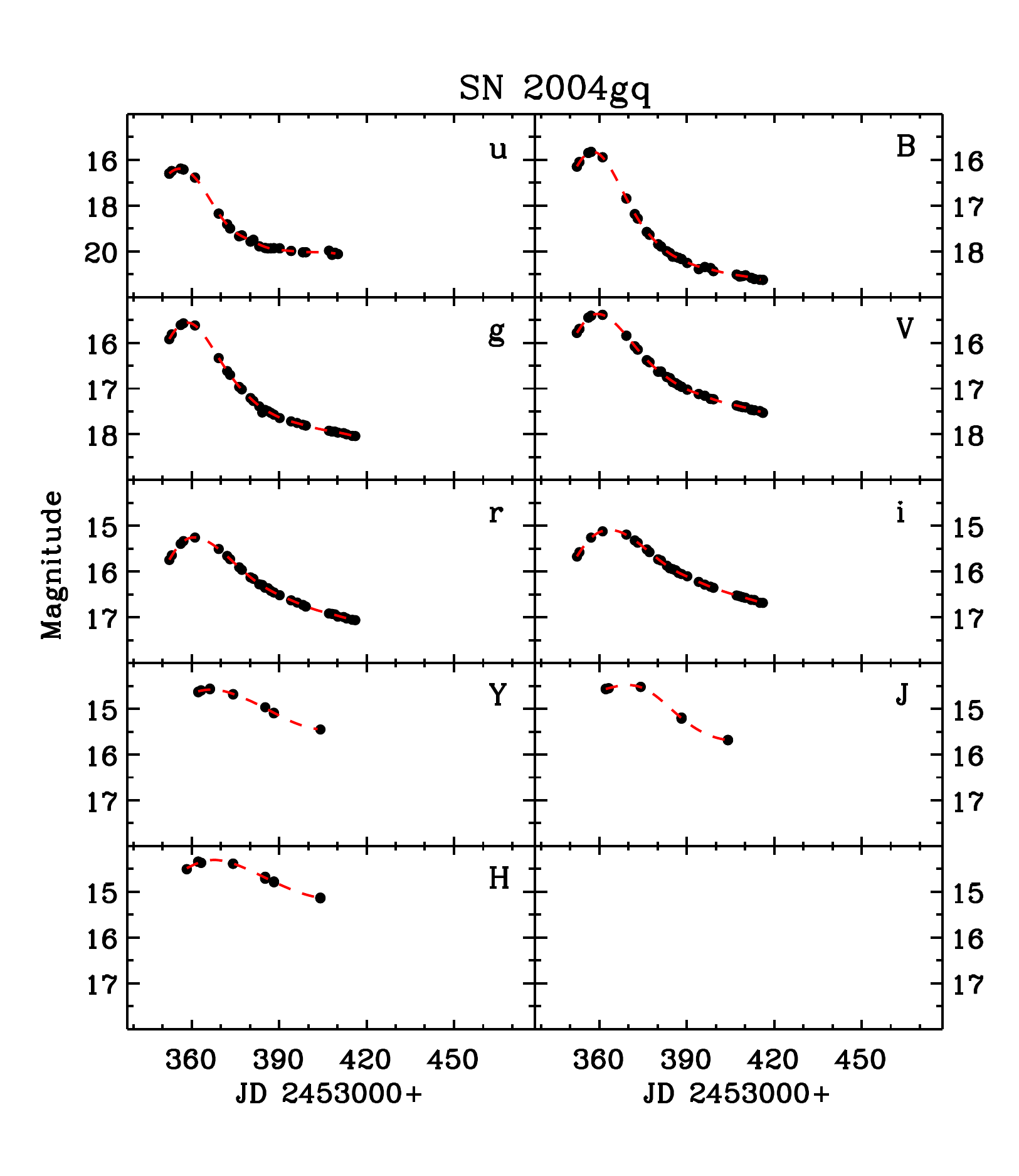}
\includegraphics[width=0.5\textwidth]{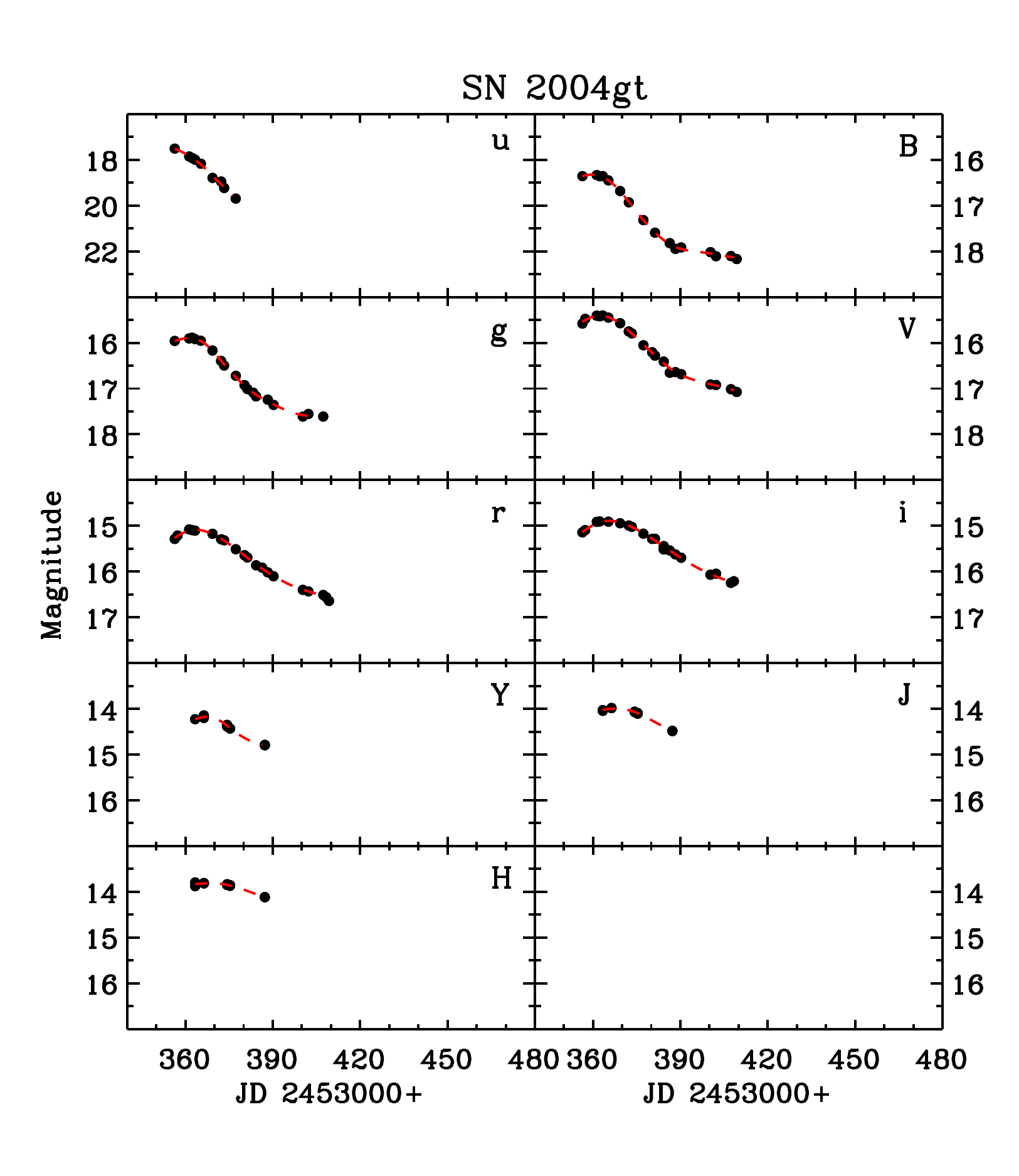}\\
\includegraphics[width=0.5\textwidth]{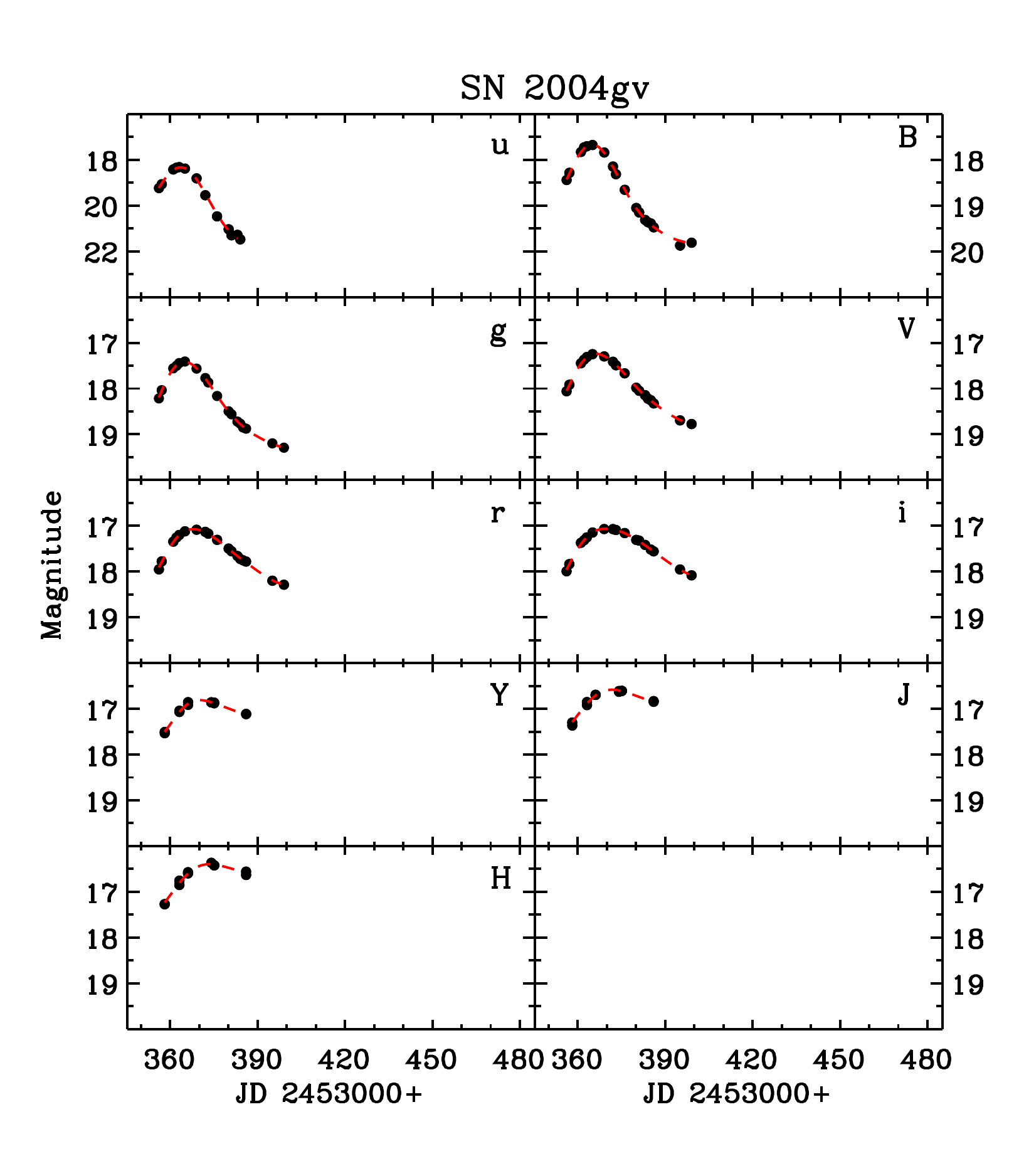}
\includegraphics[width=0.5\textwidth]{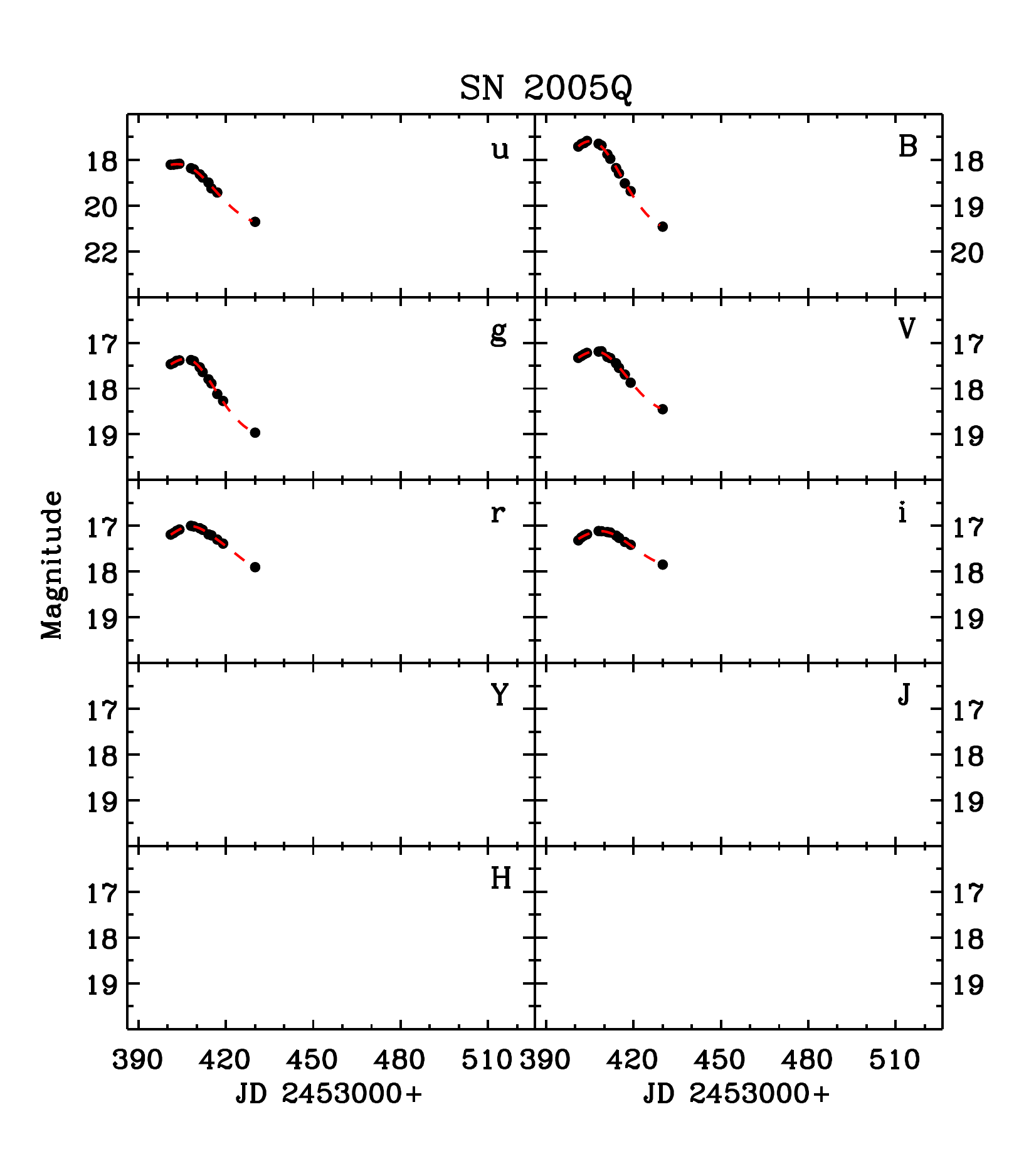}\\
\end{array}$
\end{center}
\caption{Light curves of  CSP-I SE~SNe in the 
 Swope natural photometric system.  
 Uncertainties in the photometry are smaller
  than the points, unless shown.  The smooth curves correspond to Gaussian process spline  
  fits computed within the light0curve analysis software package {\tt SNooPy} \citep{burns11}.  
}
\end{figure}
\clearpage

\setcounter{figure}{3}
\begin{figure}
\begin{center}$
\begin{array}{cc}
\includegraphics[width=0.5\textwidth]{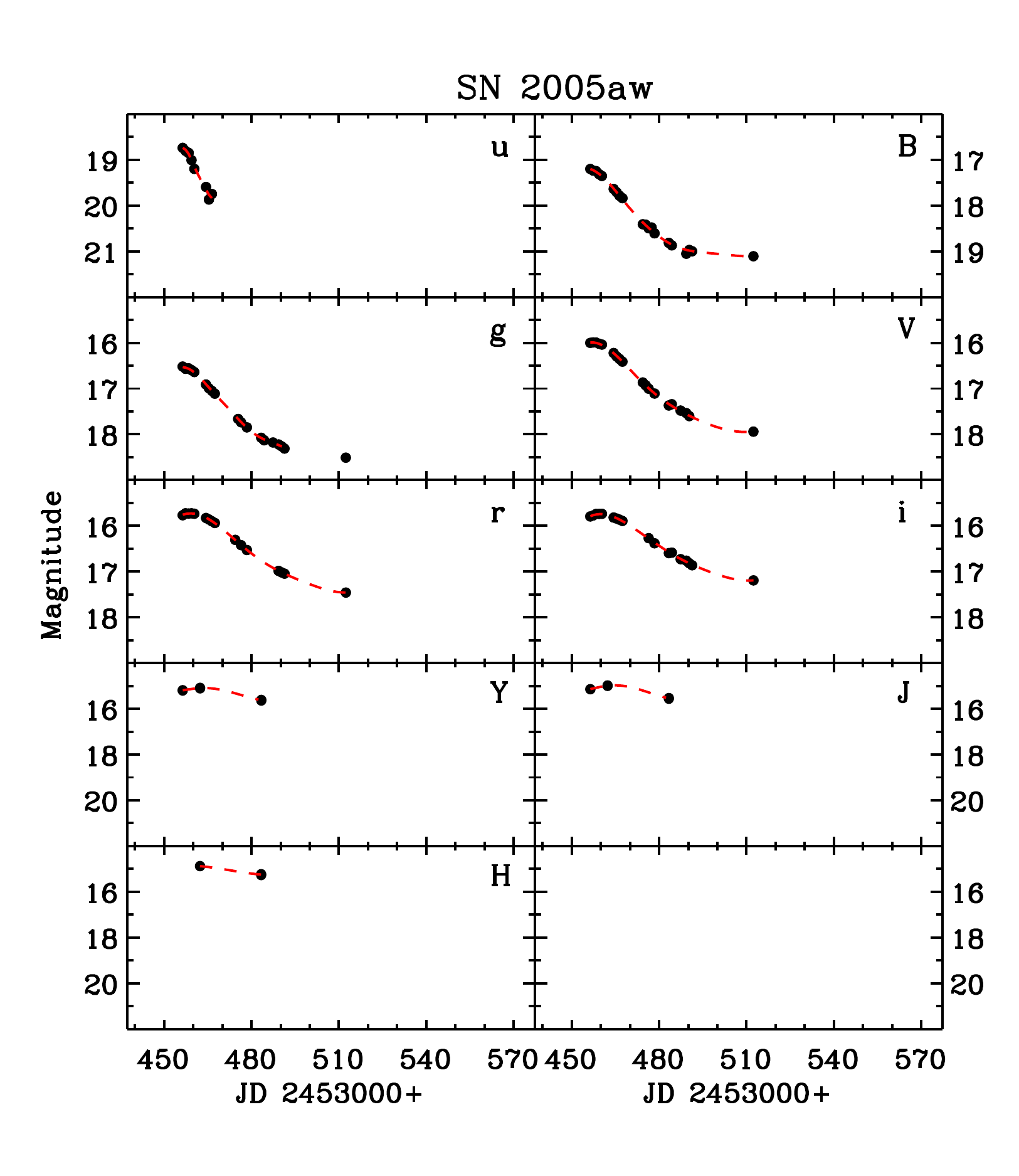}
\includegraphics[width=0.5\textwidth]{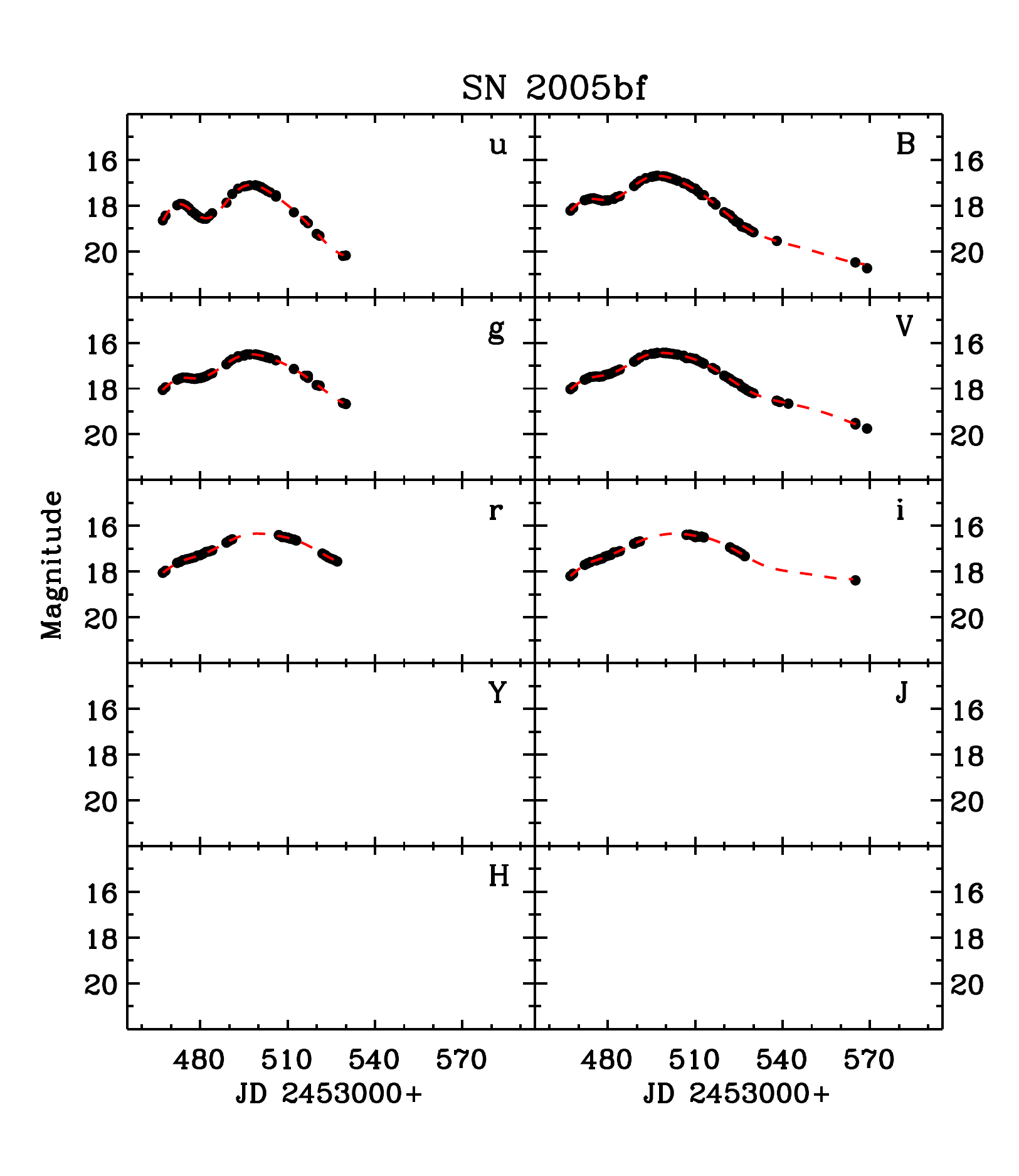}\\
\includegraphics[width=0.5\textwidth]{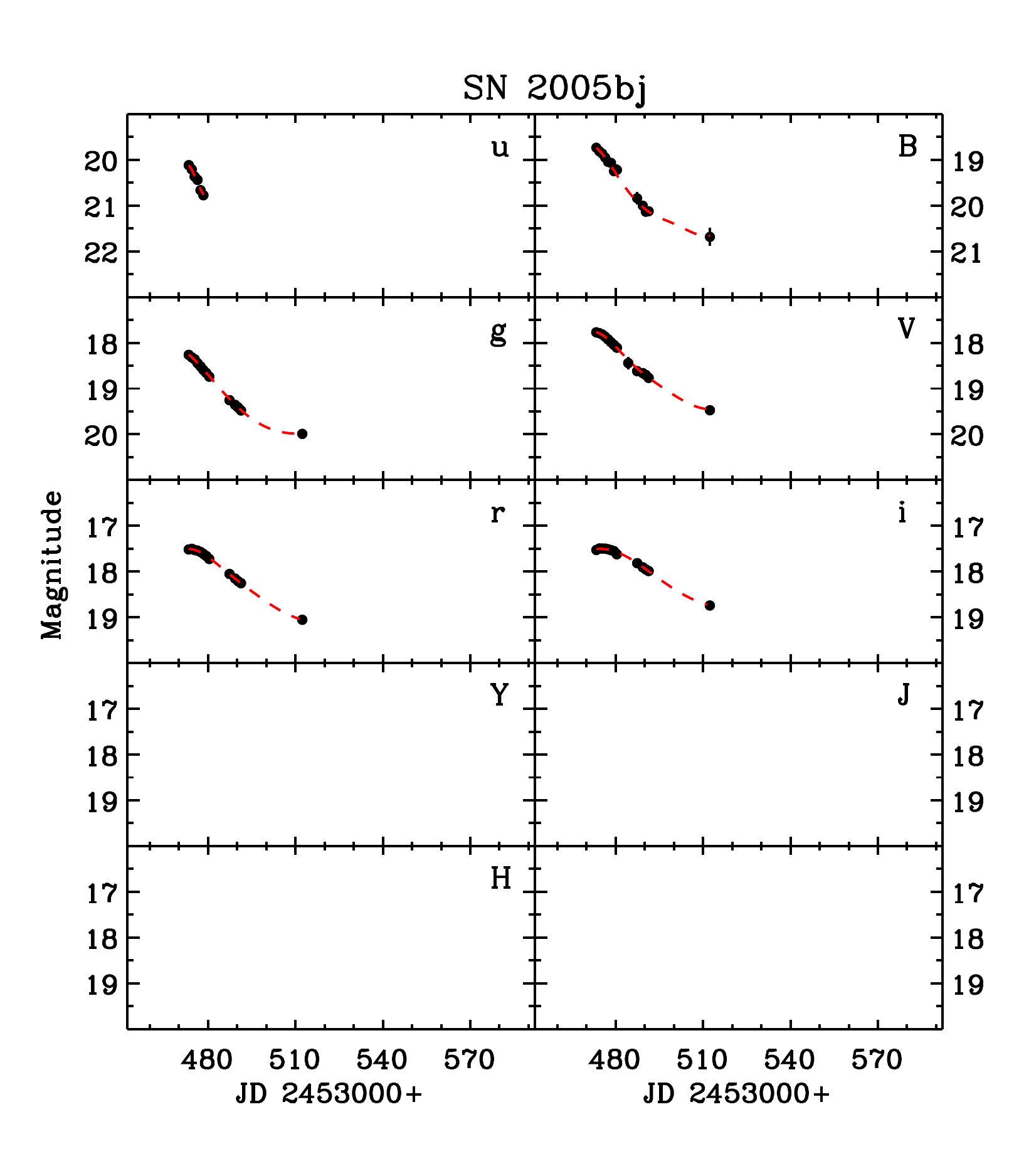}
\includegraphics[width=0.5\textwidth]{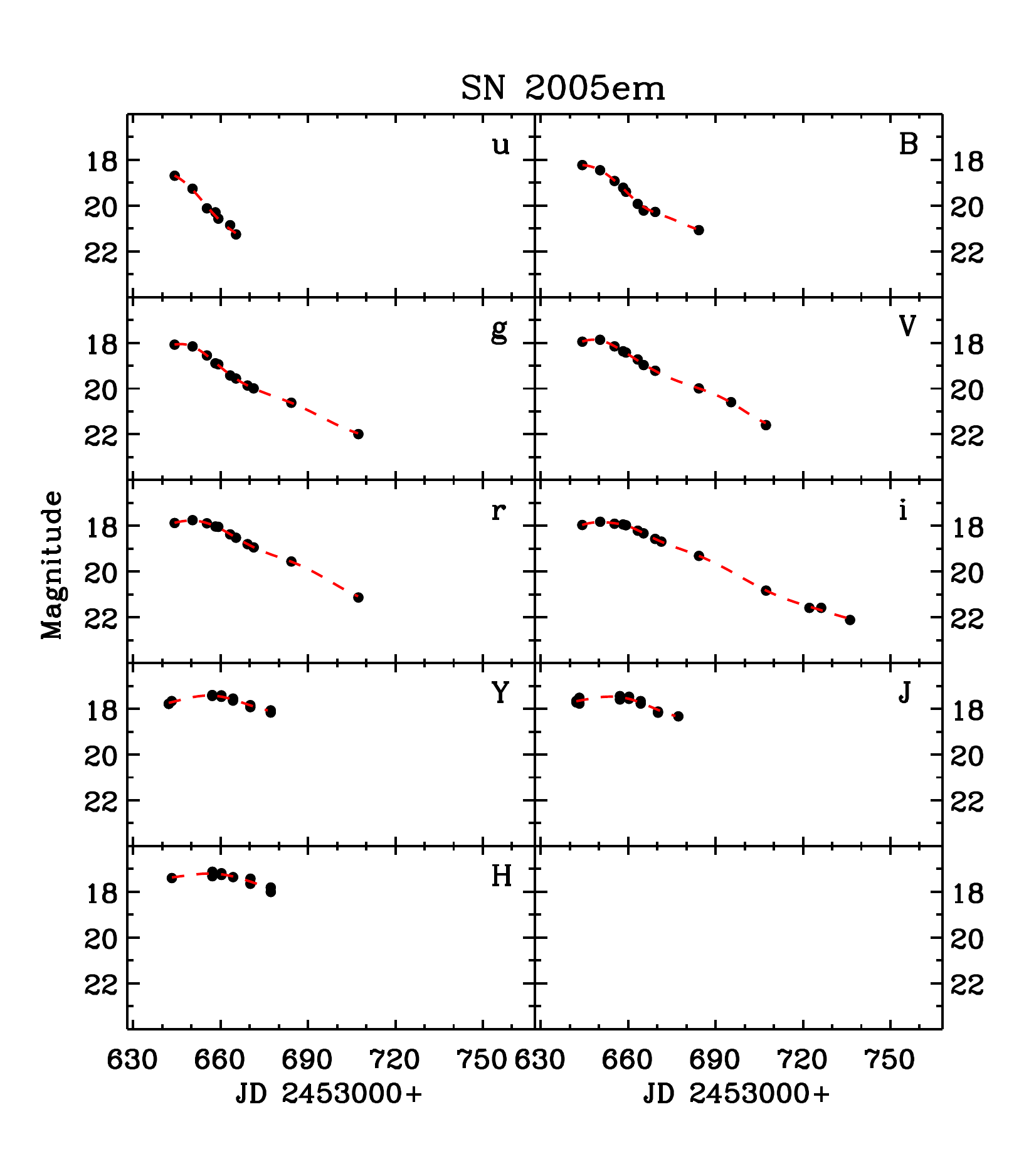}\\
\end{array}$
\end{center}
\caption{Light curves of  CSP-I SE SN in the 
 Swope natural photometric system.  
 Uncertainties in the photometry are smaller
  than the points, unless shown.  The smooth curves correspond to Gaussian process spline  
  fits computed within the light curve analysis software package {\tt SNooPy} \citep{burns11}.  
}
\end{figure}
\clearpage

\setcounter{figure}{3}
\begin{figure}
\begin{center}$
\begin{array}{cc}
\includegraphics[width=0.5\textwidth]{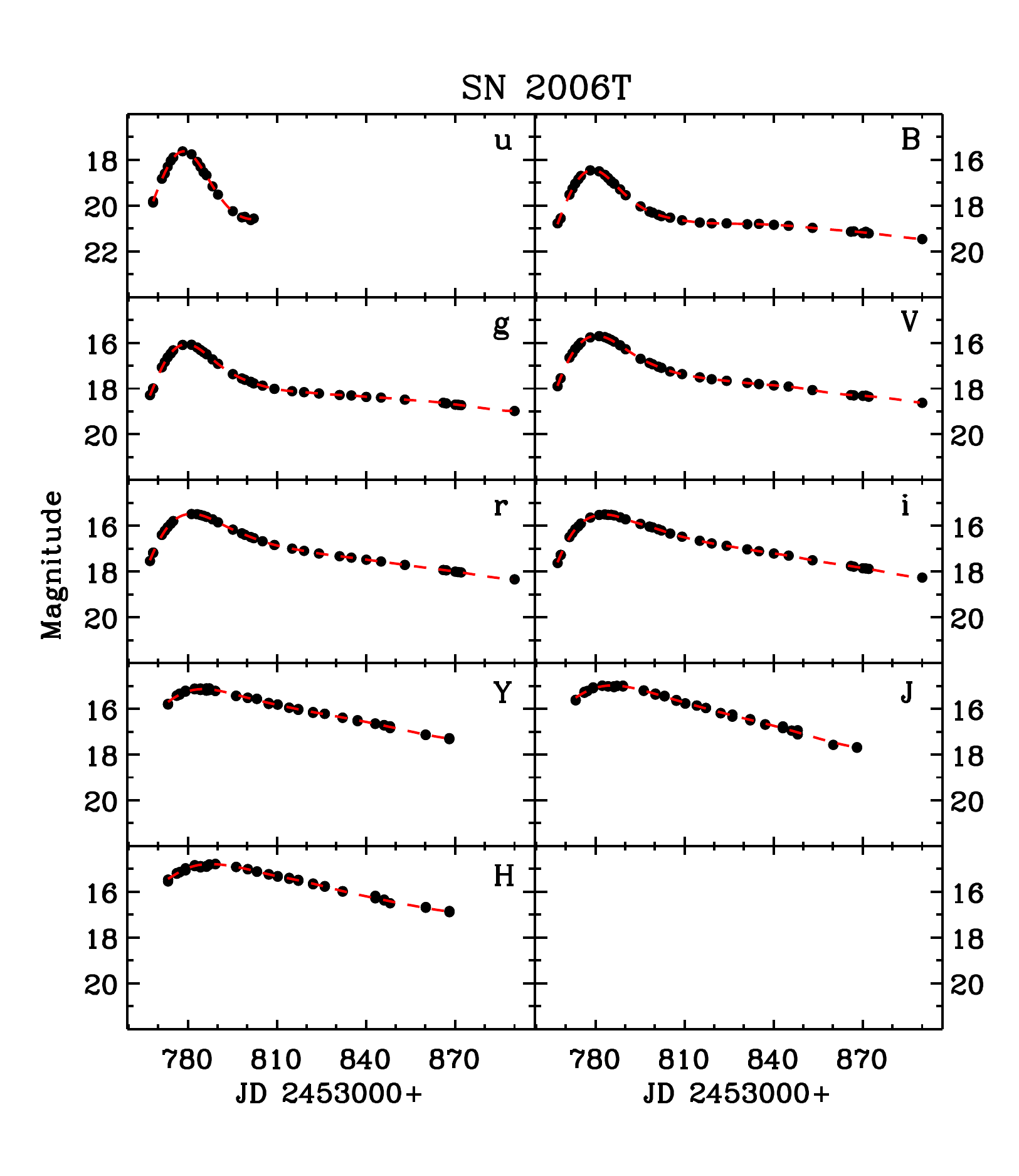}
\includegraphics[width=0.5\textwidth]{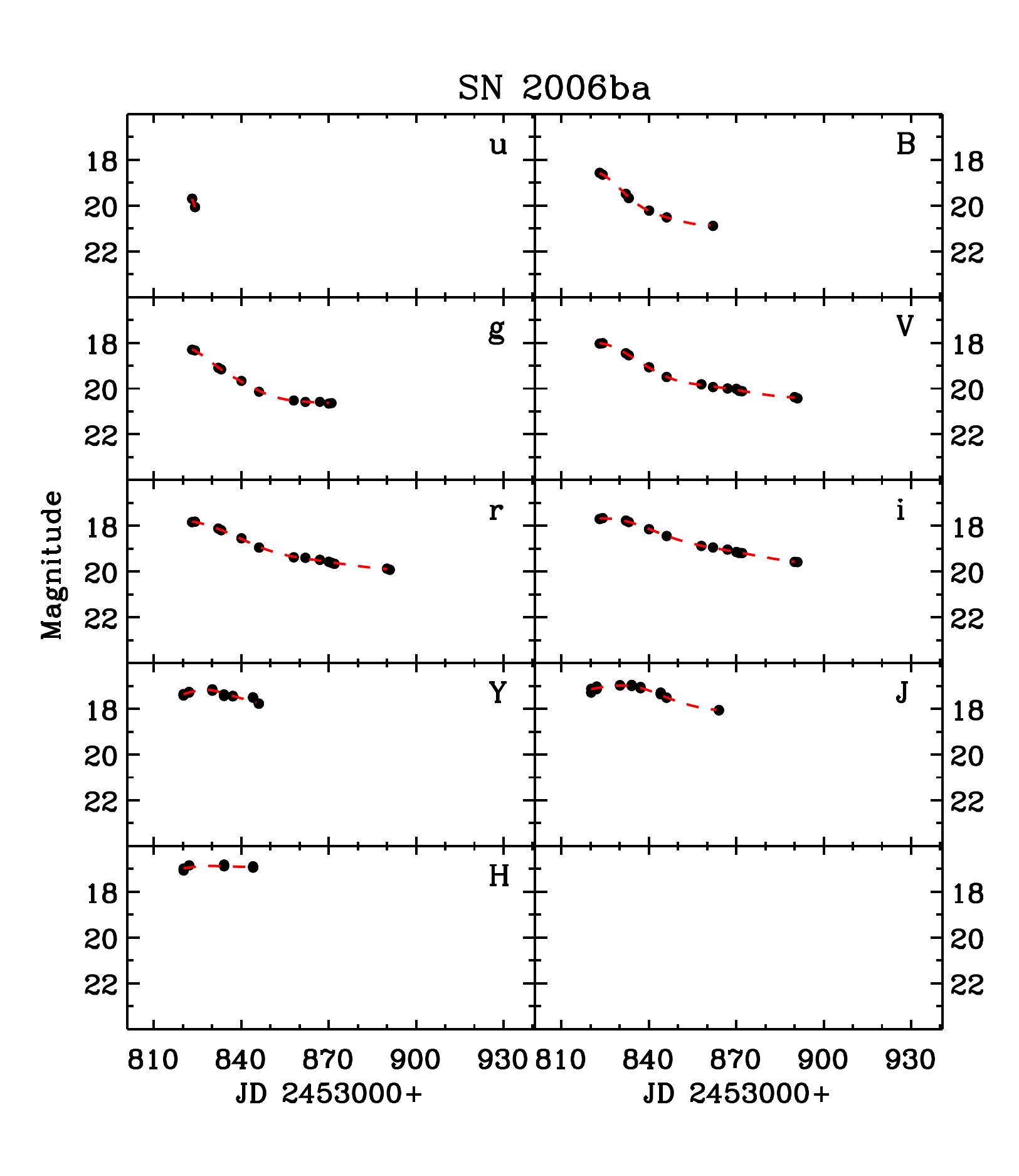}\\
\includegraphics[width=0.5\textwidth]{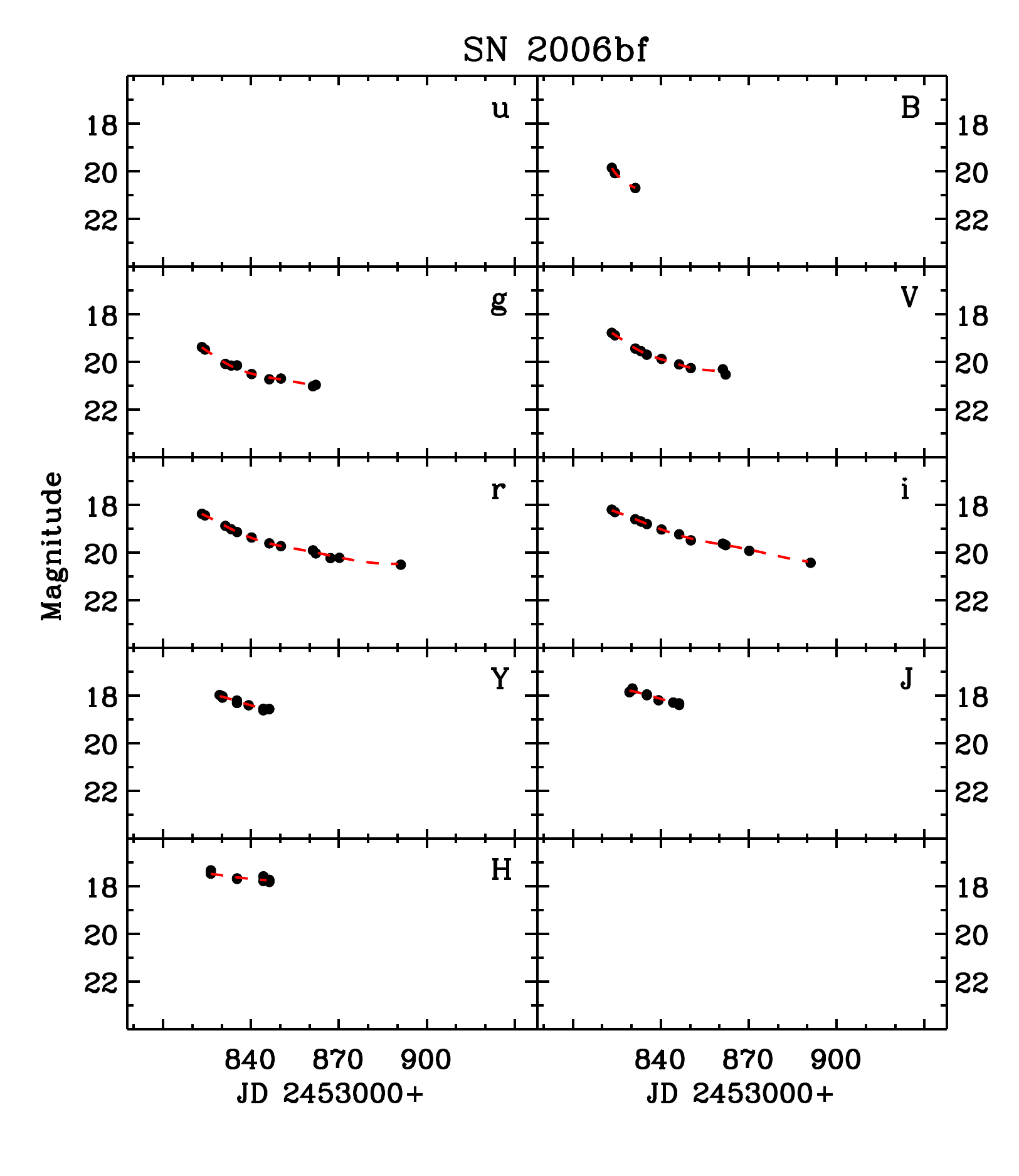}
\includegraphics[width=0.5\textwidth]{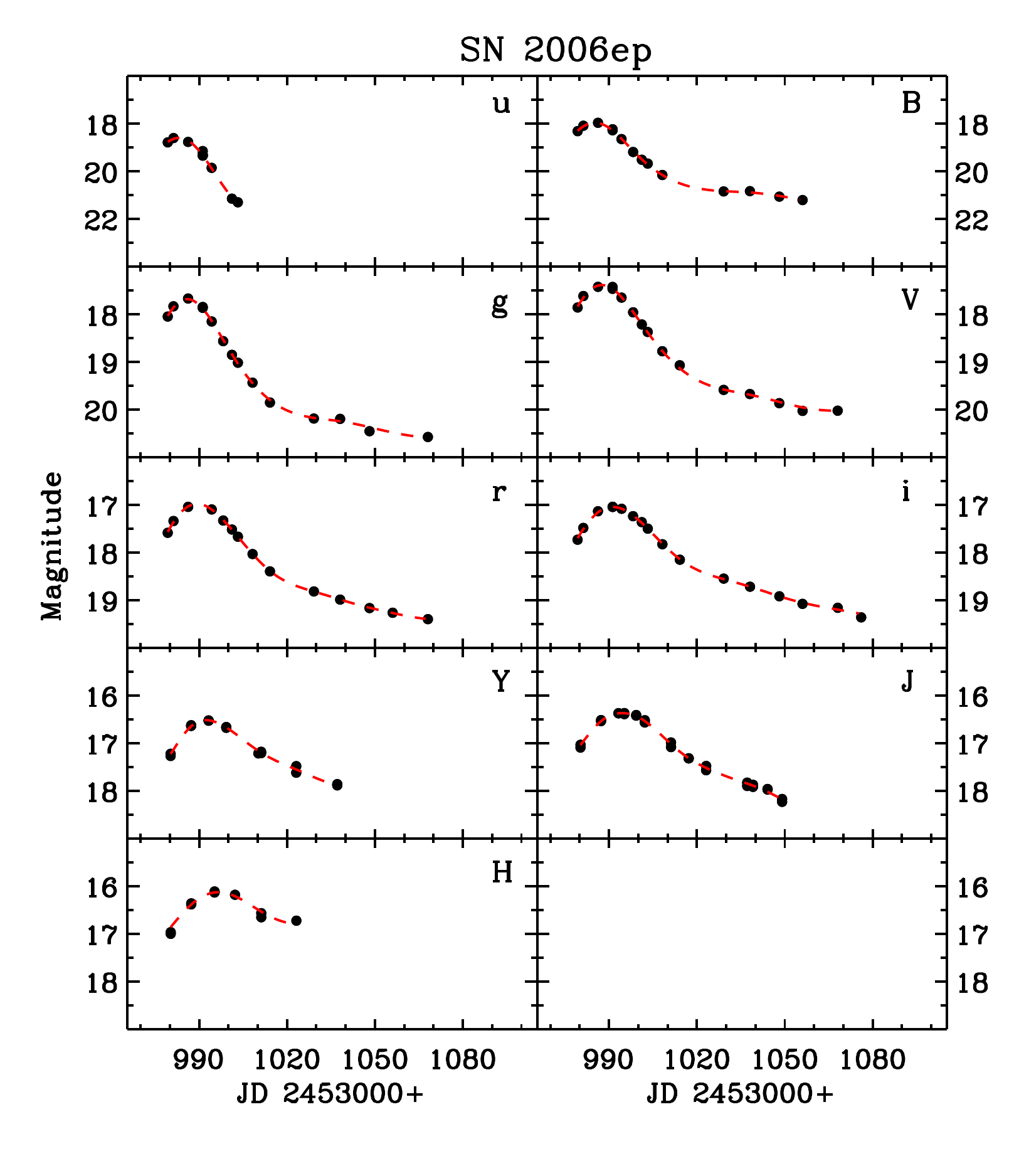}\\
\end{array}$
\end{center}
\caption{Light curves of  CSP-I SE SN in the 
 Swope natural photometric system.  
 Uncertainties in the photometry are smaller
  than the points, unless shown.  The smooth curves correspond to  Gaussian process spline  
  fits computed within the light curve analysis software package {\tt SNooPy} \citep{burns11}.  
}
\end{figure}
\clearpage

\setcounter{figure}{3}
\begin{figure}
\begin{center}$
\begin{array}{cc}
\includegraphics[width=0.5\textwidth]{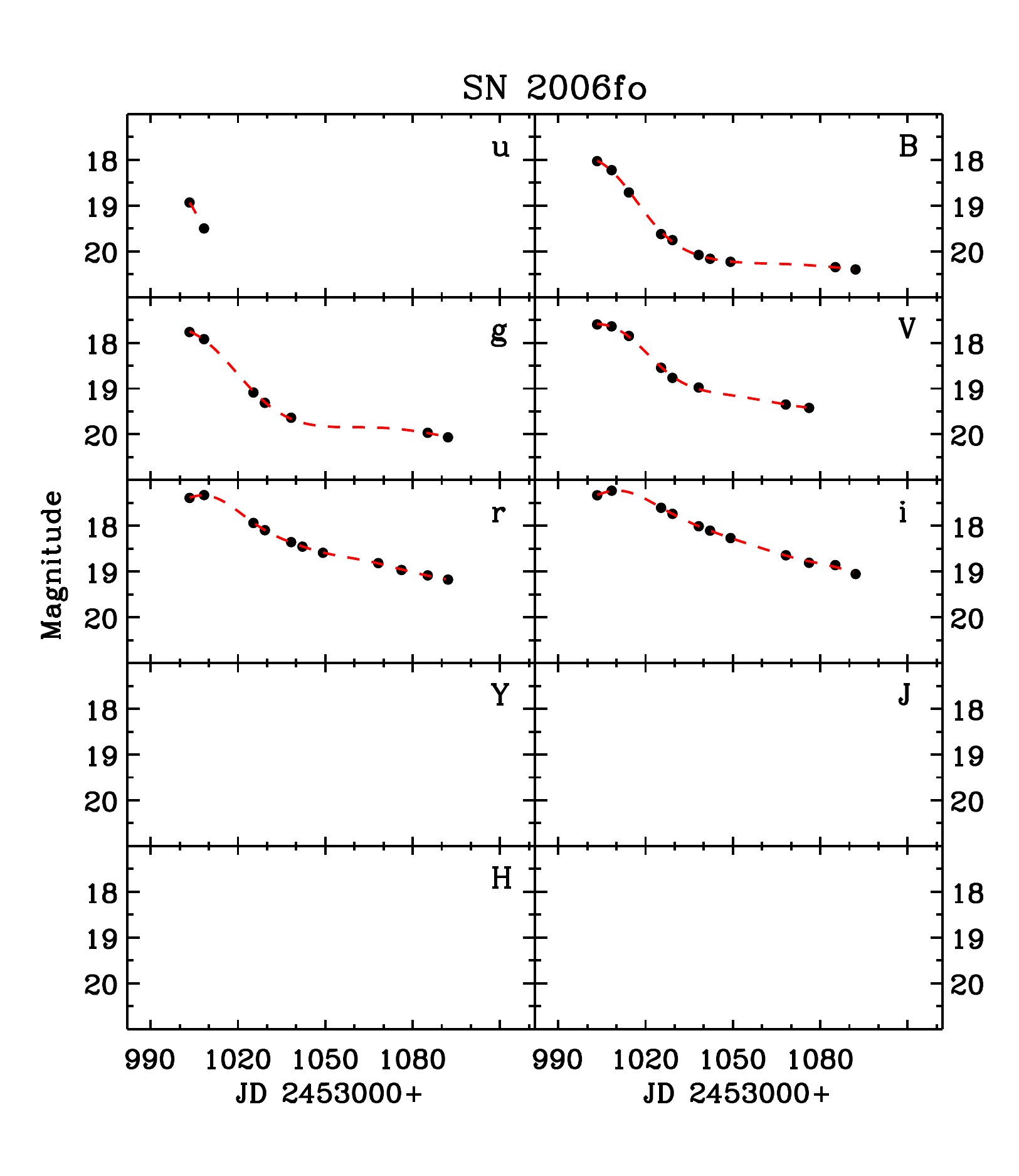}
\includegraphics[width=0.5\textwidth]{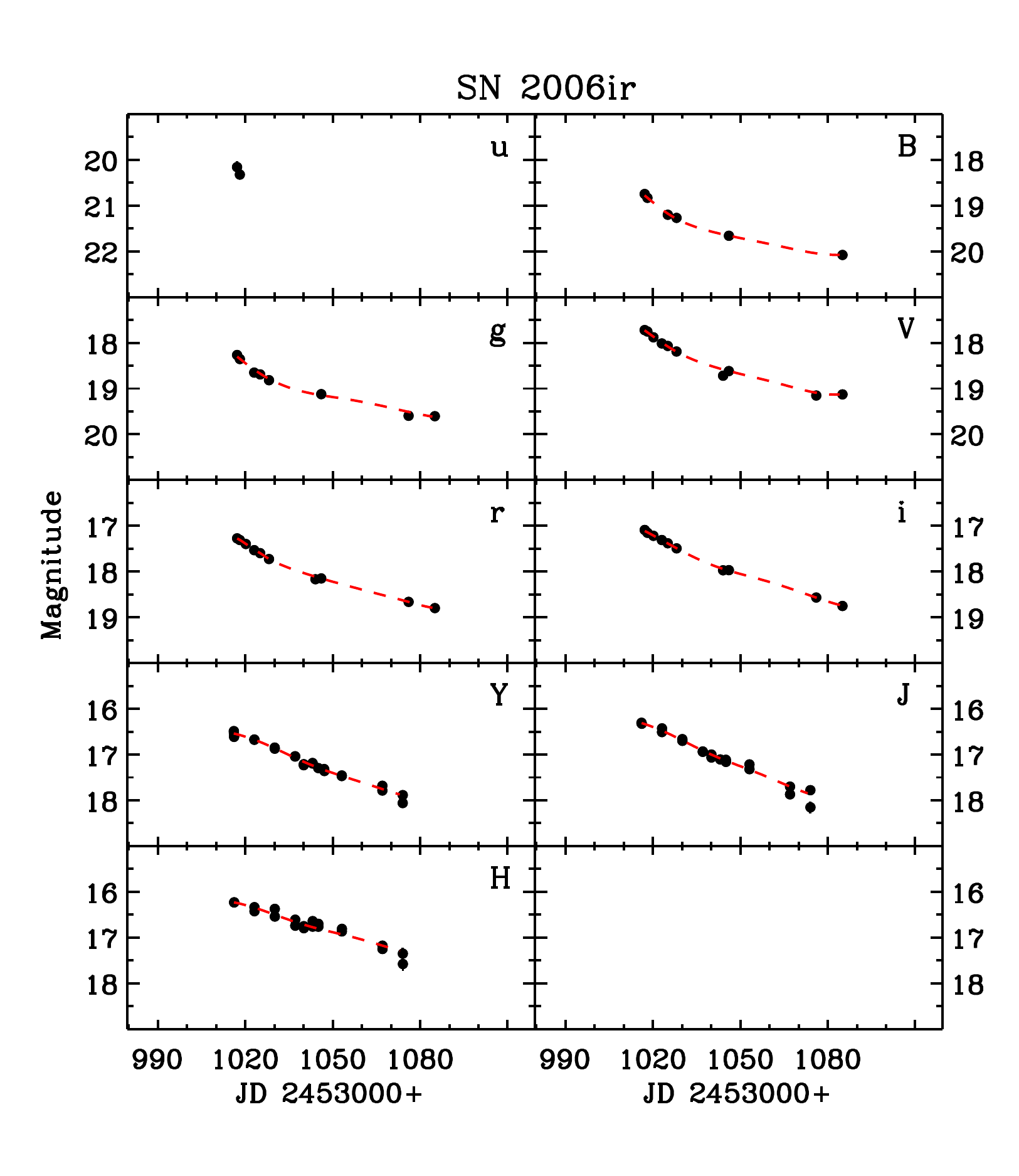}\\
\includegraphics[width=0.5\textwidth]{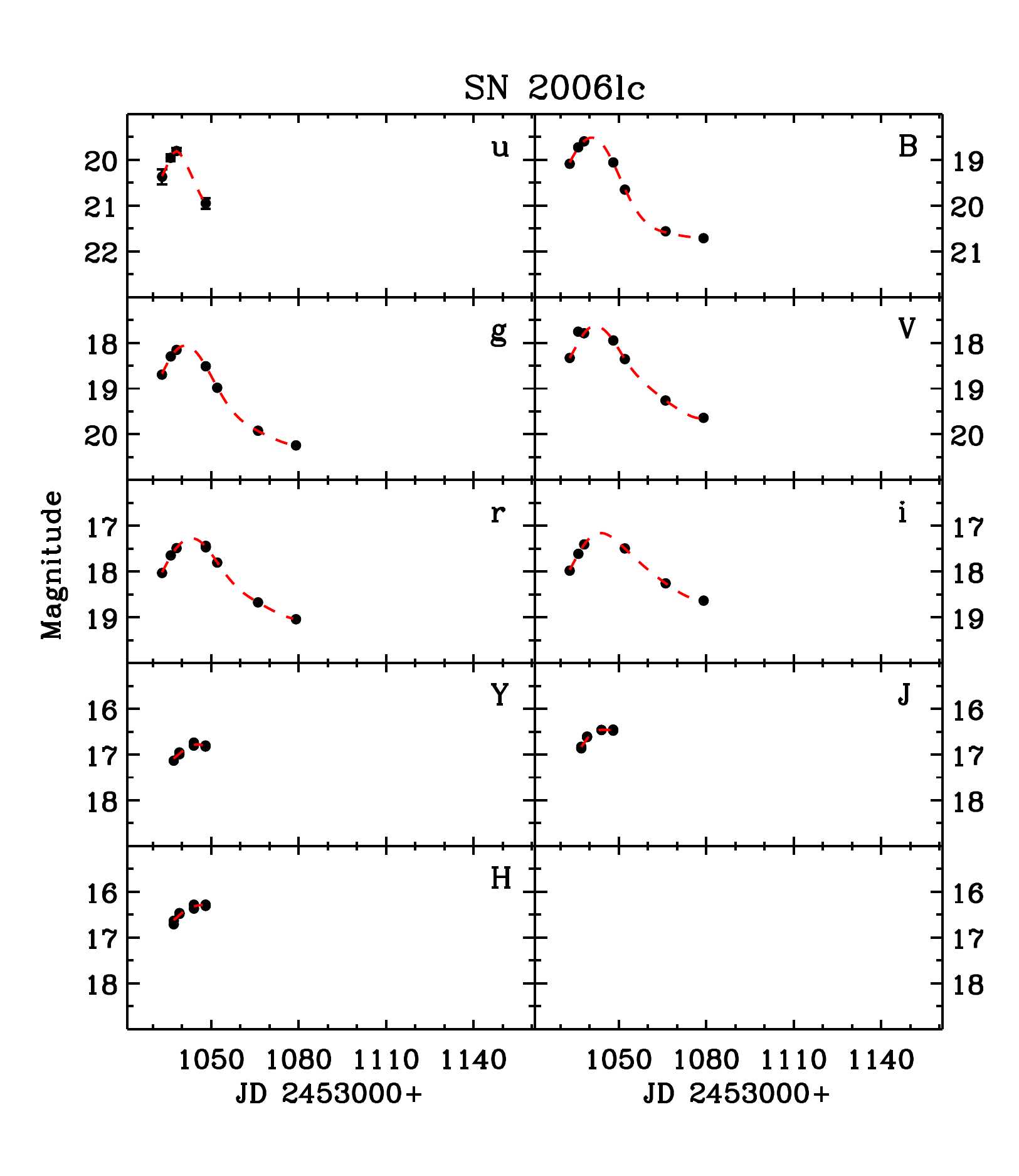}
\includegraphics[width=0.5\textwidth]{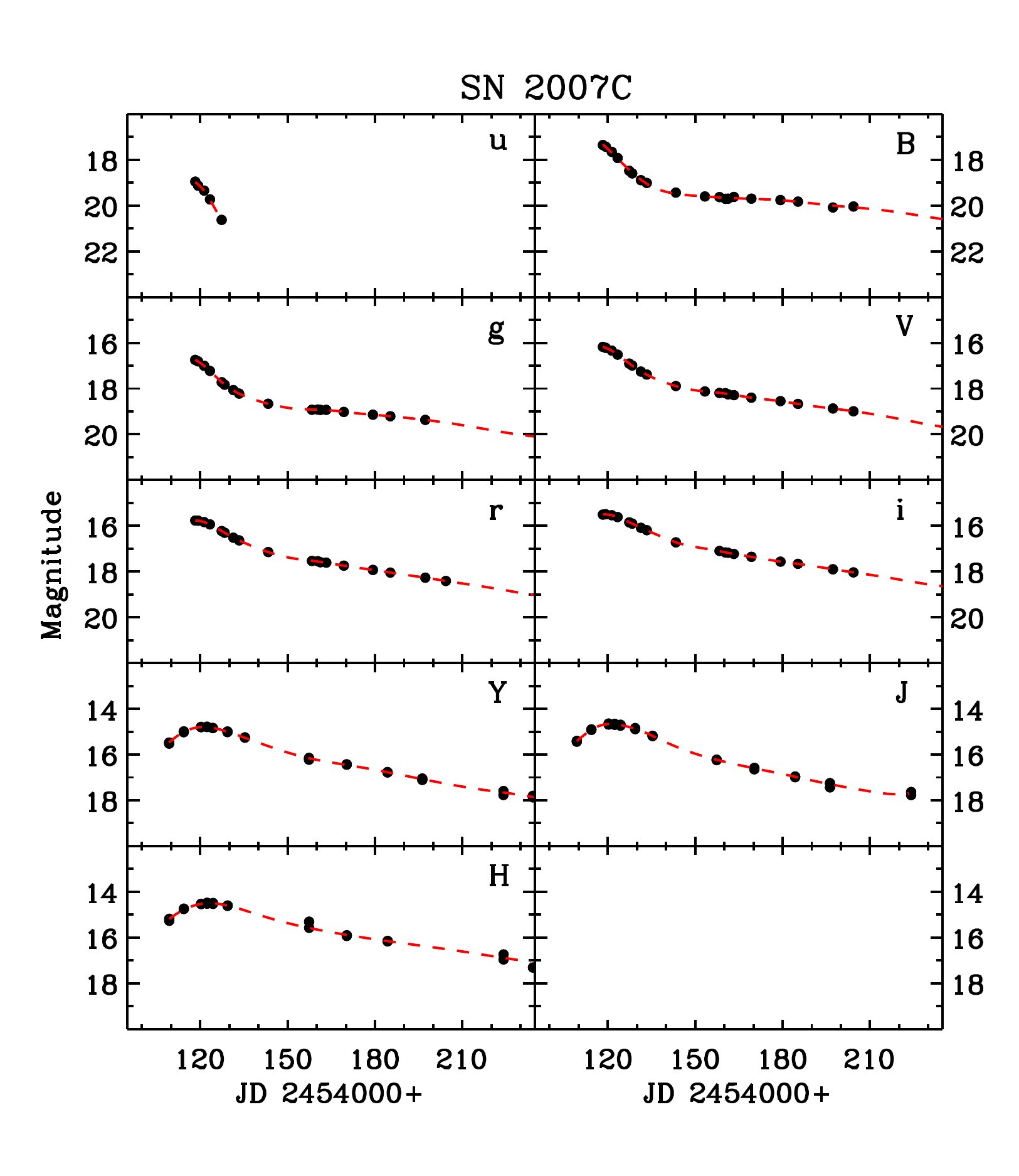}\\
\end{array}$
\end{center}
\caption{Light curves of  CSP-I SE~SN in the 
 Swope natural photometric system.  
 Uncertainties in the photometry are smaller
  than the points, unless shown.  The smooth curves correspond to Gaussian process spline  
  fits computed within the light-curve analysis software package {\tt SNooPy} \citep{burns11}.  
}
\end{figure}
\clearpage

\setcounter{figure}{3}
\begin{figure}
\begin{center}$
\begin{array}{cc}
\includegraphics[width=0.5\textwidth]{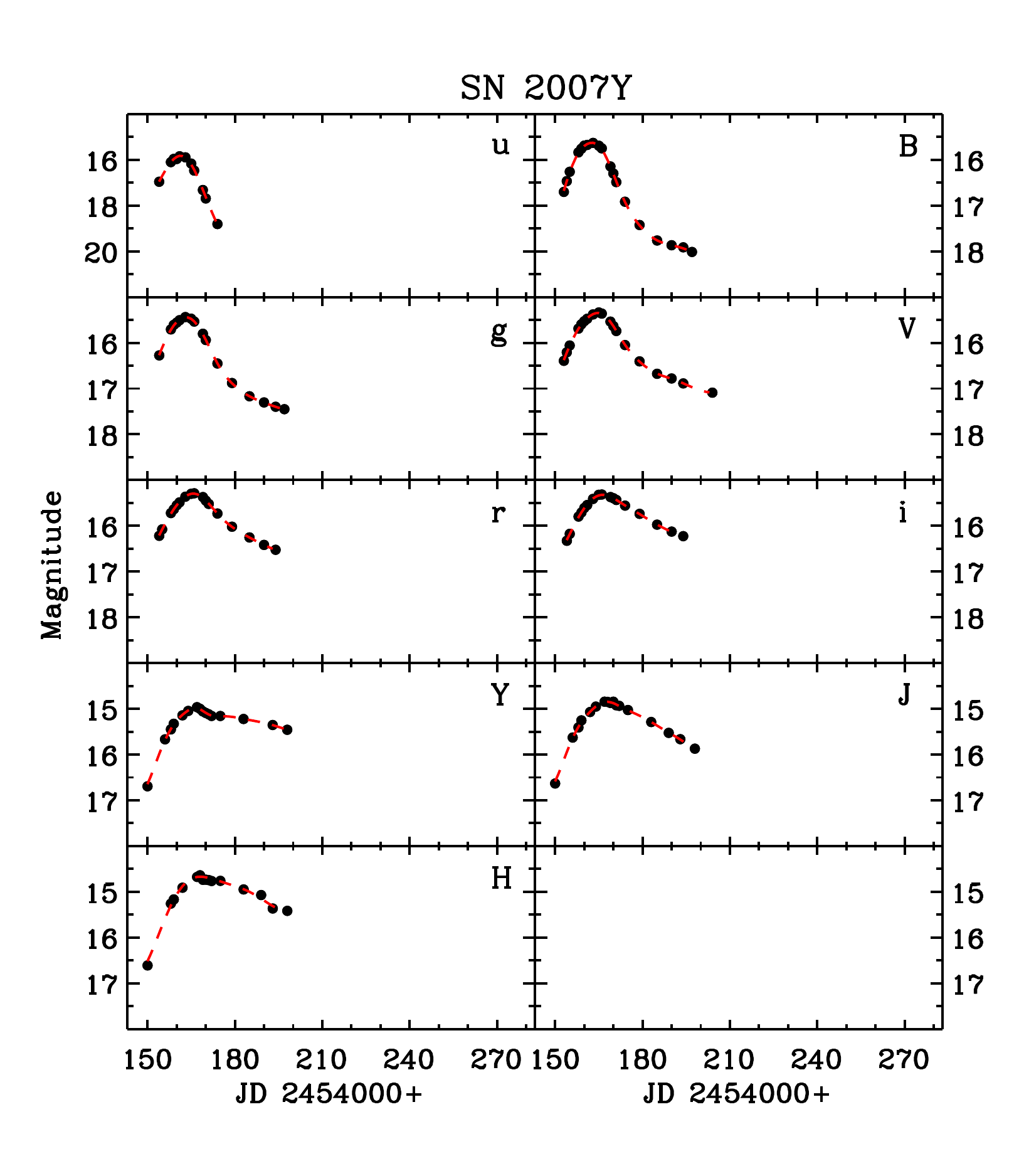}
\includegraphics[width=0.5\textwidth]{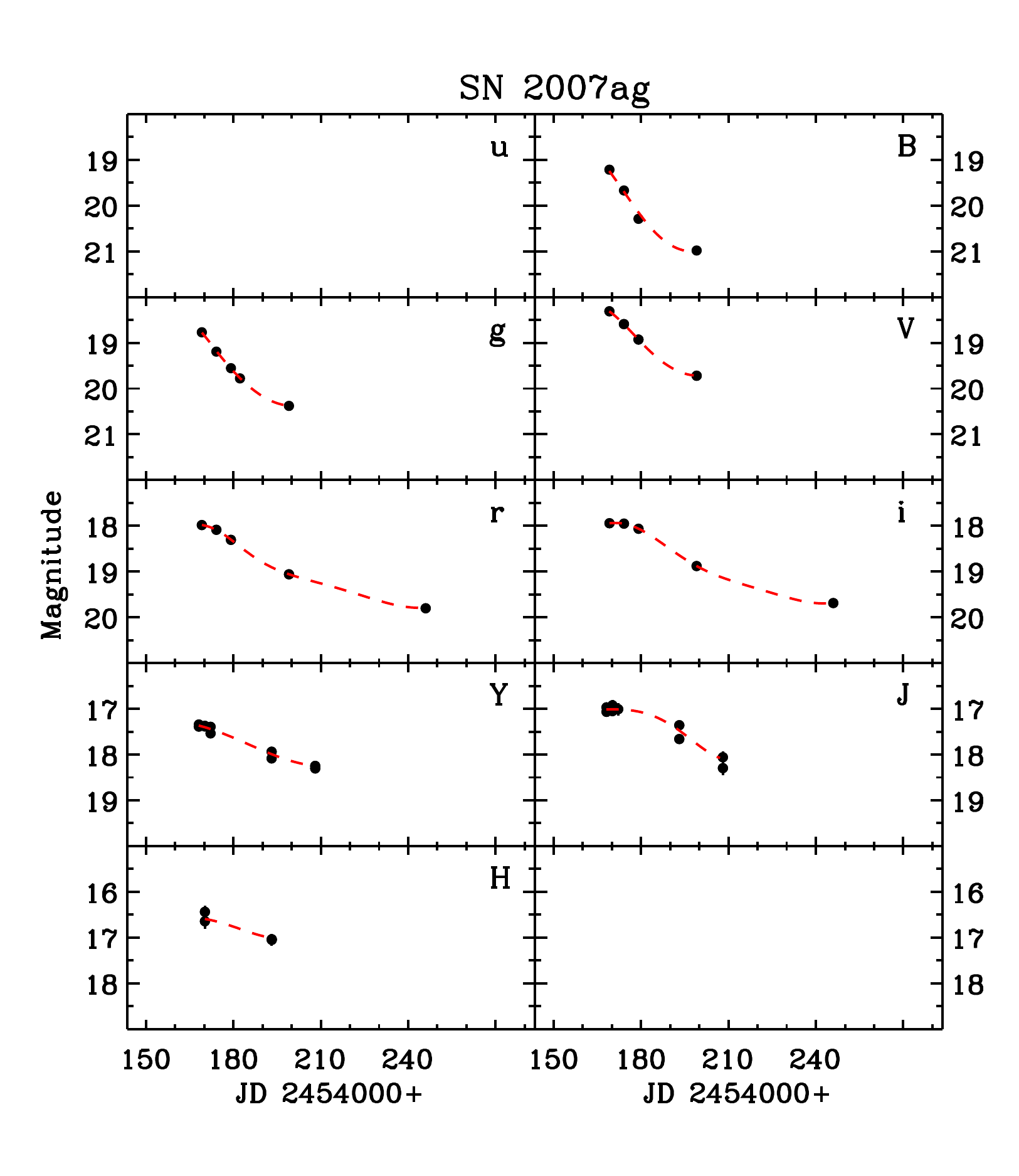}\\
\includegraphics[width=0.5\textwidth]{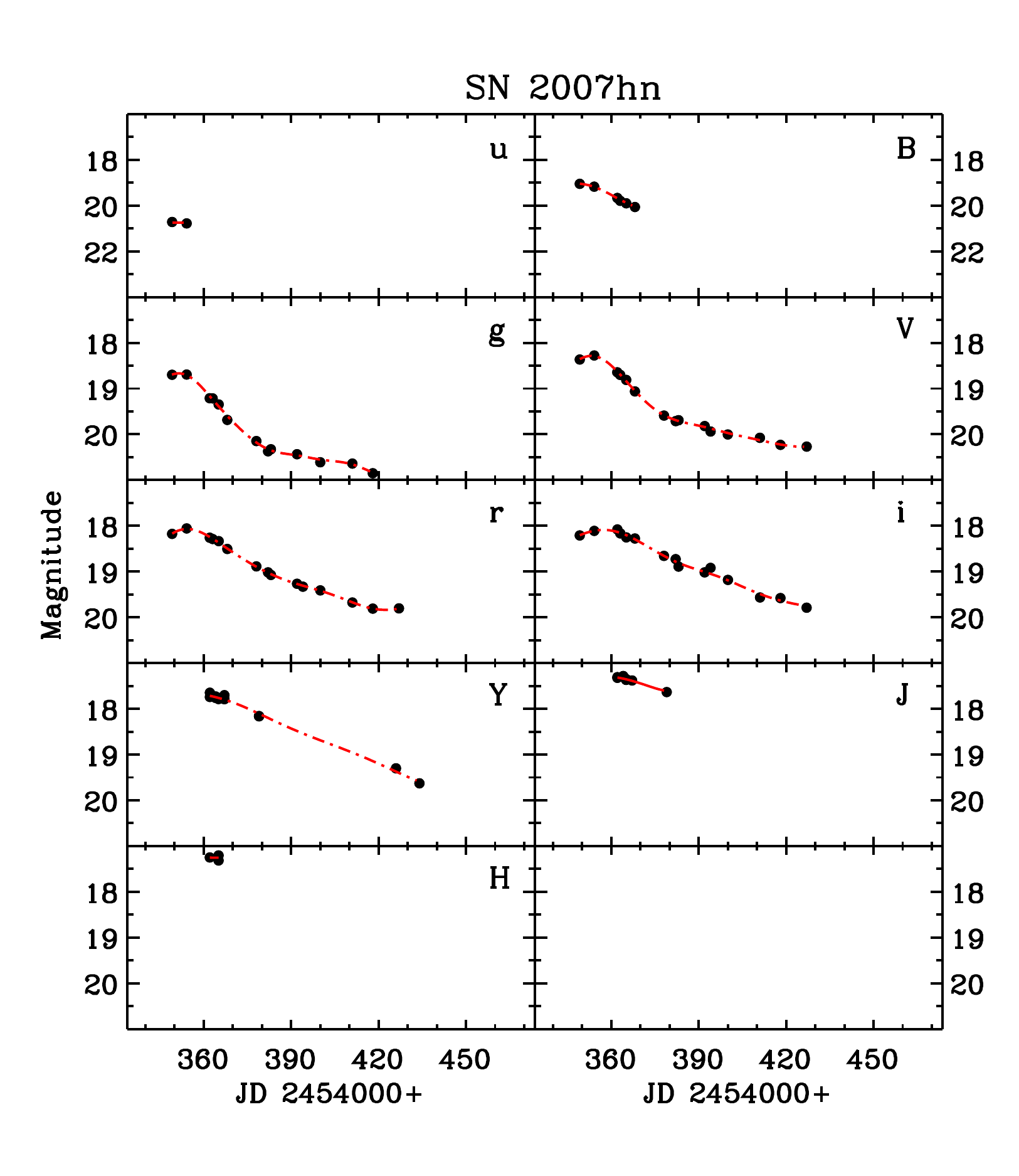}
\includegraphics[width=0.5\textwidth]{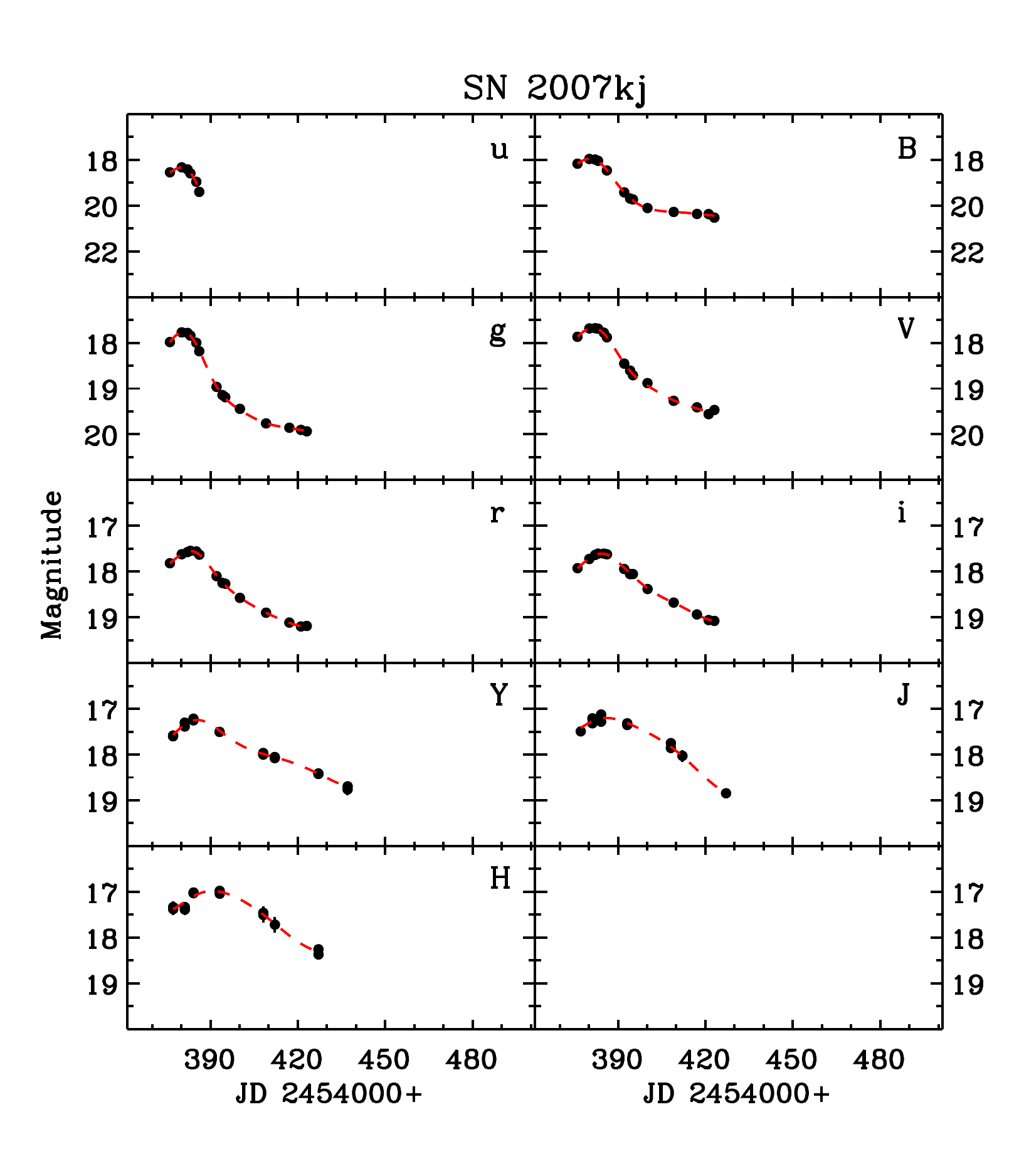}\\
\end{array}$
\end{center}
\caption{Light curves of  CSP-I SE~SN in the 
 Swope natural photometric system.  
 Uncertainties in the photometry are smaller
  than the points, unless shown.  The smooth curves correspond to  Gaussian process spline  
  fits computed within the light-curve analysis software package {\tt SNooPy} \citep{burns11}.  
}
\end{figure}
\clearpage

\setcounter{figure}{3}
\begin{figure}
\begin{center}$
\begin{array}{cc}
\includegraphics[width=0.5\textwidth]{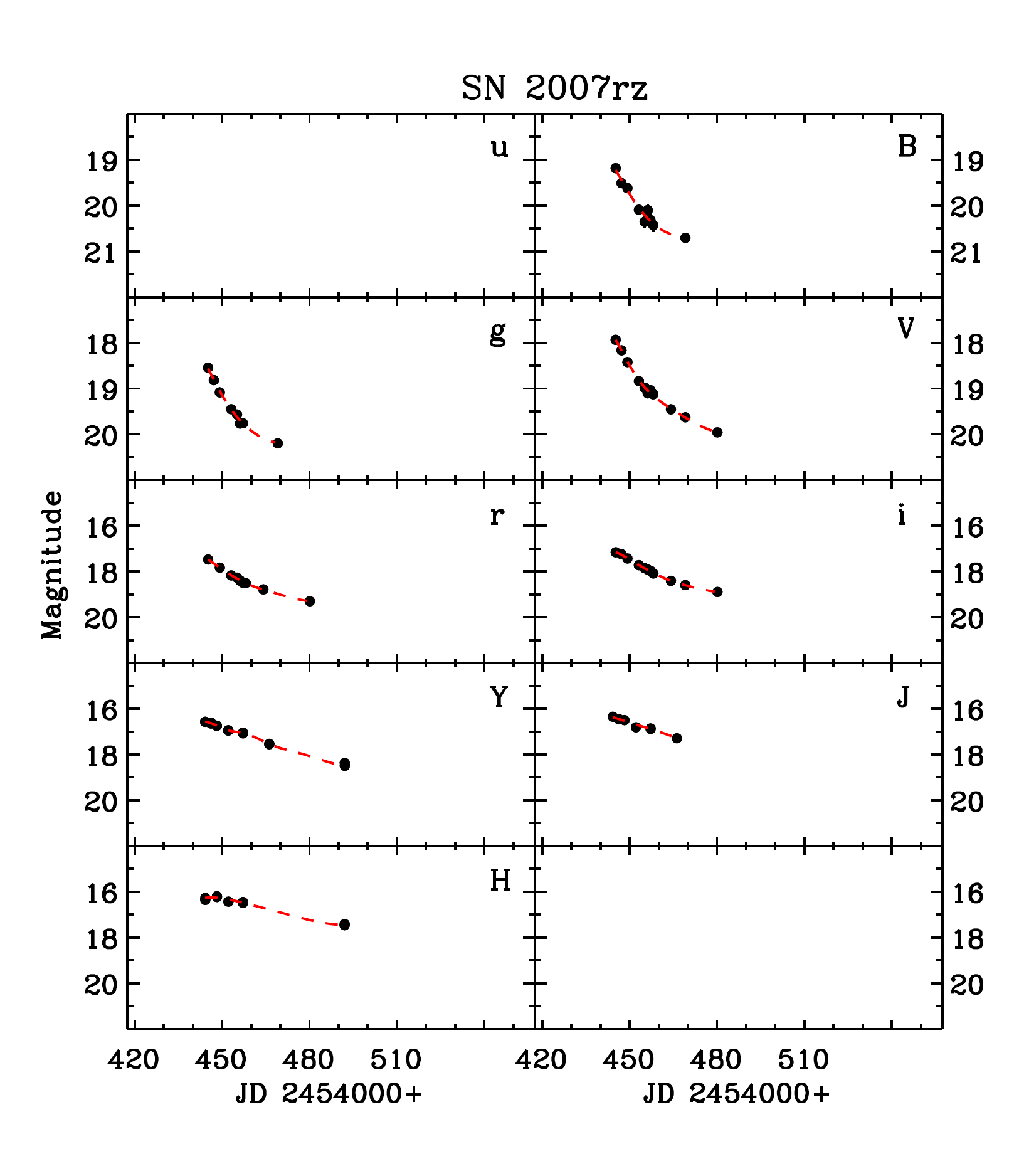}
\includegraphics[width=0.5\textwidth]{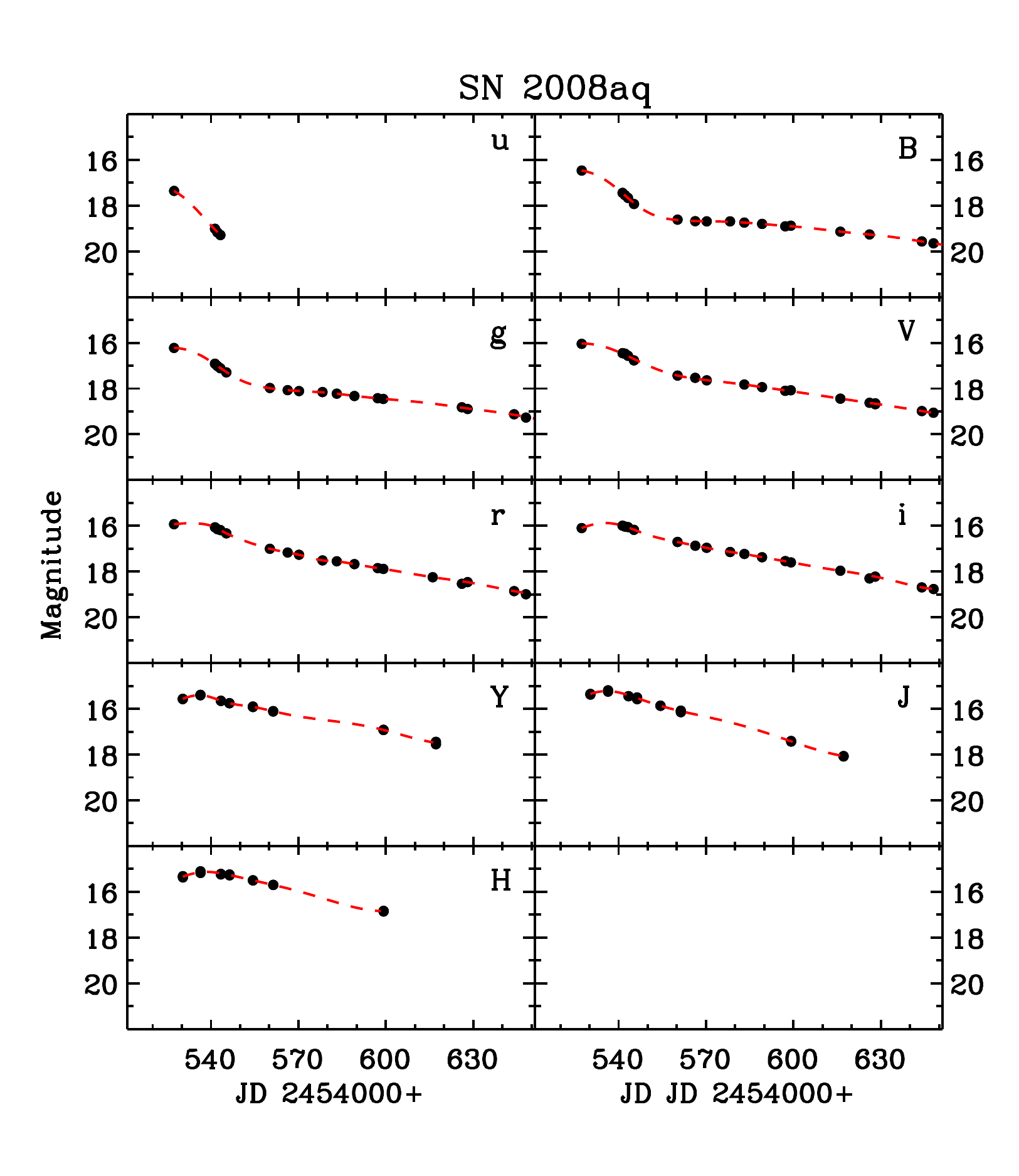}\\
\includegraphics[width=0.5\textwidth]{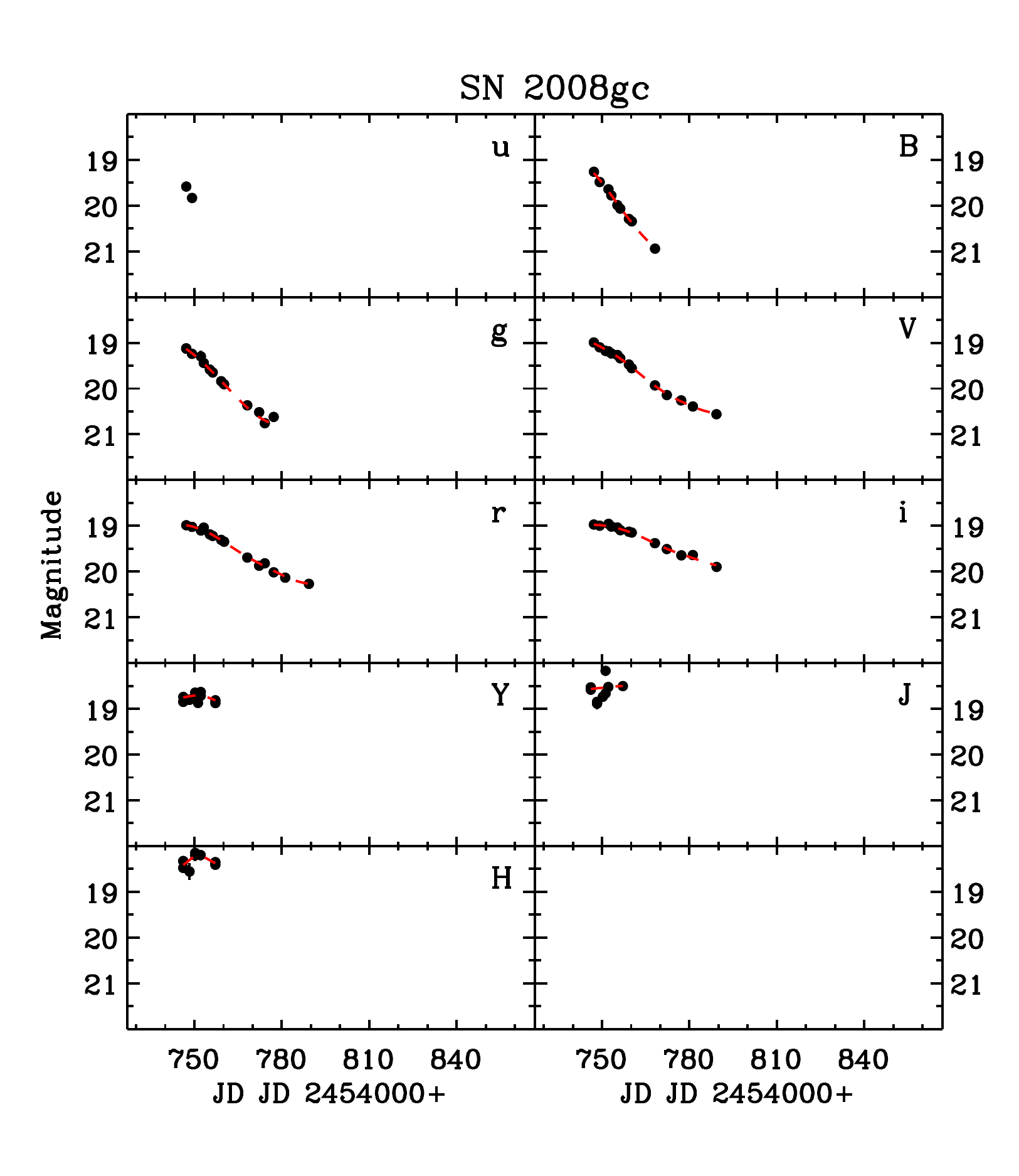}
\includegraphics[width=0.5\textwidth]{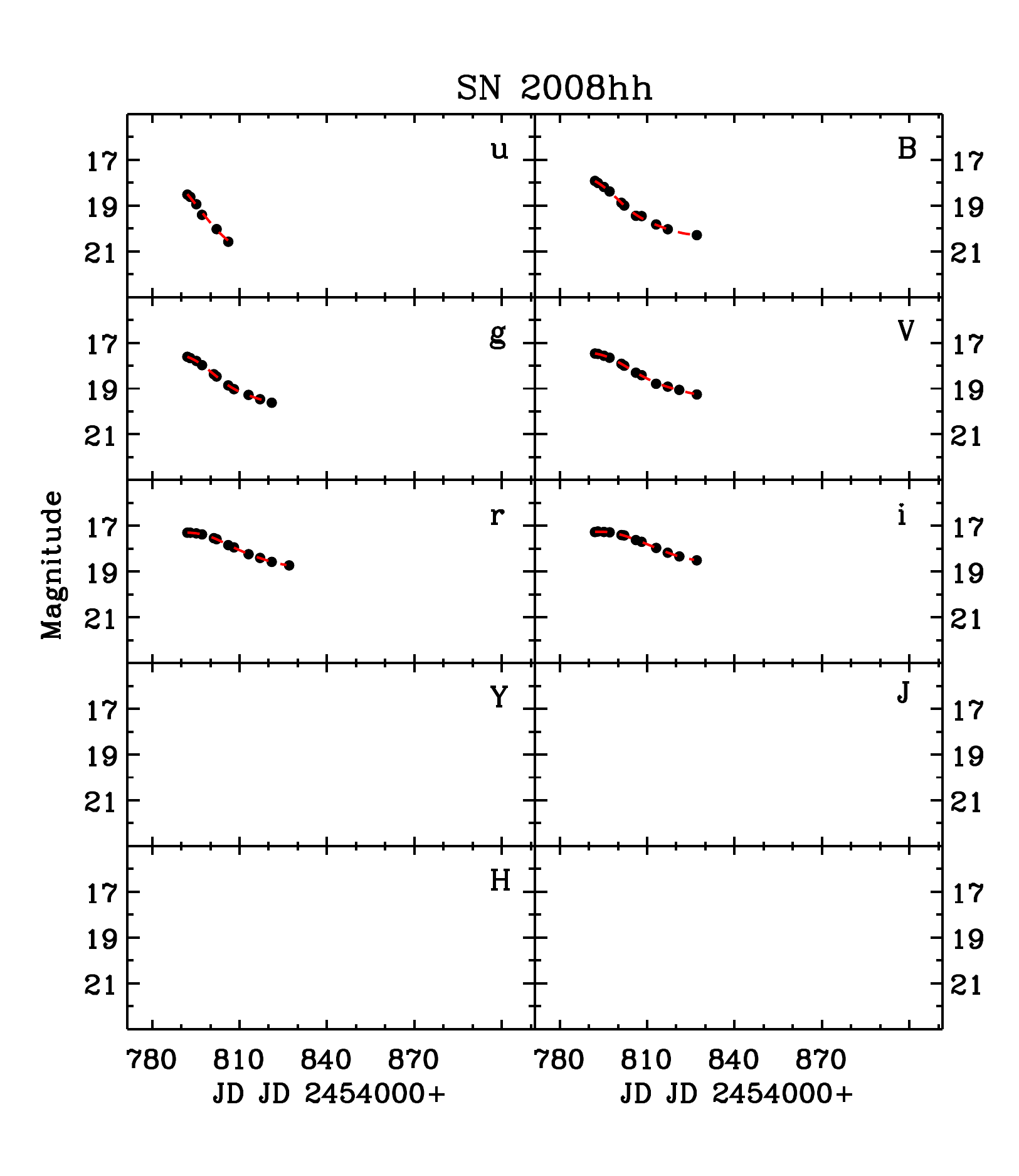}\\
\end{array}$
\end{center}
\caption{Light curves of  CSP-I SE~SN~Ia in the 
 Swope natural photometric system.  
 Uncertainties in the photometry are smaller
  than the points, unless shown.  The smooth curves correspond to   
  Gaussian process spline fits  computed within the light-curve analysis software package {\tt SNooPy} \citep{burns11}.  
}
\end{figure}
\clearpage

\setcounter{figure}{3}
\begin{figure}
\begin{center}$
\begin{array}{cc}
\includegraphics[width=0.5\textwidth]{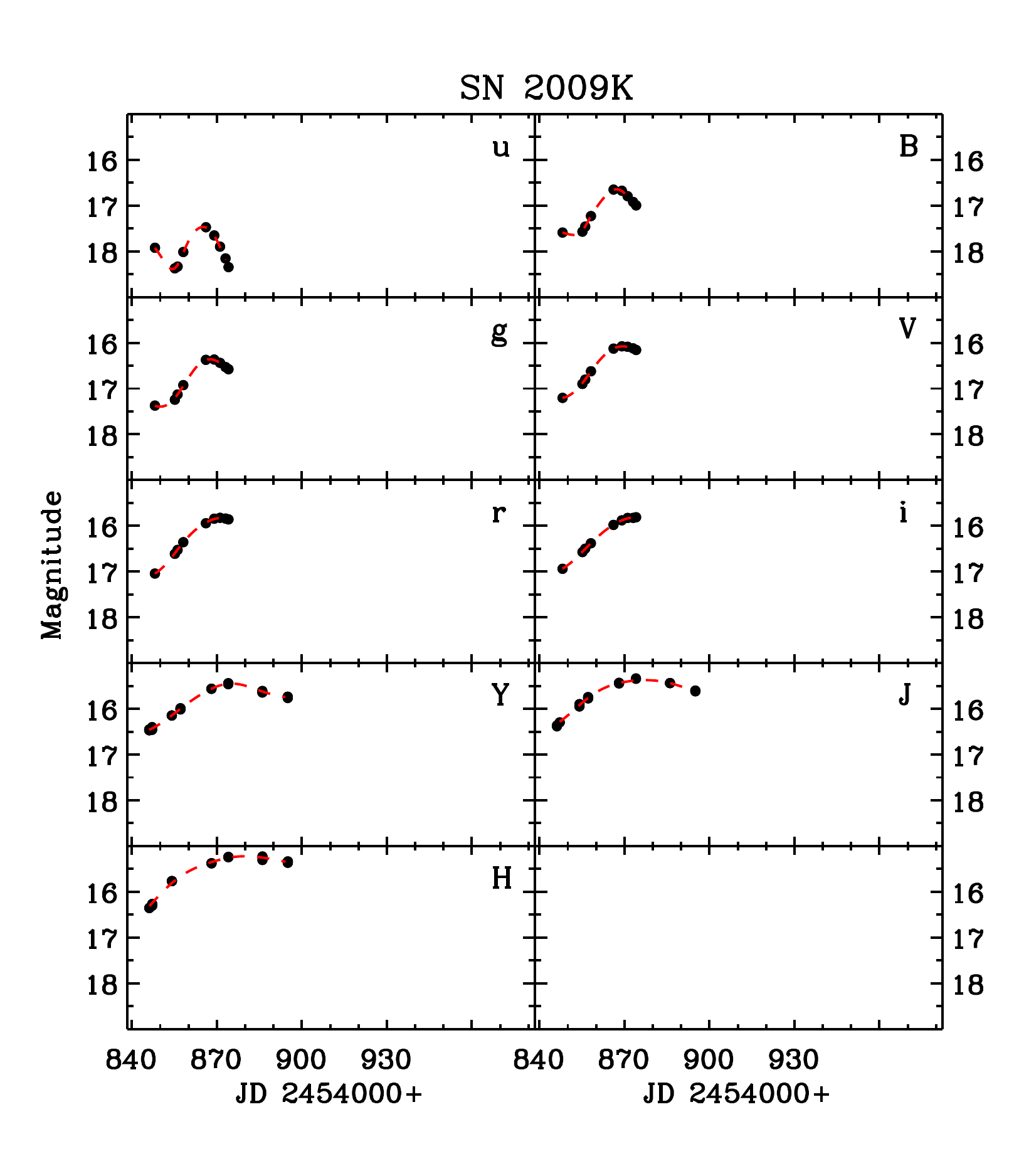}
\includegraphics[width=0.5\textwidth]{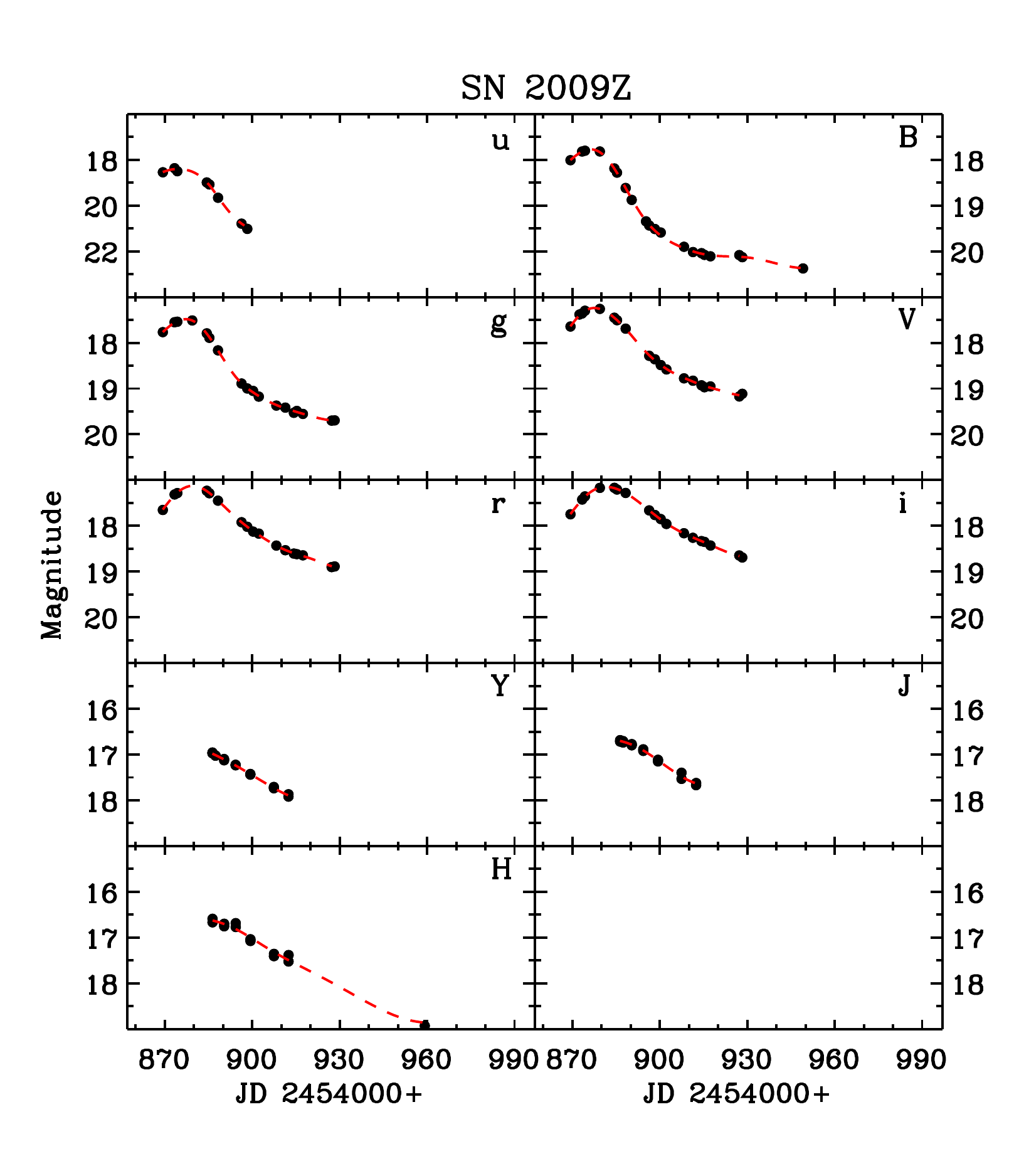}\\
\includegraphics[width=0.5\textwidth]{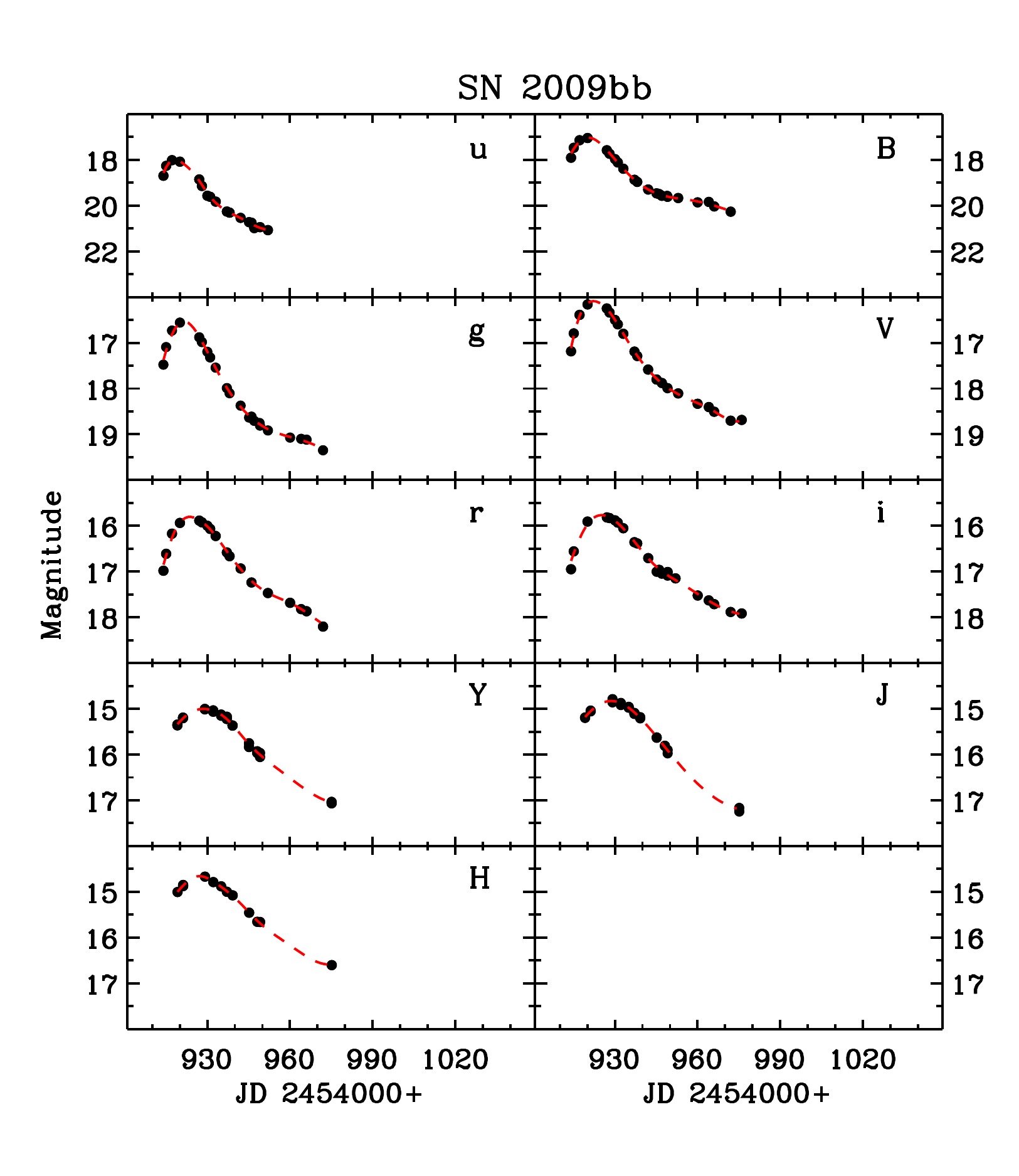}
\includegraphics[width=0.5\textwidth]{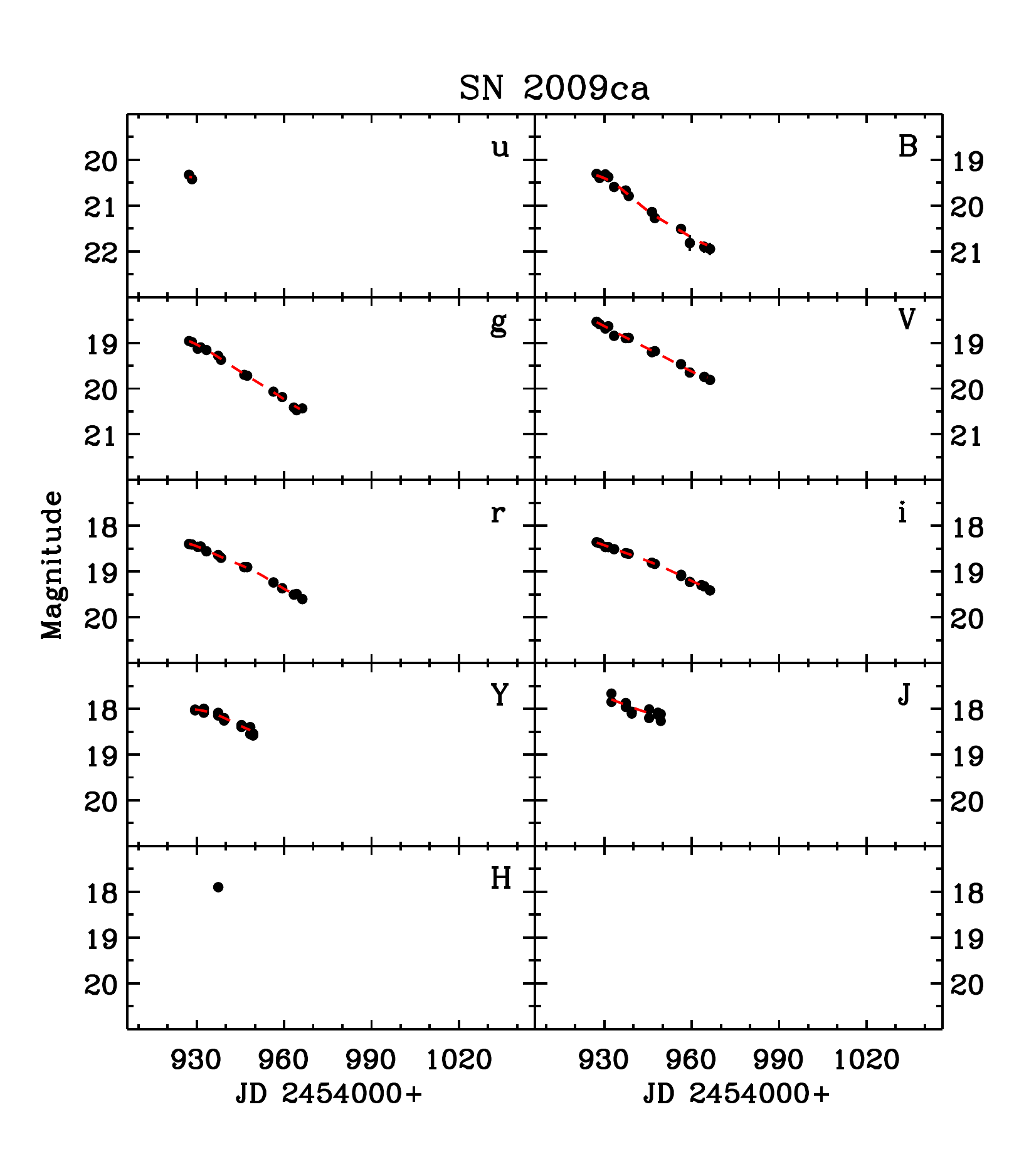}\\
\end{array}$
\end{center}
\caption{Light curves of CSP-I SE~SNe in the 
 Swope natural photometric system.  
 Uncertainties in the photometry are smaller
  than the points, unless shown.  The smooth curves correspond to   
  Gaussian process spline fits computed within the light-curve analysis software package {\tt SNooPy} \citep{burns11}.  
}
\end{figure}
\clearpage

\setcounter{figure}{3}
\begin{figure}
\begin{center}$
\begin{array}{cc}
\includegraphics[width=0.5\textwidth]{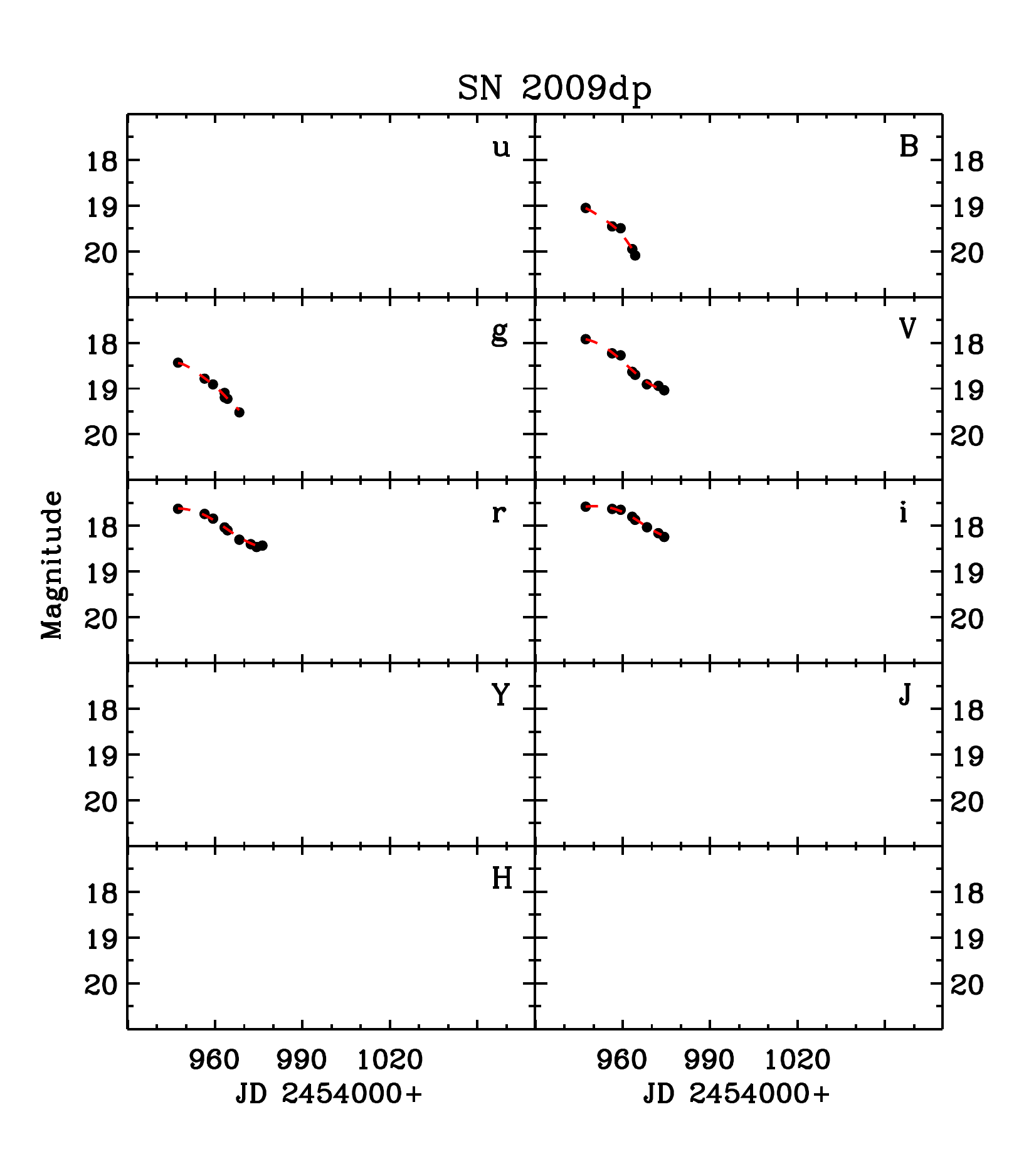}
\includegraphics[width=0.5\textwidth]{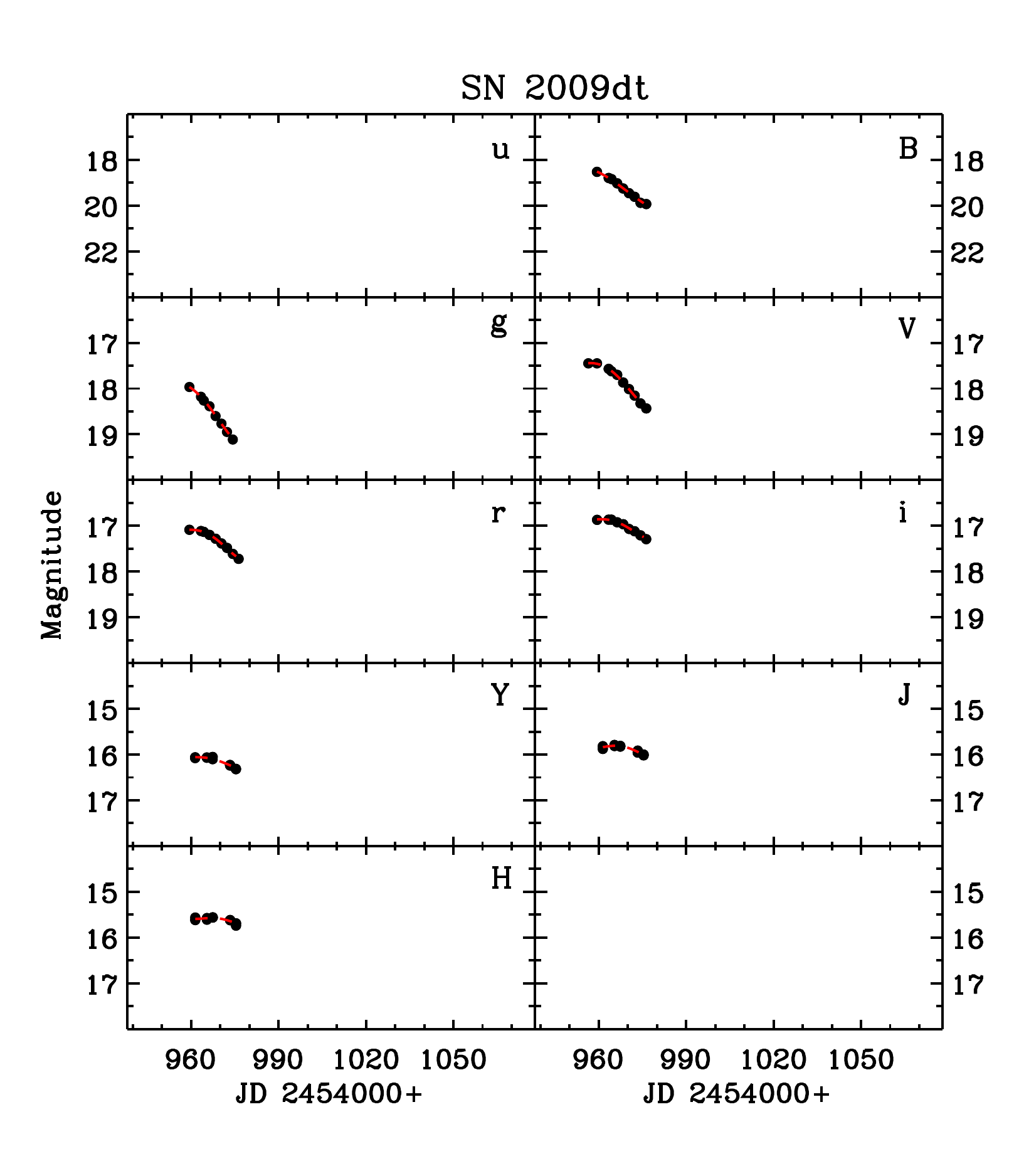}\\
\end{array}$
\end{center}
\caption{Light curves of  CSP-I SE~SNe in the 
 Swope natural photometric system.  
 Uncertainties in the photometry are smaller
  than the points, unless shown.  The smooth curves correspond to 
  Gaussian process spline fits computed within the light-curve analysis software package {\tt SNooPy} \citep{burns11}.  
}
\end{figure}

\clearpage
%

\end{document}